\documentclass[a4paper,12pt]{report}
\usepackage{pdfpages}
\usepackage{amsmath}
\usepackage{amssymb}
\usepackage{setspace}
\usepackage[toc,page]{appendix}
\usepackage{graphicx}
\usepackage{dcolumn}
\usepackage{bm}
\def\comment#1{}

\def\slashchar#1{\setbox0=\hbox{$#1$}           
   \dimen0=\wd0                                 
   \setbox1=\hbox{/} \dimen1=\wd1               
   \ifdim\dimen0>\dimen1                        
      \rlap{\hbox to \dimen0{\hfil/\hfil}}      
      #1                                        
   \else                                        
      \rlap{\hbox to \dimen1{\hfil$#1$\hfil}}   
      /                                         
   \fi}                                         %

\def\pib{{\mbox{\boldmath $\pi$}}}
\def\sigmab{{\mbox{\boldmath $\sigma$}}}

\begin{document}

\title{Introduction to the field theory of classical and quantum phase transitions}
\author{Flavio S. Nogueira}
\date{September 2010}

\maketitle


\pagenumbering{roman}
\tableofcontents


\begin{abstract}
These lecture notes provide a relatively self-contained introduction to field theoretic methods employed 
in the study of classical and 
quantum phase transitions.  
\end{abstract}

\pagenumbering{arabic}

\chapter{Introduction}
\label{ch:intro}

After the work of Wilson and others \cite{Wilson} in the 1970s 
on the renormalization group (RG) in the theory of phase transitions, field theory methods 
originally employed in high-energy physics became an indispensable tool in 
theoretical condensed matter physics. Indeed, since the seminal work of 
Wilson, many excellent textbooks on field theoretic methods in 
condensed matter physics appeared, where a variety of other topics are also discussed; 
see, for example, Refs. \cite{Kleinert-GFCM-1,Fradkin,Uzunov,ZJ,Sachdev,Nagaosa,KSF,Mucio,Wen,Herbut-Book,Schakel-Book}.  
On the other side, the use of field theoretic methods in condensed matter physics led to 
further insights in particle physics. Thus, the study of classical phase transitions in several lattice spin models 
gave birth to lattice gauge theory (for a review, see Ref. \cite{Kogut-RMP}), which provided an approach to numerically 
tackle the strong-coupling regime of gauge theories. This interchange of ideas between these two fields became even more intensive 
after the discovery of high-temperature (high-$T_c$) superconductors, where it became apparent that traditional many-body techniques are 
insufficient to understand the mechanism behind high-$T_c$ superconductivity. For instance, gauge theories, both Abelian and 
non-Abelian were employed to understand several properties of doped Mott insulators \cite{Fradkin,Nagaosa,Wen-RMP}, which eventually may become 
superconducting upon doping.

These notes have the aim of introducing some field theoretic methods widely used in the study of classical and quantum phase transitions. 
Classical phase transitions occur at a regime where quantum fluctuations do not play an important role, usually at high enough temperatures. 
In this case time, as obtained from quantum-mechanical equations of motion, does not play a role. Furthermore, in many situations is possible 
to use a continuum model Hamiltonian, which provides the stage for employing the field theory formalism. Wilson's RG was originally introduced 
in this context \cite{Wilson}. The Hamiltonians are in this case actually determined by a Landau-Ginzburg type of expansion 
\cite{Landau-vol5} where the free energy 
is expanded in powers of the order parameter and its derivatives. The result is similar to Lagrangians of quantum field theories in 
Euclidean space, i.e., a space-time where time is imaginary \cite{ZJ}. It is this fact that allows for the application of field theory in the 
study of phase transitions \cite{Kleinert-GFCM-1,ZJ,KSF,Herbut-Book}. Quantum phases transitions \cite{Sachdev}, on the other hand, occur at zero temperature, 
such that time becomes important. The quantum phase transition is usually driven by some dimensionless parameter, like for example the ratio 
between two couplings appearing in a quantum Hamiltonian. Very often the Lagrangian for a system undergoing a quantum phase transition resembles one 
for a classical phase transition, excepts that one of directions of space is the (imaginary) time. Thus, when this occurs, the same field theoretic 
methods employed in the framework of classical phase transitions can be applied in the quantum phase. Sometimes we say in such a context that a 
quantum system in $d+1$ spacetime dimensions (i.e., $d$ spatial dimensions and one time dimension) is equivalent to a classical system having $d+1$ {\it spatial} 
dimensions. Thus, if we think of a Landau-Ginzburg theory, we see that the effective Lagrangian of the quantum system has to look relativistic. There are 
several non-relativistic condensed matter systems possessing an effective description which is relativistic-like 
\cite{Fradkin,Sachdev,Nagaosa,Wen,Herbut-Book,Schakel-Book}. 
Here ``effective'' means a field theory description valid at some energy scale where the details of the underlying lattice model are irrelevant.  
There are cases, however, where time arises in the effective theory in a way which is not rotational invariant in $d+1$ spacetime dimensions 
\footnote{When we say ``rotational invariant'' here, we are thinking of $d+1$ Euclidean dimensions.}. In those situations, frequency scales 
differently from momenta near the critical point, so they differ in scaling by some power \cite{Sachdev,Mucio}. Thus, we can write the scaling relation 
$\omega\sim|{\bf p}|^z$ at the critical point, which defines the so called dynamic critical exponent, $z$. In this way, we see that a 
relativistic-like theory will have $z=1$, so that the results from classical phase transitions can be applied to the quantum case more directly. 
We will see in these notes mostly examples of theories having $z=1$.         

In these notes priority is given to the introduction of calculational methods. Thus, the reader will usually find here very detailed 
calculations, which are often done step-by-step. 

\chapter{Ferromagnetism}
\label{ch:fm}

\section{Spin in an external magnetic field}

Let us start with a very simple example spin dynamics, namely, the interaction of 
a spin with an external magnetic field. The Hamiltonian is

\begin{equation}
H=-\gamma{\bf S}\cdot{\bf B}.
\end{equation}
The Heisenberg equation of motion for the $i$-th spin 
component yields \footnote{From now on we assume that repeated indices are 
summed over.}

\begin{eqnarray}
{\rm i}\frac{\partial S_i}{\partial t}&=&[S_i,H]\nonumber\\
&=&-\gamma[S_i,S_j]B_j.
\end{eqnarray}
From the quantum mechanical commutation relation, 

\begin{equation}
[S_i,S_j]={\rm i}\epsilon_{ijk}S_k,
\end{equation}
we obtain immediately,

\begin{equation}
\frac{\partial S_i}{\partial t}=-\gamma\epsilon_{ijk}B_jS_k,
\end{equation}
or, in vector notation, 

\begin{equation}
\label{prec-1}
\frac{\partial{\bf S}}{\partial t}=\gamma({\bf S}\times{\bf B}).
\end{equation}
The above equation simply describes the precession of a spin in an external 
magnetic field. 

The above equation can also be obtained from a Lagrangian for a {\it classical} spin. 
The classical spin is given by ${\bf S}=S{\bf n}$, where ${\bf n}^2=1$. In this case 
we have the Lagrangian density,

\begin{equation}
 {\cal L}=S[{\bf {\cal A}}({\bf n})\cdot\partial_t{\bf n}+\gamma{\bf n}\cdot{\bf B}],
\end{equation}
where the vector functional ${\bf {\cal A}}({\bf n})$ has to be determined. 
This functional plays the role 
of a ``momentum'' canonically conjugate to ${\bf n}$, so that 
${\bf {\cal A}}({\bf n})\cdot\partial_t{\bf n}$ would play the role analogous to a term $pdq/dt$ 
in classical mechanics. In this context, we note that the Hamiltonian density 
${\cal H}=-S\gamma{\bf n}\cdot{\bf B}$ does not depend on ${\bf {\cal A}}$, which may appear 
to be a strange feature when compared to the most common situations in classical 
mechanics. However, we should not forget that we have a constraint, ${\bf n}^2=1$, 
which will play an important role in the derivation of the equations of motion. 

The equation of motion for each 
component of the unit vector field ${\bf n}$ is given by the Euler-Lagrange equation:

\begin{equation}
 \partial_t\frac{\partial{\cal L}}{\partial(\partial_t n_i)}-\frac{\partial{\cal L}}{\partial n_i}
=0.
\end{equation}
Thus, we have $\partial{\cal L}/[\partial(\partial_t n_i)]={\cal A}_i({\bf n})$, so that,

\begin{equation}
 \partial_t\frac{\partial{\cal L}}{\partial(\partial_t n_i)}=
S\frac{\partial {\cal A}_i}{\partial n_j}\partial_t n_j.
\end{equation}
We also have,

\begin{equation}
 \frac{\partial{\cal L}}{\partial n_i}=S\left(\frac{\partial {\cal A}_j}{\partial n_i}
\partial_tn_j+\gamma B_i\right).
\end{equation}
Altogether we obtain,

\begin{equation}
\label{prec-A}
 \left(\frac{\partial {\cal A}_i}{\partial n_j}-\frac{\partial {\cal A}_j}{\partial n_i}
\right)\partial_tn_j=\gamma B_i.
\end{equation}
The term between parentheses is an antisymmetric second-rank tensor which is a function 
of the unit vector ${\bf n}$. Therefore, it should have the form, 

\begin{equation}
 \frac{\partial {\cal A}_i}{\partial n_j}-\frac{\partial {\cal A}_j}{\partial n_i}=\epsilon_{ijk}n_k,
\end{equation}
or, equivalently, 

\begin{equation}
\label{mon}
 \epsilon_{ijk}\frac{\partial {\cal A}_k}{\partial n_j}=n_i.
\end{equation}
The LHS of Eq. (\ref{mon}) is the $i$-th component of the curl of ${\bf {\cal A}}$. We can 
interpret ${\bf {\cal A}}$ as a vector potential defined in spin space. It is actually 
called the Berry vector potential, and the corresponding term in the Lagrangian containing 
${\bf {\cal A}}$, when associated to a functional integral representation of spin systems 
\cite{Sachdev}, constitutes the so called Berry phase. We will study the Berry vector 
potential further in a while. For the moment, let us proceed by deriving the equation of 
motion for ${\bf n}$. 

Using Eq. (\ref{mon}) in Eq. (\ref{prec-A}, we obtain easily, 

\begin{equation}
 \epsilon_{ijk}n_k\partial_tn_j=\gamma B_i.
\end{equation}
Contracting both sides with $\epsilon_{lmi}n_m$ yields,

\begin{eqnarray}
 \epsilon_{lmi}\epsilon_{ijk}n_mn_k\partial_tn_j&=&\gamma\epsilon_{lmi}n_mB_i
\nonumber\\
\Longrightarrow(\delta_{lj}\delta_{mk}-\delta_{lm}\delta_{jk})n_mn_k\partial_tn_j&=&
\gamma\epsilon_{lmi}n_mB_i\nonumber\\
\Longrightarrow \partial_t n_l=\gamma({\bf n}\times{\bf B})_l,
\end{eqnarray}
where from the second line to the third we have used both ${\bf n}^2=1$ and 
${\bf n}\cdot\partial_t{\bf n}=(1/2)\partial_t{\bf n}^2=0$. Therefore, we 
have just obtained the desired result, namely, 

\begin{equation}
\label{LL-Bext}
 \partial_t{\bf n}=\gamma({\bf n}\times{\bf B})
\end{equation}
Thus, once more a precessing vector is obtained, except that this time it is not an 
operator that is precessing.

\section{The Landau-Lifshitz equation}

Eq. (\ref{LL-Bext}) is the simplest example of the so called Landau-Lifshitz (LL) equation. 
In order to study the dynamics of ferromagnetic systems, we have to go beyond the situation 
of an external magnetic field. To do this, first note that

\begin{equation}
 B_i=-\frac{1}{S}\frac{\delta H}{\delta n_i},
\end{equation}
where

\begin{equation}
 H=-\gamma S\int d^3r{\bf n}\cdot{\bf B}.
\end{equation}
If $H$ has a more complicate functional dependence on ${\bf n}$, involving 
spatial variations of the unit vector, ${\bf B}$ will be itself a function of 
${\bf n}$, constituting in this way an effective magnetic field. By assuming spatial 
isotropy, such a Hamiltonian is given by

\begin{equation}
\label{H-grad}
 {\cal H}=\int d^3r\left[\frac{JS^2}{2}\partial_i{\bf n}\cdot\partial_i{\bf n}
-\gamma S{\bf n}\cdot{\bf B}\right],
\end{equation}
where $JS^2$, with $J>0$, is the bare spin stiffness. The first term in 
the Hamiltonian above can be motivated via the continuum limit of the 
lattice spin Hamiltonian for a ferromagnet, 

\begin{equation}
 H=-J\sum_{\langle i,j\rangle}{\bf S}_i\cdot{\bf S}_j,
\end{equation}
where the lattice spin fields are assumed to be classical vectors, ${\bf S}_i=S{\bf n}_i$, 
and the lattice sites are summed over nearest neighbors. Indeed, by inserting the lattice 
Fourier transformation

\begin{equation}
 {\bf S}_i=\frac{1}{\sqrt{L}}\sum_{\bf k}e^{i{\bf k}\cdot{\bf R}_i}{\bf S}_{\bf k},
\end{equation}
where $L$ is the number of lattice sites, in the Hamiltonian above, we obtain,

\begin{equation}
 H=\sum_{\bf k}{\cal J}({\bf k})~{\bf S}_{\bf k}\cdot{\bf S}_{-{\bf k}},
\end{equation}
where 

\begin{equation}
{\cal J}({\bf k})=-2J\sum_{\alpha=1}^3\cos k_\alpha,
\end{equation}
and we have set the lattice space equal to unity. In the continuum limit, we have, 

\begin{equation}
 {\cal J}({\bf k})\approx-6J+\frac{J}{2}{\bf k}^2.
\end{equation}
The second term above contributes to the first term in (\ref{H-grad}), while the 
term $-6J$ just adds an irrelevant constant to the Hamiltonian, since 

\begin{equation}
-6J\sum_{\bf k}{\bf S}_{\bf k}\cdot{\bf S}_{-{\bf k}}=-6JS^2\sum_i{\bf n}_i^2=-6JS^2L.
\end{equation}

Let us consider now the LL equation with an effective field determined by the Hamiltonian 
(\ref{H-grad}). The LL becomes, 

\begin{equation}
 \partial_t{\bf n}=SJ({\bf n}\times\nabla^2{\bf n})+\gamma({\bf n}\times{\bf B}).
\end{equation}

Next we set ${\bf B}=0$ and study small transverse fluctuations around the ${\bf e}_3$-axis, i.e., 

\begin{equation}
 {\bf n}=\frac{1}{\sqrt{2}}{\bf e}_3+\delta{\bf n}_\perp,
\end{equation}
where

\begin{equation}
 \delta{\bf n}_\perp=\frac{1}{\sqrt{2}}[\cos({\bf k}\cdot{\bf r}-\omega t){\bf e}_1
+\sin({\bf k}\cdot{\bf r}-\omega t){\bf e}_2].
\end{equation}
This solution describes a so called spin-wave, whose quanta are known as magnons. By inserting 
the above spin-wave profile in the LL equation, we see that it solves it provided 

\begin{equation}
 \omega=\frac{JS^2}{\sqrt{2}}k^2,
\end{equation}
which yields the spin-wave spectrum for a ferromagnet. 

\section{The Berry vector potential, magnetic monopoles in spin space, and 
hedgehogs in real space}

Let us discuss the Berry vector potential further. If we compute the flux of the 
curl (in spin space) of ${\bf {\cal A}}({\bf n})$ through the unit sphere $\bar S_2$ (the 
bar reminds us that $S_2$ is embedded in spin space)  
using Eq. (\ref{mon}), we obtain, 

\begin{eqnarray}
\label{flux-berry}
 \oint_{\bar S_2} d\bar S_i ~\epsilon_{ijk}\frac{\partial {\cal A}_k}{\partial n_j}&=&
\oint_{\bar S_2} d\bar S_i ~n_i\nonumber\\
&=&\int_0^{2\pi}d\xi\int_0^\pi d \eta\left(\frac{\partial{\bf n}}{\partial \xi}
\times\frac{\partial{\bf n}}{\partial \eta}\right)\cdot{\bf n}\nonumber\\
&=&4\pi N,
\end{eqnarray}
where we have used the parametrization 
${\bf n}=(\sin\eta\cos\xi,\sin\eta\sin\xi,\cos\eta)$, and  
$N\in\mathbb{Z}$ is the winding number. 
This result shows that the Berry 
vector potential corresponds to the field strength of a magnetic monopole 
in spin space. 

Note that this result is supposed to be an intrinsic property of the Berry vector potential 
such that should hold beyond the case of a single spin in an external field.   
However, for the case where many spins are involved and have 
a gradient energy, we have a further topological aspect. In fact, when ${\bf n}$ has a nonzero 
gradient, we have the following flux through a sphere $S_2$ {\it in real space} (note the 
absence of the bar over $S_2$ and the differences in the integration measure),  

\begin{eqnarray}
\label{hedgehog-1}
\oint_{S_2}dS_i\epsilon_{ijk}{\bf n}\cdot(\partial_j{\bf n}
\times\partial_k{\bf n})&=&\int_0^{2\pi}d\varphi\int_0^\pi d\theta \epsilon_{ilm}\frac{\partial x_l}{\partial \varphi}
\frac{\partial x_m}{\partial \theta}\epsilon_{ijk}{\bf n}\cdot(\partial_j{\bf n}\times\partial_k{\bf n})\nonumber\\
&=&\int_0^{2\pi}d\varphi\int_0^\pi d\theta(\delta_{lj}\delta_{mk}-\delta_{lm}\delta_{jk})
\frac{\partial x_l}{\partial \varphi}\frac{\partial x_m}{\partial \theta}{\bf n}\cdot(\partial_j{\bf n}\times\partial_k{\bf n})
\nonumber\\
&=&2\int_0^{2\pi}d\varphi\int_0^\pi d\theta{\bf n}\cdot\left(\frac{\partial x_l}{\partial\varphi}\frac{\partial{\bf n}}{\partial x_l}
\times\frac{\partial x_m}{\partial\theta}\frac{\partial{\bf n}}{\partial x_m}\right)\nonumber\\
&=&2\int_0^{2\pi}d\varphi\int_0^\pi d\theta{\bf n}\cdot\left(\frac{\partial{\bf n}}{\partial \varphi}
\times\frac{\partial{\bf n}}{\partial \theta}\right).
\end{eqnarray}
We see that the last line of (\ref{hedgehog-1}) is twice the second line of (\ref{flux-berry}). 
Thus, we have, 

\begin{equation}
\label{hedgehog-2}
 \frac{1}{2}\oint_{S_2}dS_i\epsilon_{ijk}{\bf n}\cdot(\partial_j{\bf n}
\times\partial_k{\bf n})=4\pi N.
\end{equation}
Therefore, we see that there is not only a topological object living in spin space, but 
there is also one living in real (three-dimensional) space, a so called {\it hedgehog}. 
Just like the magnetic monopole in spin space, the hedgehog in real space is also a 
point-like topological object. These topological objects arise also 
in two-dimensional quantum antiferromagnets, which will be discussed later in 
these notes.  

\section{The classical nonlinear $\sigma$ model}
\label{sec:nlsm}

In order to study the static critical behavior of a ferromagnet, we can consider just the 
Hamiltonian (\ref{H-grad}) and try to compute the partition function via the 
functional integral representation, 

\begin{equation}
\label{Z-nlsm}
 Z=\int{\cal D}{\bf n}\delta({\bf n}^2-1)e^{-\frac{1}{T}\int d^3r{\cal H}},
\end{equation}
where the $\delta$ function enforces the constraint. We can rescale the coordinates and 
the temperature such as to have only $1/T$ as the coefficient of the gradient term 
${\cal H}$ and remove the prefactor $JS^2$. We will also generalize the model such that 
it will have $d$ dimensions and the unit vector field will have $n$ components instead of 
three. In this way, we will be able to consider two approximation schemes, namely, one 
where an expansion in $\epsilon=d-2$ is made, and another one where the number of components 
$n$ is large, such that an expansion in $1/n$ can be done. 

\subsection{The expansion in $\epsilon=d-2$}

The main idea behind all $\epsilon$-expansions is to make a perturbative analysis of 
the problem around the dimensionality where the coupling constant of the model is 
dimensionless. In such a case, at the so called critical dimension (i.e., the 
dimension that makes the coupling constant dimensionless), perturbation theory is 
relatively well behaved. The perturbation series is still divergent and 
frequently needs to be resummed, especially when higher orders in perturbation 
theory is involved. However, rigorous mathematical results exist in some cases where 
out of some (initially) perturbation series a non-perturbative expansion may be 
rigorously constructed \cite{Rivasseau}. The situation is considerably more difficult if we are not 
at the critical dimension and the theory approaches the critical point, thus becoming 
strongly coupled. In these cases there are several resummation procedures; for a review on 
these, with focus on variational perturbation theory, see the textbook by Kleinert and 
Schulte-Frohlinde \cite{KSF}. 

For the classical non-linear $\sigma$ model, the coupling constant is the temperature 
$T$. Simple dimensional analysis shows that this coupling becomes dimensionless at 
$d=2$. We will show later that for $d=2$ the model is actually asymptotically free. 
We are ultimately interested at $d=3$, so the goal will be to set $\epsilon=1$ at 
the end of the calculations.  
The case $d=2$ and $n=2$ is special and will be treated 
separately in the next chapter.

The calculations will be performed up to second order in the temperature and will 
serve as a good introduction to the calculation of loop integrals occurring in 
Feynman diagrams. Therefore, we will perform all integrals exactly 
in $d$ dimensions and in great detail below. Simple properties of analytical 
continuation of the dimension in the integrals will be assumed. This is known 
as dimensional regularization. The properties of dimensional regularization are very 
simple and are discussed in several textbooks; see for example Ref. \cite{KSF}    

By resolving the constraint ${\bf n}^2=\sigma^2+\pib^2=1$, the Hamiltonian of the nonlinear $\sigma$ 
model can be written as

\begin{eqnarray}
{\cal H}&=&\frac{1}{2T}\left[(\partial_i\pib)^2+\left(\partial_i\sqrt{1-\pib^2}
\right)^2\right]
\nonumber\\
&=&\frac{1}{2T}\left[(\partial_i\pib)^2+\frac{(\pib\cdot\partial_i\pib)^2}{1-\pib^2}\right],
\end{eqnarray}
where $\pib=(\pi_1,\dots,\pi_{n-1})$.

The Green function can be written as [${\bf n}=(\sigma,\pib)$] 

\begin{eqnarray}
G(x)&=&\langle{\bf n}(x)\cdot{\bf n}(0)\rangle
\nonumber\\
&=&\langle\sigma(x)\sigma(0)\rangle+\langle\pib(x)\cdot\pib(0)\rangle
\nonumber\\
&=&\langle\sqrt{1-\pib^2(x)}\sqrt{1-\pib^2(0)}\rangle
+\langle\pib(x)\cdot\pib(0)\rangle.
\end{eqnarray}
If one rescales $\pib$ by $\sqrt{T}$ and expand up to order $T$, we obtain

\begin{eqnarray}
G(x)&=&1-\frac{T}{2}\langle\pib^2(x)+\pib^2(0)\rangle
+T\langle\pib(x)\cdot\pib(0)\rangle+{\cal O}(T^2)
\nonumber\\
&=&1+(n-1)T\left[G_0(x)-G_0(0)\right]+{\cal O}(T^2),
\end{eqnarray}
where 

\begin{equation}
\label{G0-nlsm}
G_0(x)=\int_p\frac{e^{ip\cdot x}}{p^2},
\end{equation}
and we have used the short-hand notation

\begin{equation}
\label{shorthand}
\int_p\equiv\int\frac{d^dp}{(2\pi)^d}.
\end{equation}
Note that we have used translation invariance to write 
$\langle {\pib}^2(x)\rangle=\langle {\pib}^2(0)\rangle=(n-1)G_0(0)$.

The dimensional regularization rules demand that (see, for example, 
the textbook \cite{KSF} and Appendix \ref{app:int-d-m})

\begin{equation}
\label{rule}
\int_q\frac{1}{|q|^\alpha}=0,~~~~~~~~~~{\rm for}~~~{\rm all}~~~\alpha,
\end{equation}
which immediately leads to $G_0(0)=0$. 

In order to calculate $G_0(x)$ explicitly we use Feynman parameters:

\begin{equation}
G_0(x)=\int_0^\infty d\lambda\int_pe^{ip\cdot x-\lambda p^2}
\end{equation}
By performing the Gaussian integral in $p$ (by just completing 
the squares), we obtain

\begin{equation}
G_0(x)=\frac{1}{(4\pi)^{d/2}}\int_0^\infty d\lambda\lambda^{-d/2}
\exp\left(-\frac{x^2}{4\lambda}\right).
\end{equation}
The substitution $u=x^2/(4\lambda)$ yields

\begin{eqnarray}
\label{G0}
G_0(x)&=&\frac{2^{d-2}}{(4\pi)^{d/2}|x|^{d-2}}\int_0^\infty du
u^{(d-2)/2-1}e^{-u}\nonumber\\
&=&\frac{2^{d-2}\Gamma(d/2-1)}{(4\pi)^{d/2}|x|^{d-2}}=\frac{1}{(d-2)S_d|x|^{d-2}},
\end{eqnarray}
where $S_d=2\pi^{d/2}/\Gamma(d/2)$ is the 
surface of the unit sphere in $d$ dimensions (see Appendix \ref{app:s_d}) and we have made use of the definition of the 
gamma function:

\begin{equation}
\Gamma(z)=\int_0^\infty d\tau\tau^{z-1}e^{-\tau}.
\end{equation}
By performing the substitution $\tau=as$, with $a$ constant, the above 
definition of the gamma function gives another very useful formula 
in calculations with dimensional regularization:

\begin{equation}
\label{ident}
\frac{1}{a^z}=\frac{1}{\Gamma(z)}\int_0^\infty ds s^{z-1}e^{-as}.
\end{equation}
To illustrate the usefulness of this formula, let us calculate two 
functions that will be needed later:

\begin{equation}
\label{Galpha}
G_\alpha(x)=\int_p\frac{e^{ip\cdot x}}{(p^2)^{\alpha/2}}
\end{equation}
and

\begin{equation}
\label{Ialpha}
I_\alpha(p)=\int_q\frac{1}{q^2[(p+q)^2]^{\alpha/2}},
\end{equation}
where $\alpha\in\mathbb{R}$.

Using Eq. (\ref{ident}), we obtain for Eq. (\ref{Galpha}),

\begin{eqnarray}
\label{Galpha1}
G_\alpha(x)&=&\frac{1}{\Gamma(\alpha/2)}\int_0^\infty d\tau
\int_p\tau^{\alpha/2-1}e^{-\tau p^2+ip\cdot x}\nonumber\\
&=&\frac{1}{(4\pi)^{d/2}\Gamma(\alpha/2)}
\int_0^\infty d\tau\tau^{(\alpha-d)/2-1}\exp\left(-\frac{x^2}{4\tau}\right)
\nonumber\\
&=&\frac{2^{d-\alpha}}{(4\pi)^{d/2}\Gamma(\alpha/2)|x|^{d-\alpha}}
\int_0^\infty du u^{(d-\alpha)/2-1}e^{-u}\nonumber\\
&=&\frac{2^{d-\alpha}\Gamma\left(\frac{d-\alpha}{2}\right)}
{(4\pi)^{d/2}\Gamma(\alpha/2)|x|^{d-\alpha}}.
\end{eqnarray}
Note that by setting $\alpha=2$ we recover the expression for $G_0(x)$.  

The calculation of $I_\alpha(p)$ is more involved.  
By using Eq. (\ref{ident}), we can write

\begin{equation}
I_\alpha(p)=\frac{1}{\Gamma(\alpha/2)}\int_0^\infty d\tau_1\int_0^\infty d\tau_2\int\frac{d^dq}{(2\pi)^d}
\tau_2^{\alpha/2-1}e^{-p^2\tau_2-q^2(\tau_1+\tau_2)-2p\cdot q\tau_2}.
\end{equation}
After performing the Gaussian integral in $q$, we obtain

\begin{equation}
I_\alpha(p)=\frac{1}{(4\pi)^{d/2}\Gamma(\alpha/2)}\int_0^\infty d\tau_1\int_0^\infty d\tau_2
\frac{\tau_2^{\alpha/2-1}}{(\tau_1+\tau_2)^{d/2}}\exp\left(-\frac{p^2\tau_1\tau_2}{\tau_1+\tau_2}
\right).
\end{equation}
Now we perform the change of variables

\begin{equation}
\tau_1=\tau\sigma,~~~~~~~~~~~~~\tau_2=(1-\tau)\sigma,
\end{equation}
where $0\leq\tau\leq 1$ and $0\leq\sigma<\infty$. Note that the Jacobian of the transformation is 
$\sigma$. The result is

\begin{equation}
I_\alpha(p)=\frac{1}{(4\pi)^{d/2}\Gamma(\alpha/2)}\int_0^1d\tau\int_0^\infty d\sigma
\frac{e^{-p^2\tau(1-\tau)\sigma}}{\sigma^{(d-\alpha)/2}(1-\tau)^{1-\alpha/2}}.
\end{equation}
After introducing the change of variables $s=p^2\tau(1-\tau)\sigma$, the above integral can be rewritten as

\begin{eqnarray}
\label{Ialpha1}
I_\alpha(p)&=&\frac{|p|^{d-\alpha-2}}{(4\pi)^{d/2}\Gamma(\alpha/2)}
\int_0^1d\tau\frac{\tau^{(d-\alpha-2)/2}}{(1-\tau)^{(4-d)/2}}
\int_0^\infty dse^{-s}s^{(\alpha-d)/2}\nonumber\\
&=&\frac{|p|^{d-\alpha-2}}{(4\pi)^{d/2}\Gamma(\alpha/2)}\Gamma\left(\frac{2+\alpha-d}{2}\right)
\int_0^1d\tau\tau^{(d-\alpha-2)/2}(1-\tau)^{(d-4)/2}\nonumber\\
&=&\frac{|p|^{d-\alpha-2}}{(4\pi)^{d/2}\Gamma(\alpha/2)}\Gamma\left(\frac{2+\alpha-d}{2}\right)
\Gamma\left(\frac{d-2}{2}\right)\frac{\Gamma\left(\frac{d-\alpha}{2}\right)}{
\Gamma(d-1-\alpha/2)},
\end{eqnarray}
where from the second line to third we have used the integral

\begin{equation}
\int_0^1 dxx^a(1-x)^b=\frac{\Gamma(1+a)\Gamma(1+b)}{\Gamma(2+a+b)}.
\end{equation}

We can easily go up to third order in $T$ using the methods illustrated 
in the paper of Amit and Kotliar \cite{Amit-Kotliar}.
Up to second order in $T$, for example, we can proceed as before and 
expand further $\sqrt{1-T{\pib}^2(x)}$ to obtain

\begin{eqnarray}
\label{2nd-0}
 \langle \sqrt{1-T{\pib}^2(x)}\sqrt{1-T{\pib}^2(0)}\rangle
&=&1-(n-1)TG_0(0)-\frac{T^2}{4}\langle {\pib}^2(x){\pib}^2(0)\rangle+\dots
\nonumber\\
&=&1-\frac{T^2}{4}\langle {\pib}^2(x){\pib}^2(0)\rangle+\dots,
\end{eqnarray}
where we have used dimensional regularization to set $G_0(0)=0$. The 
remaing avarage involving four $pi$ fields can be decoupled with the help of 
Wick's theorem, i.e., 

\begin{eqnarray}
\label{2nd}
\langle {\pib}^2(x){\pib}^2(0)\rangle&=&2[\langle\pib(x)\cdot\pib(0)\rangle
-\langle\pib^2(x)\rangle\langle\pib^2(0)\rangle]\nonumber\\
&=&2(n-1)[G_0^2(x)-G_0^2(0)]=2(n-1)G_0^2(x). 
\end{eqnarray}

The third order contribution is more difficult to obtain, since in this case the 
non-linear factor

\begin{equation}
 \exp\left[-\frac{1}{2}\int d^dx\frac{(\pib\cdot\partial_i\pib)^2}{1-T\pib^2}\right]
\end{equation}
in the partition function has to be taken into account when performing the 
correlation function average. By expanding the exponential factor above in 
powers of $T$ we obtain also derivative contributions. The 
latter are more easily handled with in momentum space. Afterwards we can   
transform the result to real space. Thus, we have the following expansion, 

\begin{eqnarray}
G(p)&=&(2\pi)^d\delta^d(p)+T(n-1)/p^2+\frac{1}{2}T^2(n-1)I_b(p)
\nonumber\\
&+&T^3(n-1)\left[\frac{1}{2}(n-1)I_c(p)+I_d(p)
-\frac{1}{4}(n-1)I_e(p)-\frac{1}{2}I_f(p)\right],\nonumber\\
\end{eqnarray}
where \cite{Amit-Kotliar}

\begin{equation}
\label{Ib-0}
I_b(p)=\int_q\frac{1}{q^2(p+q)^2},
\end{equation}

\begin{equation}
I_c(p)=\frac{1}{p^4}\int_q\int_{q_1}\frac{(p+q)^4}{q^2q_1^2(p+q+q_1)^2},
\end{equation}

\begin{equation}
I_d(p)=\frac{1}{p^4}\int_q\int_{q_1}
\frac{(p+q)^2(p+q_1)^2}{q^2q_1^2(p+q+q_1)^2},
\end{equation}

\begin{equation}
I_e(p)=p^2I_b^2(p),
\end{equation}
and

\begin{equation}
I_f(p)=\int_q\int_{q_1}
\frac{(q-q_1)^2}{q^2(p+q)^2q_1^2(p+q_1)^2}.
\end{equation}

Note that the integral $I_b$ appearing in the term of second order in $T$ is just the 
Fourier transform of Eq. (\ref{2nd}), taking into account also Eq. (\ref{2nd-0}). 
However, it is instructive as one more exercise in dimensional regularization to 
perform the integral   
$I_b$. This can be done straightforwardly by considering Eq. (\ref{Ialpha1}) 
with $\alpha=2$ to obtain

\begin{equation}
\label{Ib}
I_b(p)=c(d)|p|^{d-4},
\end{equation}
where

\begin{equation}
\label{c(d)}
c(d)=\frac{\Gamma(2-d/2)\Gamma^2(d/2-1)}{(4\pi)^{d/2}\Gamma(d-2)}.
\end{equation}
Using Eq. (\ref{Galpha1}), we obtain 

\begin{equation}
I_b(x)=G_0^2(x),
\end{equation}
which corresponds to the expected result, in view of   
Eqs. (\ref{2nd-0}) and (\ref{2nd}). 

Integrating over $q_1$ in $I_c$, using the result of $I_b$, we obtain

\begin{equation}
I_c(p)=\frac{c(d)}{p^4}\int_q\frac{1}{q^2[(p+q)^2]^{-d/2}}.
\end{equation}
By using Eq. (\ref{Ialpha1}) with $\alpha=-d$, we obtain

\begin{equation}
I_c(p)=\frac{c(d)\Gamma(1-d)\Gamma(d/2-1)\Gamma(d)}{
(4\pi)^{d/2}\Gamma(-d/2)\Gamma(3d/2-1)}|p|^{2(d-3)}.
\end{equation}
Next we can use Eq. (\ref{Galpha1}) to obtain $I_c(x)$:

\begin{equation}
I_c(x)=\frac{d}{9d-12}G_0^3(x).
\end{equation}

The integral $I_d(p)$ can be written in the form

\begin{equation}
\label{aux-int-Id}
I_d(p)=J_d(p)/p^2+K_d(p)/p^4+p^{-4}\int_q\int_{q_1}\frac{(p+q)^2}{q^2(p+q+q_1)^2},
\end{equation}

\begin{equation}
J_d=\int_q\int_{q_1}\frac{(p+q)^2}{q^2q_1^2(p+q+q_1)^2},
\end{equation}
and

\begin{equation}
K_d=\int_q\int_{q_1}\frac{2p\cdot q_1(p+q)^2}{q^2q_1^2(p+q+q_1)^2}.
\end{equation}
The last integral in Eq. (\ref{aux-int-Id}) vanishes due to 
the rule (\ref{rule}) (by integrating over $q_1$). 

Using (\ref{Ialpha1}) with $\alpha=2$ to integrate over $q_1$, we obtain

\begin{equation}
J_d=c(d)\int_q\frac{1}{q^2[(p+q)^2]^{(2-d)/2}}
\end{equation}
Now we use once more (\ref{Ialpha1}), but this time with $\alpha=2-d$, to 
obtain

\begin{equation}
\tilde J_d(p)\equiv J_d(p)/p^2=
\frac{\Gamma(2-d/2)\Gamma^3(d/2-1)\Gamma(d-2)\Gamma(d-1)}
{(4\pi)^d\Gamma(d-2)\Gamma(1-d/2)\Gamma(3d/2-2)}|p|^{2(d-3)}.
\end{equation}

For $K_d$ we need first to calculate

\begin{equation}
(p+q)_\mu\tilde I=\int_{q_1}\frac{q_{1\mu}}{q_1^2(p+q+q_1)^2},
\end{equation}
i.e.,

\begin{eqnarray}
\label{Itilde}
\tilde I&=&\frac{1}{(p+q)^2}\int_{q_1}\frac{q_{1}\cdot (p+q)}
{q_1^2(p+q+q_1)^2}\nonumber\\
&=&\frac{1}{2(p+q)^2}\int_{q_1}\frac{(p+q+q_1)^2-(p+q)^2-q_1^2}
{q_1^2(p+q+q_1)^2}\nonumber\\
&=&-\frac{1}{2}\int_{q_1}\frac{1}{q_1^2(p+q+q_1)^2}=-\frac{c(d)}{2}
[(p+q)^2]^{(d-4)/2},
\end{eqnarray}
where rule (\ref{rule}) has been used in the first and last terms of 
the second line of the above equation. Using this result in the expression 
for $K_d$, we obtain

\begin{eqnarray}
K_d(p)&=&-c(d)\int_q\frac{p\cdot(p+q)}{q^2[(p+q)^2]^{(2-d)/2}}
\nonumber\\
&=&-p^2J_d-c(d)L_d(p),
\end{eqnarray}
where

\begin{eqnarray}
L_d(p)&=&\int_q\frac{p\cdot q}{q^2[(p+q)^2]^{(2-d)/2}}
\nonumber\\
&=&\frac{1}{2}\int_q\frac{(p+q)^2-p^2-q^2}{q^2[(p+q)^2]^{(2-d)/2}}
\nonumber\\
&=&\frac{p^4I_c(p)}{2c(d)}-\frac{p^2J_d(p)}{2c(d)}.
\end{eqnarray}
At the end, we obtain

\begin{eqnarray}
I_d(x)&=&\frac{1}{2}[\tilde J_d(x)-I_c(x)]\nonumber\\
&=&\frac{1}{3}\frac{d-2}{3d-4}G_0^3(x).
\end{eqnarray} 

Since $I_e(p)=p^2I_b^2(p)$, we have

\begin{equation}
I_e(x)=c^2(d)\int_p\frac{e^{ip\cdot x}}{|p|^{2(3-d)}},
\end{equation}
which can be solved using (\ref{Galpha1}) with $\alpha=2(3-d)$. We have,

\begin{equation}
I_e(x)=\frac{c^2(d)2^{3(d-2)}\Gamma(3d/2-3)}{(4\pi)^{d/2}\Gamma(3-d)|x|^{3(d-2)}}.
\end{equation}

For $I_f(p)$, it is easy to see by expanding $(q-q_1)^2$ that

\begin{equation}
I_f(p)=2\int_q\int_{q_1}\frac{1}{q^2(p+q)^2(p+q_1)^2}
-2\int_q\int_{q_1}\frac{q\cdot q_1}{q^2(p+q)^2q_1^2(p+q_1)^2}.
\end{equation}
The first integral above vanishes because of (\ref{rule}), and we obtain

\begin{equation}
I_f(p)=-2\left[\int_q\frac{q_\mu}{q^2(p+q)^2}\right]^2.
\end{equation}
We can now use (\ref{Itilde}) with $p+q$ replaced by $p$ to obtain

\begin{equation}
I_f(p)=-\frac{c^2(d)}{2}|p|^{2(d-3)}.
\end{equation}
Therefore, 

\begin{equation}
I_f(x)=-\frac{c^2(d)2^{3(d-2)}\Gamma(3d/2-3)}{2(4\pi)^{d/2}\Gamma(3-d)|x|^{3(d-2)}}.
\end{equation}

In the expression for $G(x)$ it is the following combination that appears:

\begin{equation}
\frac{1}{4}(n-1)I_e(x)+\frac{1}{2}I_f(x)
=-\frac{(n-2)}{2}\frac{\cos(\pi d/2)\Gamma(2-d/2)\Gamma(3d/2-3)}{\Gamma(d-2)}G_0^3(x).
\end{equation}

Therefore, 

\begin{eqnarray}
\label{G-d-exp}
G(x)&=&1+T(n-1)G_0(x)+\frac{T^2}{2}(n-1)G_0^2(x)\nonumber\\
&+&(n-1)T^3G_0^3(x)\left[\frac{(n+1)d-4}{6(3d-4)}+
\frac{(n-2)\cos(\pi d/2)\Gamma(2-d/2)\Gamma(3d/2-3)}{2\Gamma(d-2)}\right]
\nonumber\\
&+&{\cal O}(T^4).
\end{eqnarray}

The above is the final expression for $G(x)$ up to third order in $T$. By setting $d=2+\epsilon$ 
and expanding for small $\epsilon$, we obtain exactly the same expression as in the paper of 
Amit and Kotliar \cite{Amit-Kotliar}, i.e., 

\begin{eqnarray}
\label{G-nlsm-ep} 
 G(x)&=& 1+T(n-1)G_0(x)+\frac{T^2}{2}(n-1)G_0^2(x)\nonumber\\
&+&T^3(n-1)\left\{-\frac{n-3}{6}-\frac{(n-2)\epsilon}{6}\right.\nonumber\\
&+&\left.\frac{(n-2)\epsilon^2}{4}+\frac{\epsilon^3}{4}(n-2)\left[\zeta(3)-\frac{3}{2}\right]\right\},
\end{eqnarray}
Note that due to the expression for $G_0(x)$,  
Eq. (\ref{G-nlsm-ep}) contains poles for $\epsilon=0$. These poles can be subtracted by introducing 
two renormalization constants \cite{Stone,Amit-Kotliar}, $Z$ and $Z_t$, such that the renormalized 
correlation function reads

\begin{equation}
 G_r(x;t)=Z^{-1}G(x;T=Z_t\mu^{-\epsilon}t),
\end{equation}
having no poles for $\epsilon=0$. 

The bare correlation function should be independent of the renormalization scale $\mu$, 
which is expressed by the equation,

\begin{equation}
 \mu\frac{dG}{d\mu}=\mu\frac{d}{d\mu}[ZG_r(x,t)]=0,
\end{equation}
or, using the chain rule, 

\begin{equation}
 \left[\mu\frac{\partial}{\partial\mu}+\beta(t)\frac{\partial}{\partial t}
+\zeta(t)\right]G(x;T=Z_t\mu^{-\epsilon}t)=0,
\end{equation}
where the renormalization group (RG) functions $\beta(t)$ and $\zeta(t)$ are 
given by

\begin{equation}
 \beta(t)\equiv\mu\frac{\partial t}{\partial\mu},~~~~~~~~~~~~
\zeta(t)\equiv\mu\frac{\partial\ln Z}{\partial\mu}.
\end{equation}
Using the derived perturbative expansion for the correlation function, we 
obtain,  

\begin{equation}
\label{beta}
 \beta(t)=\epsilon t-(n-2)t^2-(n-2)t^3,
\end{equation} 

\begin{equation}
\label{zeta}
 \zeta(t)=(n-1)t+\frac{3}{4}
(n-1)(n-2)t^3.
\end{equation} 
The vanishing of the $\beta$ function determines the critical temperature as an 
expansion in powers of $\epsilon$,

\begin{equation}
 t_c=\frac{\epsilon}{n-2}\left[1-\frac{\epsilon}{n-2}+\frac{\epsilon^2}{4}\left(\frac{6-n}{n-2}
\right)\right]+{\cal O}(\epsilon^4).
\end{equation}

 In terms of the RG functions above, the exponents $\nu$ and $\eta$ are respectively given by 
\cite{Amit-Kotliar}

\begin{equation}
\label{crexp}
 \nu=-\frac{1}{\beta'(t_c)},~~~~~~~~~~~~~~\eta=\zeta(t_c)-\epsilon,
\end{equation} 
corresponding to the critical behavior

\begin{equation}
 \xi\sim (t-t_c)^{-\nu},
\end{equation} 
where $\xi$ is the correlation length, and 
\begin{equation}
 G(x)\sim\frac{1}{|x|^{d-2+\eta}}.
\end{equation} 

The correspondence between the RG functions and the critical exponents follows by making 
dimensionless variables in $G(x)$ explicit. For instance, we can write, 

\begin{equation}
 G(x)\sim\left(\frac{\mu}{\Lambda}\right)^{\epsilon+\eta}(\mu|x|)^{-\epsilon-\eta},
\end{equation}
where $\Lambda$ is the ultraviolet cutoff. Thus, we have the behavior, 

\begin{equation}
 Z\sim\left(\frac{\mu}{\Lambda}\right)^{-(\epsilon+\eta)}.
\end{equation}
Therefore, near the critical point we have, 

\begin{equation}
 \mu\frac{\partial \ln Z}{\partial\mu}\approx -\epsilon-\eta.
\end{equation}
Comparison with Eq. (\ref{zeta}) yields the second of Eqs. (\ref{crexp}). 

For the critical exponent $\nu$, we linearize the $\beta$ function near the 
critical point $t_c$, i.e., 

\begin{equation}
 \mu\frac{\partial(t-t_c)}{\partial\mu}\approx-\beta'(t_c)(t-t_c),
\end{equation}
where the prime denote a derivative with respect to $t$ and the minus sign 
reflects the negative slope of the $\beta$ function at $t_c$ (ultraviolet 
stability of the fixed point). Integrating the above equation yields

\begin{equation}
 \frac{\mu}{\Lambda}\sim(t-t_c)^{-1/\beta'(t_c)},
\end{equation}
which upon the identification $\mu=\xi^{-1}$ leads to the determination of 
the critical exponent $\nu$. 

Explicitly, we have

\begin{equation}
 \frac{1}{\nu}=\epsilon+\frac{\epsilon^2}{n-2}+{\cal O}(\epsilon^3),
\end{equation} 

\begin{equation}
\label{eta-nlsm}
 \eta=\frac{\epsilon}{n-2}\left[1-\left(\frac{n-1}{n-2}\right)\epsilon+\frac{n(n-1)}{2(n-2)^2}\epsilon^2
\right]+{\cal O}(\epsilon^4).
\end{equation} 
Note that the calculations allow to obtain $\eta$ with one order of $\epsilon$ higher than for $1/\nu$. 

\subsection{The $1/n$ expansion}

\subsubsection{The saddle-point approximation}

A common nonperturbative approach to study the non-linear $\sigma$ is the large $n$ limit \cite{Sachdev,ZJ}, where $n$ is taken to be large while 
$ng$ is kept fixed. Such an analysis is more easily done if the constraint is implemented with the help of a Lagrange multiplier field, such 
that the Lagrangian becomes

\begin{equation}
 {\cal L}=\frac{1}{2g}[(\partial_i{\bf n})^2+i\lambda({\bf n}^2-1)].
\end{equation}
This Lagrange multiplier field arises from the partition function (\ref{Z-nlsm}) when the 
functional integral representation of the delta function is used:

\begin{equation}
 \delta({\bf n}^2-1)=\int{\cal D}\lambda e^{-\frac{1}{2T}\int d^dxi\lambda({\bf n}^2-1)}.
\end{equation}

Let us integrate $n-1$ components of ${\bf n}$ exactly and call the non-integrated one $\sigma$. The resulting effective action reads

\begin{equation}
\label{Seff-nlsm-large-n}
 S_{\rm eff}=\frac{(n-1)}{2}{\rm Tr}\ln(-\nabla^2+i\lambda)+\frac{1}{2T}\int d^dx[\sigma(-\nabla^2+i\lambda)\sigma-i\lambda].
\end{equation}
The limit $n\to\infty$ is obtained from the saddle-point approximation to the above effective action. We will consider a uniform 
saddle-point with $i\lambda=m^2$ and $\sigma=s$. Thus, from

\begin{equation}
 \frac{\partial S_{\rm eff}}{\partial m^2}=0,~~~~~~~~~~~~~~\frac{\partial S_{\rm eff}}{\partial s}=0,
\end{equation}
we obtain,

\begin{equation}
 s^2=1-nT\int\frac{d^dp}{(2\pi)^d}\frac{1}{p^2+m^2},
\end{equation}

\begin{equation}
 m^2s=0.
\end{equation}
For $T<T_c$ we have $s\neq 0$, so that $m^2=0$, leading to

\begin{equation}
 s^2=1-nT\int\frac{d^dp}{(2\pi)^d}\frac{1}{p^2}.
\end{equation}
At the critical point, $T=T_c$, we have that $s=0$, so that

\begin{equation}
\label{gc}
 \frac{1}{nT_c}=\int\frac{d^dp}{(2\pi)^d}\frac{1}{p^2}=\frac{S_d\Lambda^{d-2}}{(2\pi)^d(d-2)},
\end{equation}
where $\Lambda$ is the ultraviolet cutoff. 
Therefore,

\begin{equation}
 s^2=\frac{1}{T_c}(T_c-T),
\end{equation}
which implies that the critical exponent of the order parameter is $\beta=1/2$.

For $T>T_c$, on the other hand, we have $s=0$ and $m^2\neq 0$, such that

\begin{equation}
\label{gap-m2}
 \int\frac{d^dp}{(2\pi)^d}\frac{1}{p^2+m^2}=\frac{1}{nT}.
\end{equation}
If we now write

\begin{equation}
 1=nT\left(\int\frac{d^dp}{(2\pi)^d}\frac{1}{p^2+m^2}-\int\frac{d^dp}{(2\pi)^d}\frac{1}{p^2}+\int\frac{d^dp}{(2\pi)^d}\frac{1}{p^2}\right),
\end{equation}
we obtain,

\begin{equation}
 \frac{T-T_c}{T_c}=nTm^2\int\frac{d^dp}{(2\pi)^d}\frac{1}{p^2(p^2+m^2)}.
\end{equation}
The integral above can be evaluated with help of the integral $I_1$ of Appendix \ref{app:int-d-m}. Indeed, it is proportional to the 
integral $I_1$ evaluated in $d-2$ dimensions, i.e., 

\begin{equation}
\int\frac{d^dp}{(2\pi)^d}\frac{1}{p^2(p^2+m^2)}=\frac{1}{2\pi(d-2)}I_1(d-2)=\frac{2^{1-d}\pi^{-d/2}m^{d-4}}{d-2}\Gamma\left(2-\frac{d}{2}\right).
\end{equation}
Thus, 

\begin{equation}
 \frac{2^{1-d}\pi^{-d/2}}{d-2}\Gamma\left(2-\frac{d}{2}\right)nTm^{d-2}=\frac{T-T_c}{T_c},
\end{equation}
or, after substituting the value of $T_c$, Eq. (\ref{gc}), 

\begin{equation}
\label{m-nlsm}
 m=\Lambda\left\{\frac{\pi^{d/2-2}}{\Gamma(d/2)\Gamma(2-d/2)}
\left[1-\frac{(2\pi)^d(d-2)\Lambda^{2-d}}{S_dnT}\right]\right\}^{1/(d-2)}
\end{equation}
It turns out that the mass gap $m=\xi^{-1}$, where $\xi$ is the correlation length. Therefore, the corresponding critical exponent is

\begin{equation}
 \nu=\frac{1}{d-2}.
\end{equation}

The limit $d\to 2$ of Eq. (\ref{m-nlsm}) yields \footnote{We are making use of the well-known 
result $e^x=\lim_{m\to\infty}(1+x/m)^m$.}

\begin{equation}
 m=\frac{\Lambda}{\pi}\exp\left(-\frac{2\pi}{nT}\right),
\end{equation}
which implies that for $d=2$ the system is gapped for all $T>0$, i.e., no phase 
transition occurs in this case. This is consistent with the Mermin-Wagner theorem 
\cite{MW}, which states that a continuous symmetry  
associated to quadratic dispersions cannot be broken in $d=2$. Note that the 
Mermin-Wagner theorem rules out the symmetry breaking at $d=2$, but not necessarily the 
phase transition. Indeed, we will see in the next chapter that $d=2$ and $n=2$ is special. 
In fact, although no symmetry breaking occurs in this case, a phase transition happens 
even in its absence. 

The $\beta$ function for the dimensionless coupling $t=\Lambda^{d-2}T$ 
at large $n$ is easily obtained by demanding the scale invariance of the mass gap, i.e., 

\begin{equation}
 \Lambda\frac{\partial m}{\partial \Lambda}=0,
\end{equation}
which yields, 

\begin{equation}
 \Lambda\frac{\partial t}{\partial \Lambda}=(d-2)t-\frac{S_d}{(2\pi)^d}nt^2, 
\end{equation}
and we see that for $d=2$ the theory is asymptotically free. 

\subsubsection{$1/n$ corrections to the saddle-point}

\begin{figure}
\begin{center}
\includegraphics[width=10cm]{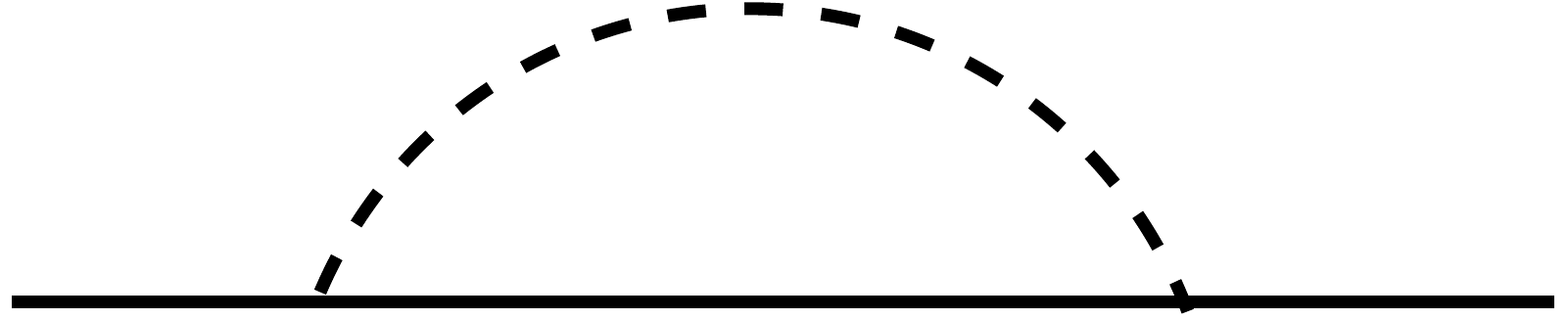}
\caption{Feynman diagram representing the $1/n$ correction to $\sigma$-propagator. The internal 
solid line corresponds to the free $\sigma$-propagator, while the dashed line is 
the $\lambda$-propagator.}\label{fig:diagram} 
\end{center}
\end{figure}

Now we want to compute the $1/n$ corrections to the saddle-point calculation. Let us consider for instance the 
theory at the critical temperature and find the anomalous dimension of the $\sigma$-field. This is achieved 
by first expanding the effective action (\ref{Seff-nlsm-large-n}) up to quadratic order in $\lambda$, such as 
to obtain the $\lambda$-propagator. This yields 

\begin{equation}
\label{Seff-1/n}
 S_{\rm eff}\approx \frac{n}{4}\int d^dx \int d^dx' G_0(x-x')G_0(x'-x)\lambda(x)\lambda(x')
+\frac{1}{2T}\int d^dx\sigma(-\nabla^2+i\lambda)\sigma,
\end{equation}
where $G_0(x)$ is given in Eq. (\ref{G0-nlsm}). Thus, it is easy to obtain the correlation function for 
the field $\lambda$ in momentum space as

\begin{equation}
 \langle\lambda(p)\lambda(-p)\rangle=\frac{(2/n)}{\int\frac{d^dq}{(2\pi)^d}\frac{1}{q^2(p+q)^2}},
\end{equation}
or, using the result (\ref{Ib}), 

\begin{equation}
 \langle\lambda(p)\lambda(-p)\rangle=\frac{2}{c(d)n|p|^{d-4}},
\end{equation}
where $c(d)$ is given in Eq. (\ref{c(d)}). In order to find the anomalous dimension of the $\sigma$-field we need to compute 
the $\sigma$-propagator up to order $1/n$. This is done by taking into account the vertex $i\lambda\sigma^2/(2T)$ in 
Eq. (\ref{Seff-1/n}). The lowest order contribution to the $\sigma$-propagator features two vertices. The corresponding 
Feynman diagram is shown in Fig. \ref{fig:diagram}. Thus, 

\begin{eqnarray}
\label{prop-sigma-1}
 \langle\sigma(p)\sigma(-p)\rangle^{-1}&=&p^2+\int\frac{d^dq}{(2\pi)^d}\frac{\langle\lambda(p+q)\lambda(-p-q)\rangle}{q^2}
\nonumber\\
&=&p^2+\frac{2}{nc(d)}\int\frac{d^dq}{(2\pi)^d}\frac{1}{|p+q|^{d-4}q^2}.
\end{eqnarray}
The above result contains an integral which can be evaluated with the help of the integral (\ref{Ialpha}) along with its 
explicit evaluation in Eq. (\ref{Ialpha1}). However, for $\alpha=d-4$, like in Eq. (\ref{prop-sigma-1}), Eq. (\ref{Ialpha1}) has a 
pole, which is actually related to a logarithmic behavior. The trick to complete the calculation is to replace $d-4$ in 
Eq. (\ref{prop-sigma-1}) by $\alpha$, evaluate the integral explicitly via Eq. (\ref{Ialpha1}), and perform an expansion in 
powers of $d-4-\alpha$, setting $\alpha=d-4$ at the end.  

The dangerous contribution in Eq. (\ref{Ialpha1}) comes from the factor $\Gamma((2+\alpha-d)/2)$, whose singularity for 
$\alpha=d-4$ can be isolated as

\begin{equation}
 \Gamma\left(\frac{2+\alpha-d}{2}\right)=\frac{2}{d-4-\alpha}+({\rm regular~terms}).
\end{equation}
Furthermore, the result of the integration is proportional to $|p|^{d-\alpha-2}$, which can be expanded as

\begin{equation}
 \Lambda^{d-4-\alpha}p^2 e^{(d-4-\alpha)\ln(p/\Lambda)}\approx\Lambda^{d-4-\alpha}p^2[1+(d-4-\alpha)\ln(p/\Lambda)+\dots].
\end{equation}
Now we note that the anomalous dimension is given by the infrared behavior of the propagator, i.e., 

\begin{equation}
  \langle\sigma(p)\sigma(-p)\rangle\sim\frac{1}{p^{2-\eta}}.
\end{equation}
Thus, before setting $\alpha=d-4$, we neglect the terms which are smalller in comparison with $p^{2-\eta}$ as $p\to 0$, anticipating 
already that $\eta>0$. Thus, by keeping only the dominant terms in the infrared, we can safely set $\alpha=d-4$ to obtain

\begin{equation}
 \langle\sigma(p)\sigma(-p)\rangle^{-1}\approx p^2\left[1+\frac{2(d-4)\Gamma(d-2)}{n\Gamma(2-d/2)\Gamma(d/2+1)\Gamma^2(d/2-1)}\ln\left(\frac{p}{\Lambda}
\right)\right],
\end{equation}
which should be compared to

\begin{equation}
 p^{2-\eta}\approx p^2[1-\eta\ln(p/\Lambda)],
\end{equation}
to obtain, after some simplifications, 

\begin{equation}
\label{eta-nlsm-large-n}
 \eta=\frac{(4-d)(d-2)^2\Gamma(d-2)}{nd\Gamma(2-d/2)\Gamma^3(d/2)}.
\end{equation}
For $d=3$ this yields  

\begin{equation}
 \eta=\frac{8}{3\pi^2n}.
\end{equation}
The above result for $\eta$ is the same as obtained for the $O(n)$ Landau-Ginzburg model at large $n$. Indeed, both models have 
in the framework of a $1/n$ expansion the same critical exponents, belonging henceforth to the same universality class  
\cite{ZJ}. 

In order to see the non-perturbative character of Eq. (\ref{eta-nlsm-large-n}), let us compare this expression with 
the perturbative expansion in powers of $\epsilon=d-2$ in Eq. (\ref{eta-nlsm}) for $n$ large. If we expand Eq. (\ref{eta-nlsm-large-n}) in 
powers of $\epsilon$, we obtain

\begin{equation}
 \eta=\frac{\epsilon}{n}\left(1-\epsilon+\frac{\epsilon^2}{2}+\dots\right),
\end{equation}
which is precisely Eq. (\ref{eta-nlsm}) in the limit $n\gg 1$. The large $n$ result agrees with the large $n$ regime of the 
$\epsilon$-expansion even up to order four. Indeed, we have that up to this order the anomalous dimension is given by 
(see for example Ref. \cite{ZJ} and references therein)

\begin{eqnarray}
\label{eta-nlsm-hot}
 \eta&=&\tilde \epsilon+(n-1)\tilde \epsilon^2\left\{-1+\frac{n}{2}\tilde \epsilon\right.
\nonumber\\
&+&\left.\left[-b+(n-2)\left(\frac{2-n}{3}+\frac{3-n}{4}\zeta(3)\right)\right]\tilde \epsilon^2\right\}+{\cal O}(\epsilon^5),
\end{eqnarray}
where we have defined

\begin{equation}
 \tilde \epsilon=\frac{\epsilon}{n-2}, 
\end{equation}
and 

\begin{equation}
 b=-\frac{1}{12}(n^2-22n+34)+\frac{3}{2}\zeta(3)(n-3). 
\end{equation}
In the limit $n\gg 1$ Eq. (\ref{eta-nlsm-hot}) becomes

\begin{equation}
 \eta=\frac{\epsilon}{n}\left(1-\epsilon+\frac{\epsilon^2}{2}-\frac{1+\zeta(3)}{4}\epsilon^3\right)+{\cal O}(\epsilon^5),
\end{equation}
which agrees precisely with the expansion of Eq. (\ref{eta-nlsm-large-n}) in powers of $\epsilon$ up to the same order. This comparison 
clearly exhibits the non-perturbative character of the $1/n$ expansion, since the lowest non-trivial $1/n$ correction to $\eta$ is 
already able to reproduce the large $n$ limit of perturbation theory up to fourth order.

\chapter{The Kosterlitz-Thouless phase transition}
\label{ch:KT}

\section{The XY model}

When $n=2$, the local constraint ${\bf n}_i^2=1$ in the classical Heisenberg model is solved by writing 

\begin{equation}
 {\bf n}_i=(\cos\theta_i,\sin\theta_i), 
\end{equation}
such that the Heisenberg Hamiltonian becomes

\begin{equation}
\label{XY-lattice}
 H=-JS^2\sum_{\langle i,j\rangle}\cos(\theta_i-\theta_j).
\end{equation}
Such a planar magnetic system is known as XY model. As we will see in a later Chapter, this model is in the same universality class as a 
superfluid. The reason is not difficult to understand, since universality classes are usually determined by the underlying symmetry group, 
which in the present case is $O(2)$, and this is the same as the unitary group $U(1)$. The spontaneous symmetry breaking of the 
$U(1)$ group will be studied in several ways for $d>2$ in the next Chapters. Here we will explore the case $n=2$ and $d=2$, where 
no spontaneous symmetry breaking occurs, as we have 
already mentioned in the last Chapter. While for $n>2$ no phase transition can occur at $d=2$, the situation is different for $n=2$. When 
$n=2$ and $d=2$ a phase transition occurs {\it without} spontaneous symmetry breaking. 

\section{Spin-wave theory}

Let us consider the classical non-linear $\sigma$ model for $n=2$. By writing ${\bf n}=(\cos\theta,\sin\theta)$, we obtain,

\begin{equation}
 {\cal H}=\frac{1}{2T}(\nabla\theta)^2. 
\end{equation}
The above equation is just the continuum limit of the XY model in Eq. (\ref{XY-lattice}), up to a trivial renaming of the couplings. 

It is easy to see that the correlation function $G(x)=\langle{\bf n}(x)\cdot{\bf n}(0)\rangle$ of the non-linear $\sigma$ model can be 
written in this case as \footnote{Note that translation invariance implies $\langle\cos\theta(x)\sin\theta(0)\rangle=\langle\sin\theta(x)\cos\theta(0)\rangle$.}

\begin{equation}
 G(x)=\langle e^{i[\theta(x)-\theta(0)]}\rangle.
\end{equation}
More explicitly, 

\begin{equation}
 G(x)=\frac{1}{Z}\int{\cal D}\theta e^{-\int d^dx'\left[\frac{1}{2T}(\nabla'\theta)^2-J(x')\theta(x')\right]},
\end{equation}
where $Z$ is the partition function and 

\begin{equation}
 J(x')=i[\delta^d(x-x')-\delta^d(x')].
\end{equation}
Note that we are not setting $d=2$ yet. We will see soon why it is convenient to do so. 
We can perform the Gaussian integral over $\theta$ exactly to obtain

\begin{eqnarray}
 G(x)&=&\exp\left[\frac{T}{2}\int d^dx'\int d^dx''J(x'){\cal G}(x'-x'')J(x'')\right]
\nonumber\\
&=&\exp\left\{T\left[{\cal G}(x)-{\cal G}(0)\right]\right\},
\end{eqnarray}
where

\begin{equation}
 {\cal G}(x)=\int\frac{d^dp}{(2\pi)^d}\frac{e^{ip\cdot x}}{p^2}.
\end{equation}
We are not going to use dimensional regularization in this Chapter. It is actually essential to us not to have ${\cal G}(0)=0$. This is 
important, in order to take the limit $d\to 2$ in a clean way. Thus, by evaluating  ${\cal G}(0)$ explicitly using a cutoff, we obtain,

\begin{equation}
 {\cal G}(0)=\frac{S_d\Lambda^{d-2}}{(2\pi)^d(d-2)}.
\end{equation}

From Eq. (\ref{G0}), we obtain

\begin{equation}
 {\cal G}(x)=\frac{|x|^{2-d}}{S_d(d-2)}.
\end{equation}
Thus, 

\begin{equation}
\label{g-g0}
 {\cal G}(x)-{\cal G}(0)=\frac{S_d\Lambda^{d-2}}{(2\pi)^d(d-2)}\left[\frac{(2\pi)^d}{S_d^2}(\Lambda|x|)^{2-d}-1\right],
\end{equation}
which has a straightforward limit $d\to 2$, 

\begin{equation}
 {\cal G}(x)-{\cal G}(0)=-\frac{1}{2\pi}\ln(\Lambda|x|).
\end{equation}
Therefore, we have 

\begin{equation}
\label{G-2-sw}
 G(x)=\frac{1}{(\Lambda|x|)^{\eta(T)}},
\end{equation}
where 

\begin{equation}
 \eta(T)=\frac{T}{2\pi},
\end{equation}
is the anomalous dimension of the theory. Interestingly, in contrast with the case $d>2$, the correlation function for $n=2$ and $d=2$ features a 
temperature-dependent anomalous dimension. This is a particularity of the two-dimensional XY model, as shown first by Kosterlitz and Thouless \cite{KT}. 

The above result is exact within spin-wave theory, which corresponds to a situation where the fact that $\theta$ is an angle is not taken into account. 
The periodicity of $\theta$ is, however, crucial for characterizing the phase structure of the model. This will be the subject of the next Section. 
 
Before closing this Section, let us point out the agreement between the perturbation theory for the non-linear $\sigma$ model for $n=2$ and the 
results of this Section. Indeed, if we set $n=2$ in Eq. (\ref{G-d-exp}), we obtain, 

\begin{equation}
 G(x)=1+TG_0(x)+\frac{T^2}{2}G_0(x)+\frac{T^3}{6}G_0^3(x)+{\cal O}(T^4).
\end{equation}
The above expansion contains the first terms of the expansion of $\exp[TG_0(x)]$. Note that $G_0(x)$ is the same as ${\cal G}(x)$ in this Section. 
Furthermore, since we are not using dimensional regularization, we must make the replacement $G_0(x)\to {\cal G}(x)-{\cal G}(0)$. 

Another point worth mentioning in the context of the perturbation theory of the previous chapter is that for $n=2$ and $d=2$ the $\beta$ function 
(\ref{beta}) vanishes, i.e., 

\begin{equation}
 \mu\frac{\partial t}{\partial\mu}=0.
\end{equation}
Thus, for $n=2$ and $d=2$ perturbation theory is just reproducing the result of spin-wave theory, i.e., a free theory.  

\section{Two-dimensional vortices}

We have already mentioned in the previous Section that the spin-wave theory neglects the periodicity of $\theta$. Thus, the spin-wave 
analysis implies

\begin{equation}
 \oint_C dx\cdot\nabla\theta=0.
\end{equation}
The above is not always true in the case of a periodic field \cite{KT,Kleinert-GFCM-1,Kleinert-MVF}. This is precisely the case if we 
interpret $\nabla\theta$ as the superfluid velocity \footnote{The superfluid velocity is given in terms of the phase of the order 
parameter as \cite{HM} ${\bf v}_s=(\hbar/m)\nabla\theta$.}. In the presence of vortices, the circulation of the superfluid velocity is 
quantized, 

\begin{equation}
\label{BS}
 \oint_C dx\cdot\nabla\theta=2\pi n,
\end{equation}
where $n\in\mathbb{Z}$ is the winding the number counting how many times the closed curve $C$ goes around the vortex. The equation above 
is just the Bohr-Sommerfeld quantization condition applied to a superfluid. 

In two dimensions vortices are just points. Thus, the problem of studying the statistical mechanics of many vortices in two dimensions is 
much easier as in three dimensions. In three dimensions vortices are either infinite lines or loops. The statitiscal mechanics of vortex 
loops is much more complicated and requires often the use of special duality techniques involving 
multivalued fields \cite{Kleinert-GFCM-1,Kleinert-MVF}.

Let us give a concrete example of a single vortex at the origin in two dimensions. Such a vortex has to be singular at $x=(x_1,x_2)=(0,0)$. Such 
a vortex configuration is simply given by

\begin{equation}
 \nabla\theta=\frac{1}{x^2}(-x_2,x_1).
\end{equation}
Thus, 

\begin{equation}
 dx\cdot\nabla\theta=\frac{x_1dx_2-x_2dx_1}{x^2}.
\end{equation}
Using polar coordinates, 

\begin{equation}
 x_1=r\cos\theta,~~~~~~~~~~x_2=r\sin\theta,
\end{equation}
where $r=\sqrt{x_1^2+x_2^2}$, we obtain simply, 

\begin{equation}
 \oint_C dx\cdot\nabla\theta=\int_0^{2\pi}d\theta=2\pi, 
\end{equation}
which is just Eq. (\ref{BS}) with $n=1$. It is easy to see that the $\theta$ corresponding to such a singular configuration is given by

\begin{equation}
 \theta=\arctan\left(\frac{x_2}{x_1}\right).
\end{equation}

We are of course interested in a configuration involving many vortices. We will consider only vortices with vorticity $n=\pm 1$, which are energetically 
more favorable. A many-vortex configuration can be easily constructed. We have, 

\begin{equation}
 \nabla\theta_V=\sum_i\frac{q_i}{(x-x_i)^2}[-(x-x_i)_2{\bf e}_1+(x-x_i)_1{\bf e}_2],
\end{equation}
where $x_i$ is the position of the $i$-th vortex and $q_i=\pm 1$. Note that $\nabla\theta_V$ looks like an electrostatic field 
caused by a potential $\theta_V$ in two dimensions. In this analogy, the vorticities $q_i$ play the role of point charges in 
two dimensions. Thus, we are going to study the statistical mechanics of a two-dimensional Coulomb gas. Since 

\begin{equation}
 \nabla\ln(\Lambda|x-x_0|)=\frac{x-x_0}{(x-x_0)^2},
\end{equation}
we have that the vortex contribution of the Hamiltonian is given by

\begin{eqnarray}
 H_V&=&\frac{1}{2T}\int d^2x(\nabla\theta_V)^2
\nonumber\\
&=&\frac{1}{2T}\sum_{i,j}q_iq_j\int d^2x\nabla\ln(\Lambda|x-x_i|)\cdot\nabla\ln(\Lambda|x-x_j|).
\end{eqnarray}
A partial integration yields

\begin{equation}
 H_V=\frac{1}{2T}\sum_{i,j}q_iq_j\int d^2x\ln(\Lambda|x-x_i|)\cdot[-\nabla^2\ln(\Lambda|x-x_j|),
\end{equation}
and the boundary term vanishes by imposing the ``neutrality'' of the two-dimensional Coulomb gas, 

\begin{equation}
 \sum_iq_i=0.
\end{equation}
Since

\begin{equation}
 \nabla^2\ln|x-x_0|=2\pi\delta^2(x-x_0), 
\end{equation}
we have, 

\begin{equation}
 H_V=-\frac{\pi}{T}\sum_{i,j}q_iq_j\ln(\Lambda|x_i-x_j|).
\end{equation}

The electric susceptibility of this Coulomb gas is obtained by coupling the vortex Hamiltonian to an external ``electric'' field and 
taking the second derivative of the free energy with respect to it. The result for a single dipole is a susceptibility proportional to 
$\langle x^2\rangle$. In calculating this average the Boltzmann factor involving the energy of a vortex-antivortex pair has to be used 
as weight for the averaging. Thus, 

\begin{equation}
 \langle x^2\rangle\sim\int_{\Lambda^{-1}}^\infty dr r^3 e^{-(\pi/T)\ln(\Lambda r)}.
\end{equation}
Note that the integral features a short-distance cutoff. The integral converges only if 

\begin{equation}
 T<\frac{\pi}{2}.
\end{equation}
We obtain in this case, 

\begin{equation}
 \langle x^2\rangle\sim\frac{1}{\pi/T-2}.
\end{equation}
Therefore, the susceptibility will diverge at the critical temperature

\begin{equation}
 T_c=\frac{\pi}{2}.
\end{equation}
At this critical temperature the system changes the phase in a singular way (the susceptibility diverges). For $T<T_c$ the system is in a 
dielectric phase of vortices. For $T>T_c$ the vortices unbind and we have a plasma of vortex ``charges'' $\pm 1$. This phase transition between 
vortex states is the celebrated Kosterlitz-Thouless phase transition \cite{KT}. 

At the critical temperature the anomalous dimension has the universal value

\begin{equation}
 \eta(T_c)=\frac{1}{4}.
\end{equation}

\section{The renormalization group for the Coulomb gas}

In this Section we will derive the RG equations governing the KT phase transition. These equations were derived 
for the first time by Kosterlitz \cite{KT-1} and are important in order to derive one of the most remarkable 
results of the KT phase transition, namely, the existence of a universal jump in the superfluid density at $T_c$ 
\cite{NelKost}. This result was later confirmed by experiments \cite{Reppy}. 

In this Section we will derive the RG equations 
for the Coulomb gas 
using a scale-dependent Debye-H\"uckel 
screening theory. The derivation will be done in $d$ dimensions, as the three-dimensional 
result will be useful to us later on in a different context.The RG equations for a $d$-dimensional Coulomb gas were first derived by Kosterlitz 
\cite{Kosterlitz} using the so called ``poor man scaling'' \cite{Anderson-poorman}. 
The Debye-H\"uckel method used here to analyze the $d$-dimensional Coulomb gas 
follows Refs. \cite{KNS1,Nogueira-Kleinert-2008}, which 
is inspired from a paper by Young \cite{Young}, who analyzed the two-dimensional case. 

To begin with, let us consider the bare Coulomb potential in $d$ dimensions, which 
can be written using the result (\ref{g-g0}) as

\begin{equation}
\label{bare-U}
 U_0(r)=-4\pi^2K_0V(r),
\end{equation}
where we have defined $r\equiv|x|$ and $K_0=1/T$, and 

\begin{eqnarray}
 V(r)&=&{\cal G}(r)-{\cal G}(0)\nonumber\\
&=&\frac{S_da^{2-d}}{(d-2)(2\pi)^d}\left[\frac{(2\pi)^d}{S_d^2}
\left(\frac{r}{a}\right)^{2-d}-1\right],
\end{eqnarray}
with $a=\Lambda^{-1}$. From the bare Coulomb interaction (\ref{bare-U}) we obtain the 
bare electric field, 

\begin{equation}
 E_0(r)=-\frac{\partial U_0}{\partial r}=-\frac{4\pi^2K_0}{S_dr^{d-1}}.
\end{equation}
The crucial step for our analysis is the introduction of a scale-dependent 
dielectric function defining an effective medium for the Coulomb system, so that 
the renormalized electric field determines a renormalized Coulomb potential $U(r)$, i.e., 

\begin{equation}
\label{E-ren}
 E(r)=-\frac{\partial U_0}{\partial r}=-\frac{4\pi^2K_0}{S_d\varepsilon(r)r^{d-1}}
=-\frac{\partial U}{\partial r}.
\end{equation}
In order to establish a selfconsistent equation, we need to specify the dielectric 
constant via the electric susceptibility of the system. Thus, from the standard theory 
of electricity, we know that

\begin{equation}
 \varepsilon(r)=1+S_d\chi(r),
\end{equation}
where the scale-dependent electric susceptibility is given by

\begin{equation}
 \chi(r)=S_d\int_a^r ds s^{d-1}\alpha(s)n(s), 
\end{equation}
and the polarizability for small separation of a dipole pair is

\begin{equation}
 \alpha(r)\approx\frac{4\pi^2K_0}{d}r^2,
\end{equation}
while the density of dipoles is given by a Boltzmann distribution in terms of the 
renormalized Coulomb potential, 

\begin{equation}
 n(r)=z_0^2e^{-U(r)},
\end{equation}
with $z_0$ being the bare fugacity. Therefore, by integrating Eq. (\ref{E-ren}) 
the selfconsistent equation for the renormalized Coulomb potential is obtained:

\begin{equation}
 U(r)=U(a)+\frac{4\pi^2K_0}{S_d}\int_a^r\frac{ds}{\varepsilon(s)s^{d-1}}.
\end{equation}
Now we set $l\equiv\ln(r/a)$ and define the effective coupling via

\begin{equation}
 K^{-1}(l)=\frac{\varepsilon(ae^l)}{K_0}e^{(d-2)l}.
\end{equation}
Since 

\begin{equation}
 r\frac{dU}{dr}=\frac{4\pi^2K_0}{S_d\varepsilon(ae^l)r^{(d-2)}},
\end{equation}
we obtain

\begin{equation}
 \frac{dU}{dl}=\frac{4\pi^2}{S_da^{d-2}}K(l).
\end{equation}
Furthermore, 

\begin{eqnarray}
\frac{dK^{-1}}{dl}&=&(d-2)K^{-1}+\frac{S_d}{K_0}e^{(d-2)l}\frac{d\chi}{dl} 
\nonumber\\
&=&(d-2)K^{-1}+\frac{4\pi^2S_d^2a^{d+2}z_0^2}{d}e^{2dl-U(ae^l)}
\nonumber\\
&=&(d-2)K^{-1}+z^2,
\end{eqnarray}
where $z^2(l)$ is obviously defined by the second term of the second line in the 
equation above. Thus, the renormalized 
fugacity $z(l)$ satisfies the differential equation, 

\begin{equation}
 \frac{dz}{dl}=\left(d-\frac{2\pi^2K}{S_da^{d-2}}\right)z.
\end{equation}
By introducing the dimensionless couplings $\kappa=a^{2-d}K$ and 
$y=a^dz$, we finally obtain the desired RG equations for the $d$-dimensional Coulomb gas, 

\begin{equation}
\label{cgrg-1}
 \frac{d\kappa^{-1}}{dl}=(d-2)\kappa^{-1}+y^2,
\end{equation}

\begin{equation}
\label{cgrg-2}
 \frac{dy}{dl}=\left(d-\frac{2\pi^2\kappa}{S_d}\right)y.
\end{equation}
For $d=2$ the above equations yield the celebrated RG equations for the KT phase 
transition \cite{NelKost}, 

\begin{equation}
\label{ktrg-1}
 \frac{d\kappa^{-1}}{dl}=y^2,
\end{equation}

\begin{equation}
\label{ktrg-2}
 \frac{dy}{dl}=\left(2-\pi\kappa\right)y.
\end{equation}

\begin{figure}
 \begin{center}
  \includegraphics[width=12cm]{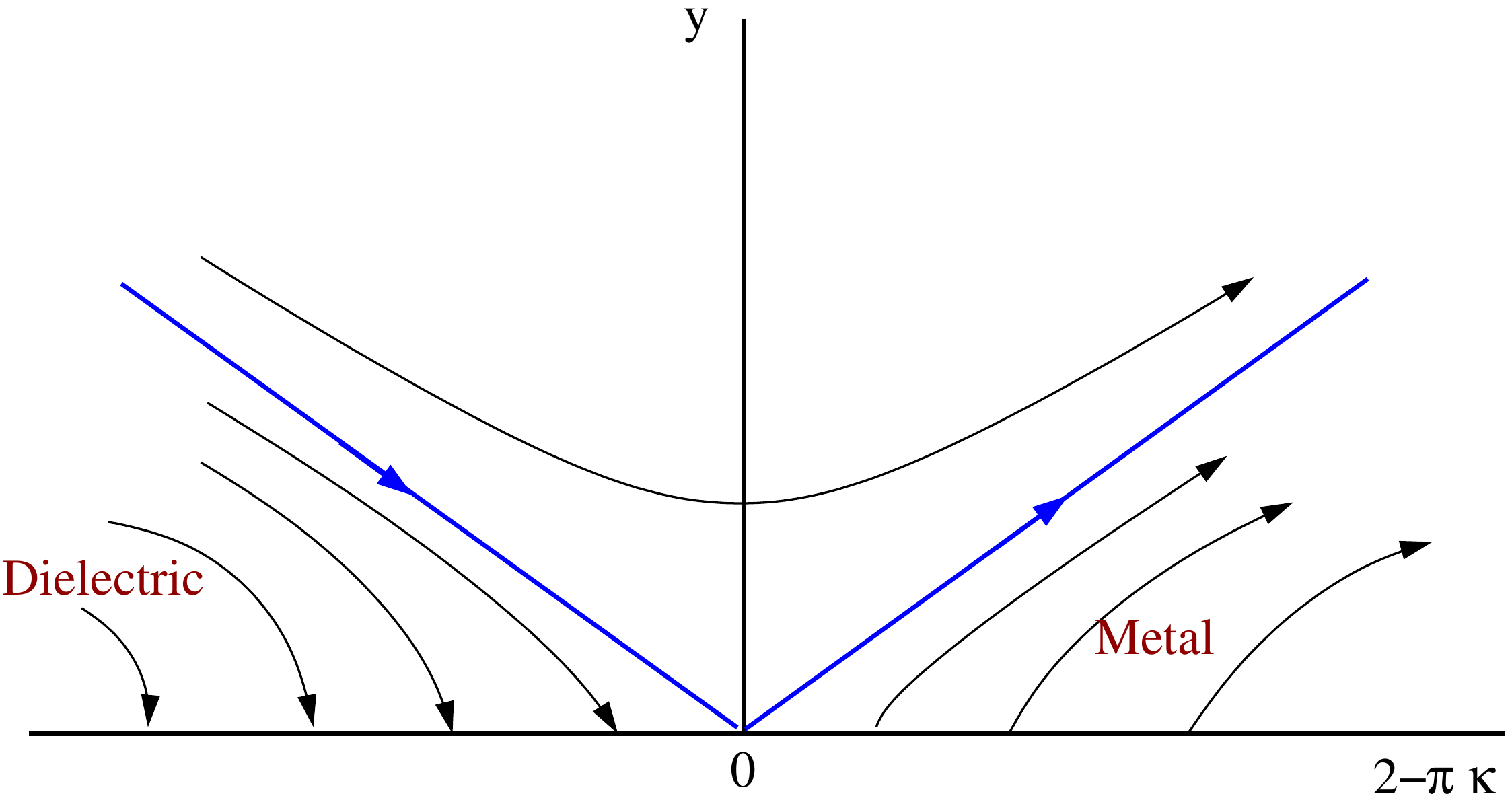}
\caption{Schematic flow diagram for the KT phase transition.}\label{fig:KT}
 \end{center}
\end{figure}

Eq. (\ref{ktrg-2}) has a fixed point at $\kappa_c=2/\pi$, which corresponds precisely to 
the critical temperature $T_c=\pi/2$ obtained before. However, Eqs. (\ref{ktrg-1}) and 
(\ref{ktrg-2}) contain additional information. The flow diagram actually features a 
line of fixed points for $\kappa>\kappa_c$ at zero fugacity. Usually we say that 
the KT flow diagram has a {\it fixed line} rather than a fixed point. The RG flow diagram 
is shown in Fig. \ref{fig:KT}. 
Note that the we set $2-\pi\kappa$ as the horizontal axis, such that the critical 
point occurs for a vanishing abscisse. 
The blue line shown in the figure is a separatrix 
delimitating three different regimes. Below the separatrix and for $2-\pi\kappa<0$ we 
have a dielectric phase of vortices, which corresponds to the low temperature regime 
where a vortex-antivortex pair is tightly bound, forming in this way a dipole. 
The dielectric phase is gapless because dipoles do not screen \cite{Spencer}. That is 
the reason why for $2-\pi\kappa<0$ there is a fixed line. At a fixed point all modes 
are gapless, so if there is a line of fixed points, like in the KT case, the 
excitation spectrum is completely gapless along this line. For  $2-\pi\kappa>0$, 
on the other hand, Debye screening occurs and a gap arises. This is a (classical) 
metallic phase (or plasma phase) of vortices.  

Let us study in more detail the RG equations. To this end we introduce the new variables, 

\begin{equation}
 X=2-\pi\kappa,~~~~~~~~~Y=\frac{2}{\sqrt{\pi}}y.
\end{equation}
The RG equations become

\begin{equation}
 \frac{dX}{dl}=\left(1-\frac{X}{2}\right)^2Y^2\approx Y^2,
\end{equation}

\begin{equation}
 \frac{dY}{dl}=XY.
\end{equation}
Thus, the family of hyperbola

\begin{equation}
 X^2-Y^2={\rm const}
\end{equation}
are RG invariants. We introduce further new variables, 

\begin{equation}
 u=X+Y,
\end{equation}
and

\begin{equation}
 v=Y-X,
\end{equation}
such that the RG invariance equation becomes

\begin{equation}
 u(l)v(l)=u_0v_0,
\end{equation}
where $u_0=u(0)$ and $v_0=v(0)$ are given initial conditions. Thus, 

\begin{eqnarray}
 \frac{du}{dl}&=&\frac{1}{2}(uv+u^2)\nonumber\\
&=&\frac{1}{2}(u_0v_0+u^2).
\end{eqnarray}
It is straightforward to solve the above equation by direct integration. The result is

\begin{equation}
 \arctan\left(\frac{u}{\sqrt{u_0v_0}}\right)-\arctan\left(\frac{u_0}{\sqrt{u_0v_0}}\right)=2\sqrt{u_0v_0}\ln\left(\frac{r}{a}\right).
\end{equation}
Now we set $r=\xi$, i.e., we let the distance scale be equal to the correlation length. In this case we have that the 
mass gap is approximately given by 

\begin{equation}
\label{gap-KT}
 \xi^{-1}\approx\exp\left(-\frac{{\rm const}}{\sqrt{T-T_c}}\right),
\end{equation}
which is a result characteristic of the KT transition. Note that we do not obtain in this case a power law for the mass gap.

For $d>2$ the situation is completely different. First of all, the Coulomb gas cannot 
be interpreted as vortices any longer, since in three dimensions vortices are one-dimensional 
objects, lines or loops \cite{Kleinert-GFCM-1}. However, there are physical 
systems in three dimensions with point-like topological defects where an analysis similar to the one made here is 
applicable. For example, there are systems where magnetic monopole-like defects occur in three spacetime dimensions  
\cite{KNS1,Nogueira-Kleinert-2008,Polyakov,IL,KNS,Herbut1,Herbut3,Kragset,Borkje,Hermele,NK,Nogueira-Nussinov}. Second, for $d>2$ the RG equations (\ref{cgrg-1}) and 
(\ref{cgrg-2}) do not have nontrivial fixed points. Thus, no phase transition occurs 
in this case. The excitation spectrum is always gapped, so that the system remains 
permanently in the plasma phase.     

\chapter{Bose-Einstein condensation and superfluidity}
\label{bec-sf}

\section{Bose-Einstein condensation in an ideal gas}

The Lagrangian for an ideal Bose gas is written in the imaginary time formalism as

\begin{equation}
 {\cal L}=b^*\partial_\tau b-\mu |b|^2+\frac{1}{2m}\nabla b^*\cdot\nabla b.
\end{equation}
All thermodynamic properties of the ideal Bose gas can be derived from the 
partition function, which is given by the functional integral representation, 

\begin{equation}
 Z=\int{\cal D}b^*{\cal D}b~e^{-S},
\end{equation}
where

\begin{equation}
 S=\int_0^\beta d\tau\int d^dr {\cal L}.
\end{equation}
The functional integral above is to be solved using periodic boundary conditions 
$b(0)=b(\beta)$ and $b^*(0)=b^*(\beta)$. 

We see from the action for the ideal Bose gas that the case $\mu=0$ is special. Indeed, 
if $\mu=0$, the action is invariant by a transformation where 
the Bose field is shifted by a constant, $b\to b+c$. Note that the periodic boundary 
conditions make the contribution $c^*\partial_\tau b$ vanish. Thus, the special 
role of the $\mu=0$ regime can be accounted for by shifting the Bose field by 
a constant, i.e., 

\begin{equation}
 b=b_0+\tilde b,
\end{equation}
and we require that $b_0$ minimizes the action. This requirement implies that no term 
linear in $\tilde b$ or $\tilde b^*$ appears in the action. This is only true provided 

\begin{equation}
 \mu b_0=0.
\end{equation}
This equation is fulfilled either for $b_0=0$ and $\mu\neq 0$, or $b_0\neq 0$ and 
$\mu=0$. 
The Lagrangian is rewritten as

\begin{equation}
 {\cal L}=-\mu|b_0|^2+\tilde b^*\left(\partial_\tau-\mu-\frac{1}{2m}\nabla^2\right)\tilde b,
\end{equation}
where the Laplacian term is obtained through partial integration in the action. By 
performing the Gaussian functional integral over $\tilde b$, we obtain, up to a constant, 
the result, 

\begin{equation}
 Z\sim \frac{\exp(\beta V\mu|b_0|^2)}{\det\left(\partial_\tau-\mu-\frac{1}{2m}\nabla^2
\right)},
\end{equation}
where $V$ is the (infinite) volume. 
Thus, the free energy density is given by,

\begin{eqnarray}
\label{f-ibec}
 f&=&-\frac{1}{\beta V}\ln Z
\nonumber\\
&=&-\mu|b_0|^2+\frac{1}{\beta V}\ln\det\left(\partial_\tau-\mu-\frac{1}{2m}\nabla^2
\right).
\end{eqnarray}
Since the determinant of an operator is given by the product of the eigenvalues of 
the operator, we have to solve the differential equation, 

\begin{equation}
 \left(\partial_\tau-\mu-\frac{1}{2m}\nabla^2
\right)\psi=E\psi,
\end{equation}
where $E$ is the eigenvalue. The equation above should be solved with periodic 
boundary conditions $\psi(0)=\psi(\beta)$. In order to solve the eigenvalue 
problem we perform a Fourier transformation in the spatial variables,  

\begin{equation}
 \psi(\tau,{\bf r})=\int\frac{d^dr}{(2\pi)^d}e^{i{\bf p}\cdot{\bf r}}\psi(\tau,{\bf p}).
\end{equation}
In this way the partial differential equation becomes an ordinary differential equation 
of first order,

\begin{equation}
 \left(\partial_\tau-\mu+\frac{p^2}{2m}\right)\psi(\tau,{\bf p})=E\psi(\tau,{\bf p}),
\end{equation}
which can be easily solved to obtain, 

\begin{equation}
 \psi(\tau,{\bf p})=\psi(0,{\bf p})\exp\left[\tau\left(E+\mu-\frac{p^2}{2m}\right)\right].
\end{equation}
Due to the periodic boundary condition, the above equation for $\tau=\beta$ becomes

\begin{equation}
 1=\exp\left[\beta\left(E+\mu-\frac{p^2}{2m}\right)\right].
\end{equation}
This implies, 

\begin{equation}
 E_n({\bf p})=-i\omega_n-\mu+\frac{p^2}{2m},
\end{equation}
where $\omega_n=2\pi n/\beta$ with $n\in \mathbb{Z}$ is the so called Matsubara frequency. 
Inserting these eigenvalues in Eq. (\ref{f-ibec}) yields, 

\begin{equation}
 f=-\mu|b_0|^2+\frac{1}{\beta}\sum_{n=-\infty}^\infty\int\frac{d^dp}{(2\pi)^d}\ln\left(
-i\omega_n-\mu+\frac{p^2}{2m}\right).
\end{equation}

Our interest is to compute the particle density, $n$, which is the variable conjugated to 
the chemical potential. We have, 

\begin{equation}
 n=-\frac{\partial f}{\partial \mu},
\end{equation}
which yields, 

\begin{equation}
 n=|b_0|^2-\frac{1}{\beta}\sum_{n=-\infty}^\infty\int\frac{d^dp}{(2\pi)^d}
\frac{1}{i\omega_n+\mu-\frac{p^2}{2m}}.
\end{equation}
In order to have $b_0\neq 0$ we need $\mu=0$, so that the above equation becomes 
for $b_0\neq 0$,  

\begin{equation}
 n=|b_0|^2-\frac{1}{\beta}\sum_{n=-\infty}^\infty\int\frac{d^dp}{(2\pi)^d}
\frac{1}{i\omega_n-\frac{p^2}{2m}}.
\end{equation}
The Matsubara sum appearing above is performed in the Appendix \ref{app:sums}. Thus,

\begin{equation}
 n=|b_0|^2+\int\frac{d^dp}{(2\pi)^d}\frac{1}{e^{\frac{\beta p^2}{2m}}-1}..
\end{equation}
The remaining integral is over a Bose distribution for free bosons in $d$ dimensions. 
An integral involving a more general spectrum is evaluated in Appendix \ref{app:be-int}. 
Using Eq. (\ref{int-bec-1}) with $z=2$ and $c=1/(2m)$ and making some simplifications, we 
obtain, 

\begin{equation}
\label{condens-1}
 |b_0|^2=n\left[1-\frac{\zeta(d/2)}{n}\left(\frac{mT}{2\pi}\right)^{d/2}\right].
\end{equation}
Note that we have solved for $|b_0|^2$, which is the so called condensate density. 
Its physical meaning is that for $\mu=0$ the particle density 
{\it zero momentum} gets depleted due to temperature effects. The density at zero 
momentum emerges because in momentum and frequency space, 

\begin{equation}
 b(\omega_n,{\bf p})=b_0\delta^d({\bf p})\delta_{n,0}+\tilde b(\omega_n,{\bf p}),
\end{equation}
such that $b_0$ is associated to the zero momentum and zero Matsubara mode contribution of 
the Bose field. The Bose-Einstein condensation is thus the macroscopic occupation of 
the zero momentum state, in which case $\tilde b$ represents the fluctuation around 
the condensate. 

Eq. (\ref{condens-1}) can be rewritten as

\begin{equation}
\label{condens-2}
 |b_0|^2=n\left[1-\left(\frac{T}{T_c}\right)^{d/2}\right],
\end{equation}
where

\begin{equation}
\label{Tc-1}
 T_c=\frac{2\pi}{m}\left[\frac{n}{\zeta(d/2)}\right]^{2/d},
\end{equation}
is the critical temperature. For $T=T_c$ the condensate vanishes. Note that for 
$T=0$ all the particles are condensed. This is a feature of the ideal Bose gas. We 
will see that in the interacting case the condensate is also depleted at $T=0$ due to 
the interaction. 

From the expression for the critical temperature we see that it vanishes for 
$d=2$, implying that no condensate exists in a two-dimensional ideal Bose gas at 
finite temperature. This result is actually more general and holds even in the 
interacting case. It is known as Hohenberg's theorem \cite{Hohenberg1}. 

\section{The dilute Bose gas in the large $N$ limit}

In Chapter \ref{ch:fm} we have studied the $O(n)$ classical non-linear $\sigma$ model in the large 
$n$ limit. We will now use this knowledge to perform a $1/N$ expansion for an interacting 
Bose gas. Such an expansion actually corresponds to the so called random phase approximation (RPA)  
for the dilute Bose gas introduced long time ago \cite{Tser,Kondor1,Kondor2,Kondor3,Kondor-4}. That 
the $1/N$ expansion for the dilute Bose gas corresponds to RPA was recognized by Kondor and 
Szepfalusy \cite{Kondor3} long time ago. Their analysis will be revisited here from a 
functional integral point of view \cite{Nogueira-pi-book}.

\subsection{The saddle-point approximation}

Let us consider the following action for a $N$-component 
interacting Bose gas:

\begin{equation}
\label{action}
S=\int_0^\beta d\tau\int d^dr\left[\sum_{\alpha=1}^N b_\alpha^*
\left(\partial_\tau-\mu-\frac{\nabla^2}{2m}\right)b_\alpha
+\frac{g}{2}\left(\sum_{\alpha=1}^N |b_\alpha|^2\right)^2\right],
\end{equation}
where $b_\alpha$ and $b_\alpha^*$ are complex commuting fields.
The partition function is then given by

\begin{equation}
\label{partition}
Z=\int\left[\prod_\alpha{\cal D}b_\alpha^*{\cal D}b_\alpha
\right] e^{-S}.
\end{equation} 
In order to perform the $1/N$-expansion we introduce an auxiliary 
field $\lambda(\tau,{\bf r})$ via a Hubbard-Stratonovich 
transformation:

\begin{equation}
\label{action1}
S'=\int_0^\beta d\tau\int d^dr\left[\sum_{\alpha=1}^N b_\alpha^*
\left(\partial_\tau-\mu-\frac{\nabla^2}{2m}+i\lambda\right)b_\alpha
+\frac{1}{2g}\lambda^2\right].
\end{equation}
Now we integrate out $N-1$ Bose fields to obtain the effective action 

\begin{eqnarray}
\label{Seff}
S_{\rm eff}&=&(N-1){\rm Tr}\ln
\left(\partial_\tau-\mu-\frac{\nabla^2}{2m}+i\lambda\right)
\nonumber\\
&+&
\int_0^\beta d\tau\int d^dr\left[b^*
\left(\partial_\tau-\mu-\frac{\nabla^2}{2m}+i\lambda\right)b+
\frac{1}{2g}\lambda^2\right],
\end{eqnarray}
where we have called $b$ the unintegrated Bose field.

Next we extremize the action according to the saddle-point 
approximation (SPA), a procedure that becomes exact for $N\to\infty$. 
This part of the calculation is practically identical with the one for the ideal 
Bose gas. 
This is 
done by making the replacement $i\lambda\to\lambda_0$ and $b\to b_0$, with 
$\lambda_0$ and $b_0$ being constant fields, followed by extremization 
with respect to these constant background fields. From this SPA we 
obtain the equations

\begin{equation}
\label{SPA1}
(\lambda_0-\mu)b_0=0,
\end{equation}

\begin{equation}
\label{SPA2}
\lambda_0=g|b_0|^2-\frac{Ng}{\beta}\sum_{n=-\infty}^\infty
\int\frac{d^dp}{(2\pi)^d}\frac{1}{i\omega_n+\mu-\lambda_0-\frac{{\bf p}^2}
{2m}}.
\end{equation}

The large $N$ limit is taken with $Ng$ fixed. 
Below the critical temperature $T_c$ we have $b_0\neq 0$, and thus 
from Eq. (\ref{SPA1}) $\lambda_0=\mu$. Therefore, Eq. (\ref{SPA2}) 
becomes

\begin{equation}
\label{cdens}
|b_0|^2=\frac{\mu}{g}-N\left(\frac{m}{2\pi\beta}\right)^{d/2}\zeta(d/2),
\end{equation}
provided $d>2$. The second term on the RHS of thee above equation is, up to the prefactor $N$, 
the same as the one in Eq. (\ref{condens-1}) for the condensate density of the ideal Bose gas. 
The particle density is obtained as usual $n=-\partial f/\partial\mu$, 
where $f=-\ln Z/(NV\beta)$ is the free energy density. This gives us 

\begin{equation}
\label{n}
n=\frac{|b_0|^2}{N}+\int\frac{d^dp}{(2\pi)^d}\frac{1}{
\exp\left[\beta\left(\frac{{\bf p}^2}{2m}+\lambda_0-\mu
\right)\right]-1}.
\end{equation}
By setting $\lambda_0=\mu$ in Eq. (\ref{n}) and using Eq. (\ref{cdens}), 
we obtain

\begin{equation}
n=\frac{\mu}{Ng},
\end{equation}
and therefore the condensate density becomes,  

\begin{equation}
\label{cdens1}
n_0\equiv
\frac{|b_0|^2}{N}=n\left[1-\left(\frac{T}{T_c}\right)^{d/2}\right],
\end{equation}
where 

\begin{equation}
\label{Tc}
T_c=\frac{2\pi}{m}\left[\frac{n}{\zeta(d/2)}\right]^{2/d}.
\end{equation}
We see that the SPA does not change the value of $T_c$ with respect 
to the non-interacting Bose gas. Indeed, the SPA corresponds to the 
Hartree approximation and it is well known that it gives a zero 
$T_c$ shift.

\subsection{Gaussian fluctuations around the 
saddle-point approximation: Bogoliubov theory and beyond}

\begin{figure}
\begin{center}
\includegraphics[width=10cm]{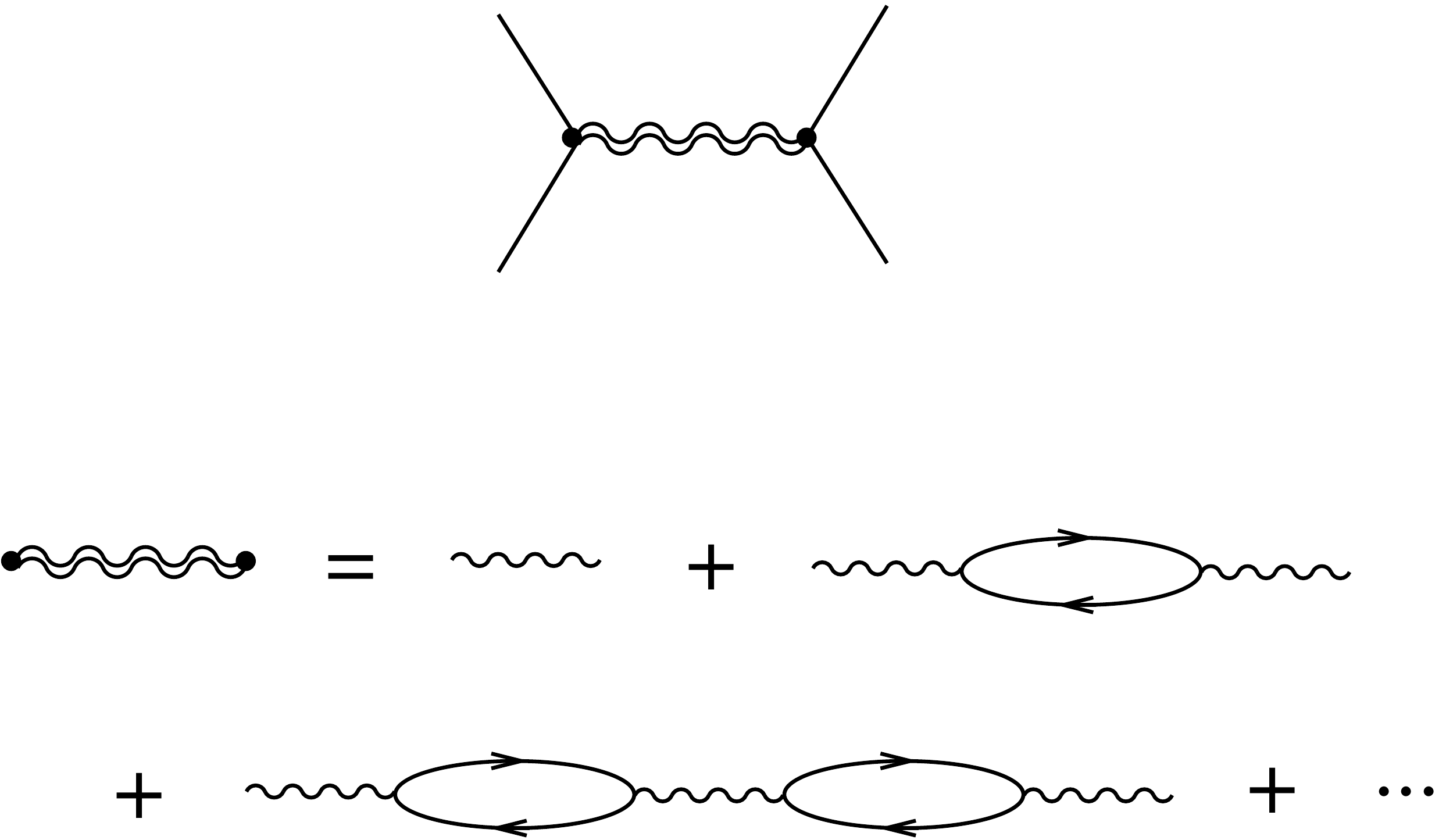}
\caption{Feynman diagram representation of the effective interaction 
Eq. (\ref{vertex}). The dashed line represents the bare $\lambda$-field 
propagator while the double dashed line represents the dressed 
$1/N$-corrected $\lambda$-field propagator. Continuous lines 
represent $\tilde{b}$-fields and each loop is the polarization 
bubble Eq. (\ref{bubble}) formed by two $\tilde{b}$-field propagators 
in convolution. The effective interaction is obtained as a geometric 
series of polarization bubbles.}\label{fig:bubblesum}
\end{center}
\end{figure}

In order to integrate out $\lambda$ approximately, we consider 
the $1/N$-corrections to the SPA by computing 
the fluctuations around the constant background fields $b_0$ and 
$\lambda_0$. 
By setting

\begin{equation}
b=b_0+\tilde{b}, ~~~~~~i\lambda=\lambda_0+i\tilde{\lambda},
\end{equation}
and expanding the effective action (\ref{Seff}) up to quadratic 
order in the $\tilde{\lambda}$ field, we obtain

\begin{eqnarray}
\label{quad}
S_{\rm eff}=S_{\rm eff}^{\rm SPA}
+\int_0^\beta d\tau\int d^dr\left[\tilde{b}^*
\left(\partial_\tau-\mu+\lambda_0-\frac{\nabla^2}{2m}\right)
\tilde{b}+i\tilde{\lambda}(b_0^*\tilde{b}+b_0\tilde{b}^*
+|\tilde{b}|^2)
+\frac{1}{2g}\tilde{\lambda}^2\right]\nonumber\\
-\frac{N}{2}\int_0^\beta d\tau\int_0^\beta d\tau'
\int d^dr\int d^dr'\tilde{\lambda}(\tau,{\bf r})
G_0(\tau-\tau',{\bf r}-{\bf r}')G_0(\tau'-\tau,{\bf r}'-{\bf r})
\tilde{\lambda}(\tau',{\bf r}'),~~~~
\end{eqnarray}
where $S_{\rm eff}^{\rm SPA}$ is the effective action (\ref{Seff}) 
in the SPA and 

\begin{equation}
G_0(\tau,{\bf r})=\frac{1}{\beta}\sum_{n=-\infty}^\infty
\int\frac{d^d p}{(2\pi)^d}e^{i({\bf p}\cdot{\bf r}
-\omega_n\tau)}\hat{G}_0(i\omega_n,{\bf p}),
\end{equation}
with

\begin{equation}
\hat{G}_0(i\omega_n,{\bf p})=\frac{1}{i\omega_n+\mu-\lambda_0
-\frac{{\bf p}^2}{2m}}.
\end{equation}      
After integrating out $\tilde{\lambda}$ the effective action can be 
cast in the form

\begin{eqnarray}
\label{quadnew}
S_{\rm eff}=S_{\rm eff}^{\rm SPA}
+\frac{1}{2}{\rm Tr}\ln\left[\delta(\tau-\tau')\delta^d({\bf r}
-{\bf r}')-Ng~G_0(\tau-\tau',{\bf r}-{\bf r}')G_0
(\tau'-\tau,{\bf r}'-{\bf r})\right]\nonumber\\
+\frac{1}{2}
\int_0^\beta d\tau\int d^dr\int_0^\beta d\tau'\int d^dr'
~\Psi^\dagger(\tau,{\bf r})
{\bf M}(\tau-\tau',{\bf r}-{\bf r}')\Psi(\tau,{\bf r}')
~~~~~~~~~~~~~~~~~~~~
\nonumber\\
+\frac{b_0^*}{2}\int_0^\beta d\tau\int d^dr\int_0^\beta d\tau'\int d^dr'~
\left[
\begin{array}{cc}
1~ & ~0
\end{array}
\right]
\Psi(\tau,{\bf r})
\Gamma(\tau-\tau',{\bf r}-{\bf r}')
\Psi^\dagger(\tau',{\bf r}')\Psi(\tau',{\bf r}')
\nonumber\\
+\frac{b_0}{2}\int_0^\beta d\tau\int d^dr\int_0^\beta d\tau'\int d^dr'~
\Psi^\dagger(\tau,{\bf r})
\left[
\begin{array}{c}
1\\
\noalign{\medskip}
0
\end{array}
\right]
\Gamma(\tau-\tau',{\bf r}-{\bf r}')
\Psi^\dagger(\tau',{\bf r}')\Psi(\tau',{\bf r}')
\nonumber\\
+\frac{1}{8}\int_0^\beta d\tau\int d^dr\int_0^\beta d\tau'\int d^dr'
\Psi^\dagger(\tau,{\bf r})\Psi(\tau,{\bf r})\Gamma(\tau-\tau',
{\bf r}-{\bf r}')\Psi^\dagger(\tau',{\bf r}')\Psi(\tau',{\bf r}'),
\end{eqnarray}
where we have introduced the two-component fields

\begin{equation}
\Psi^\dagger(\tau,{\bf r})=\left[
\begin{array}{cc}
\tilde{b}^*(\tau,{\bf r}) & \tilde{b}(\tau,{\bf r})
\end{array}
\right],~~~~
\Psi(\tau,{\bf r})=
\left[
\begin{array}{c}
\tilde{b}(\tau,{\bf r})\\
\noalign{\medskip}
\tilde{b}^*(\tau,{\bf r})
\end{array}
\right],
\end{equation}
which satisfy $\Psi^\dagger\Psi=2|\tilde{b}|^2$. 
The matrix ${\bf M}(\tau-\tau',{\bf r}-{\bf r}')$  
has a Fourier transform given by

\begin{equation}
\label{matrix-1}
\hat{\bf M}(i\omega_n,{\bf p})=
\left[
\begin{array}{cc}
-i\omega_n+ {\cal E}(\omega_n,{\bf p}) & b_0^2~\hat{\Gamma}(i\omega_n,{\bf p})\\
\noalign{\medskip}
(b_0^*)^2\hat{\Gamma}(i\omega_n,{\bf p}) & 
i\omega_n+{\cal E}(\omega_n,{\bf p})
\end{array}
\right],
\end{equation} 
where  

\begin{equation}
 {\cal E}(\omega_n,{\bf p})=-\mu+\lambda_0+\frac{{\bf p}^2}{2m}+|b_0|^2
\hat{\Gamma}(i\omega_n,{\bf p}),
\end{equation}
with

\begin{equation}
\label{vertex}
\hat{\Gamma}(i\omega_n,{\bf p})=\frac{g}{1-Ng\hat{\Pi}(i\omega_n,{\bf p})},
\end{equation}
which is the Fourier transform of the effective interaction  
$\Gamma(\tau-\tau',{\bf r}-{\bf r}')$, and 

\begin{equation}
\label{bubble}
\hat{\Pi}(i\omega_n,{\bf p})=
\frac{1}{\beta}\sum_{m=-\infty}^\infty
\int\frac{d^dq}{(2\pi)^d}\hat{G}_0(i\omega_n+i\omega_m,
{\bf p}+{\bf q})\hat{G}_0(i\omega_m,{\bf q}),
\end{equation}
is the polarization bubble. The effective interaction can be 
represented in terms of Feynman diagrams as in Fig. \ref{fig:bubblesum}.   
Physically $\tilde{\lambda}$ corresponds to the fluctuation of the 
particle density and thus the effective interaction (\ref{vertex}) 
gives in fact the density-density correlation function. An  
effective interaction like the one in Eq. (\ref{vertex}) 
was already obtained some time ago by a number of authors 
\cite{Tser,Kondor1,Cheung,Kondor2}. Thus, the $1/N$-expansion is actually 
equivalent to a random phase approximation 
(RPA) considered previously in the literature \cite{Tser,Kondor2}. 
Explicit evaluation of the Matsubara sum in Eq. (\ref{bubble}) yields

\begin{eqnarray}
\label{bubble1}
\hat{\Pi}(i\omega_n,{\bf p})&=&\int\frac{d^dq}{(2\pi)^d}\frac{1}{
i\omega_n-\frac{1}{2m}({\bf p}^2+2{\bf p}\cdot{\bf q})}
\left\{n_{\rm B}\left(\frac{{\bf q}^2}{2m}+\lambda_0-\mu\right)
\right.\nonumber\\
&-&\left.n_{\rm B}\left[\frac{({\bf p}+{\bf q})^2}{2m}+\lambda_0-\mu\right]
\right\},
\end{eqnarray}
where $n_{\rm B}(x)=1/(e^{\beta x}-1)$ is the Bose distribution 
function. 


By inverting the matrix (\ref{matrix}) we obtain the propagator

\begin{equation}
\label{prop}
\hat{\bf G}(i\omega_n,{\bf p})=
\left[
\begin{array}{cc}
\hat{\bf G}_{\tilde{b}^*\tilde{b}}(i\omega_n,{\bf p}) 
& \hat{\bf F}_{\tilde{b}^*\tilde{b}^*}(i\omega_n,{\bf p}) \\
\noalign{\medskip}
\hat{\bf F}_{\tilde{b}\tilde{b}}(i\omega_n,{\bf p}) & 
\hat{\bf G}_{\tilde{b}^*\tilde{b}}(-i\omega_n,{\bf p})
\end{array}
\right],
\end{equation}
where 

\begin{equation}
\label{G11}
\hat{\bf G}_{\tilde{b}^*\tilde{b}}(i\omega_n,{\bf p})
=\frac{i\omega_n+\lambda_0-\mu+\frac{{\bf p}^2}{2m}+|b_0|^2
\hat{\Gamma}(i\omega_n,{\bf p})}{\omega_n^2+
\left[\frac{{\bf p}^2}{2m}+\lambda_0-\mu+
|b_0|^2\hat{\Gamma}(i\omega_n,{\bf p})\right]^2
-|b_0|^4\hat{\Gamma}^2(i\omega_n,{\bf p})},
\end{equation}
and

\begin{equation}
\label{Fs}
\hat{\bf F}_{\tilde{b}^*\tilde{b}^*}(i\omega_n,{\bf p})=
-\frac{(b_0^*)^2\hat{\Gamma}(i\omega_n,{\bf p})}{
\omega_n^2+\left[\frac{{\bf p}^2}{2m}+\lambda_0-\mu+
|b_0|^2\hat{\Gamma}(i\omega_n,{\bf p})\right]^2
-|b_0|^4\hat{\Gamma}^2(i\omega_n,{\bf p})},
\end{equation}
is the anomalous propagator. 

For $T>T_c$ we have that $b_0=0$ and the matrix propagator 
becomes diagonal. In this case if we neglect the interaction terms 
from the effective action (\ref{quadnew}), the $\tilde{b}$ 
propagator corresponds to the Hartree approximation. Thus, above 
$T_c$ we have to consider the effective interaction between the 
bosons in Eq. (\ref{quadnew}) in order to obtain a nontrivial 
result for the excitation spectrum. This is achieved by 
computing the $1/N$-correction to the propagator. 
Below $T_c$, however, a nontrivial result for the excitation spectrum 
is obtained from the pole of the propagator even 
without considering the $1/N$ correction to it. This is easily seen 
from the structure of the propagators (\ref{G11}) and 
(\ref{Fs}) where 
the effective interaction $\hat{\Gamma}(i\omega_n,{\bf p})$ 
appears explicitly. {\it Above $T_c$ the effective interaction appears 
in the propagator 
only in the next to the leading order in $1/N$}.


\subsection{The excitation spectrum below $T_c$} 

As we have discussed in the previous Subsection, below $T_c$ 
a nontrivial result for the excitation spectrum is obtained  in 
an approximation where the 
interaction term of the effective action is neglected. Thus, 
we now undertake a study of the spectrum of the system in 
such a Gaussian approximation. Later we shall see that 
the effective action in the Gaussian approximation gives the 
free energy density up to the order $1/N$. 
  
From the pole of the matrix propagator (\ref{prop}) we obtain 
that the energy spectrum $E({\bf p})$ satisfy the equation 

\begin{eqnarray}
\label{excspec}
E^2({\bf p})&=&{\rm Re}\left\{\frac{{\bf p}^2}{2m}+\lambda_0-\mu+
|b_0|^2\hat{\Gamma}[E({\bf p})+i\delta,{\bf p}]\right\}^2
-|b_0|^4{\rm Re}~\hat{\Gamma}^2[E({\bf p})+i\delta,{\bf p}]\nonumber\\
&=&\left(\frac{{\bf p}^2}{2m}+\lambda_0-\mu\right)^2
+2\left(\frac{{\bf p}^2}{2m}+\lambda_0-\mu\right)
|b_0|^2{\rm Re}~\hat{\Gamma}[E({\bf p})+i\delta,{\bf p}],
\end{eqnarray}
where $\delta\to 0^+$.
Note that Eq. (\ref{excspec}) can be written as 
the product of two elementary excitations, 
$E^2({\bf p})=E_{l}({\bf p})E_{t}({\bf p})$, where 

\begin{equation}
E_{l}({\bf p})=\frac{{\bf p}^2}{2m}+\lambda_0-\mu+
2|b_0|^2{\rm Re}~\hat{\Gamma}[E({\bf p})+i\delta,{\bf p}],
\end{equation}

\begin{equation}
E_{t}({\bf p})=\frac{{\bf p}^2}{2m}+\lambda_0-\mu,
\end{equation}
are the spectrum of the longitudinal and transverse modes, respectively. 
When $\lambda_0=\mu$ we obtain that the transverse mode is gapless, 
consistent with Goldstone theorem.

By inserting the saddle-point value 
$\lambda_0=\mu$, we obtain the following self-consistent equation for 
the excitation spectrum

\begin{equation}
\label{spectrum}
\widetilde{E}({\bf p})=\sqrt{\frac{{\bf p}^4}{4m^2}+\frac{|b_0|^2}{m}
{\bf p}^2~{\rm Re}\hat{\Gamma}[\widetilde{E}({\bf p})+i\delta,{\bf p}]},
\end{equation}
where the notation $\widetilde{E}({\bf p})\equiv E({\bf p})|_{\lambda_0=\mu}$ 
is used.  
Note that the above spectrum corresponds to a generalization of 
the well known Bogoliubov spectrum \cite{Bogoliu}. The difference 
lies in the fact that in the $1/N$-expansion the coupling constant 
$g$ is replaced by the effective interaction 
$\hat{\Gamma}[E({\bf p})+i\delta,{\bf p}]$ \cite{Tser,Kondor1}. 
At zero temperature $\Pi(i\omega,{\bf p})$ vanishes and 
the excitation spectrum corresponds to the usual Bogoliubov 
spectrum. As we shall see, the modification 
of the spectrum by the effective 
interaction accounts for thermal fluctuation effects in higher 
temperatures and allows us a consistent treatment of critical 
fluctuations near $T_c$.        

In Eq. (\ref{spectrum}) we can legitimately replace $|b_0|^2$ by 
$Nn$, since the error committed in such a replacement is of higher order 
in $1/N$. Thus, Eq. (\ref{spectrum}) becomes

\begin{equation}
\label{spectrum1}
\widetilde{E}({\bf p})=\sqrt{\frac{{\bf p}^4}{4m^2}+\frac{nN}{m}
{\bf p}^2~{\rm Re}\hat{\Gamma}[\widetilde{E}({\bf p})+i\delta,{\bf p}]}.
\end{equation}
Note that since $\hat{\Gamma}[\widetilde{E}({\bf p})+i\delta,{\bf p}]\sim{\cal O}(1/N)$, 
Eq. (\ref{spectrum1}) is independent of $N$ for $N\to\infty$. 


In order to obtain the spectrum of elementary excitations we need 
to evaluate the polarization bubble (\ref{bubble1}). Unfortunately, 
it cannot be evaluated exactly, although many of its properties 
and asymptotic limits can be worked out exactly   
\cite{Kondor2,Abe}. 
For instance, for distances much larger than the thermal wavelength 
the polarization bubble can be evaluated exactly \cite{Kondor2,Abe}. 
This is called the {\it classical} limit in the early literature 
of the field. In the classical limit we 
can approximate the Bose distribution in Eq. (\ref{bubble1}) 
by $n_{\rm B}(x)\approx 1/\beta x$. In such a limit we can 
write

\begin{equation}
\hat{\Pi}(i\omega_n,{\bf p})=\Pi_0({\bf p})+\Pi_1(i\omega_n,{\bf p}),
\end{equation}
where 

\begin{equation}
\label{Pi0}
\Pi_0({\bf p})=-4m^2T\int\frac{d^dq}{(2\pi)^d}
\frac{1}{[({\bf p}+{\bf q})^2+2m(\lambda_0-\mu)]
[{\bf q}^2+2m(\lambda_0-\mu)]},
\end{equation} 

\begin{eqnarray}
\label{Pi1}
\Pi_1(i\omega_n,{\bf p})&=&8m^3Ti\omega_n
\int\frac{d^dq}{(2\pi)^d}\frac{1}{
2mi\omega_n-{\bf p}^2-2{\bf p}\cdot{\bf q}}
\nonumber\\
&\times&\frac{1}{[({\bf p}+{\bf q})^2+
2m(\lambda_0-\mu)][{\bf q}^2+2m(\lambda_0-\mu)]}.
\end{eqnarray}
Setting $\lambda_0=\mu$, we 
obtain the following result for the effective interaction:

\begin{equation}
\label{ECFTint}
\hat{\Gamma}_{0}({\bf p})=\frac{g}{
1+\alpha_dTm^2Ngp^{d-4}},
\end{equation}
where $p\equiv|{\bf p}|$ and 

\begin{equation}
\alpha_d=\frac{\Gamma(2-d/2)\Gamma^2(d/2-1)}{2^{d-2}\pi^{d/2}
\Gamma(d-2)}.
\end{equation}
When $2<d<4$ we obtain for small $p$ that 
$\hat{\Gamma}_{0}({\bf p})\approx p^{4-d}/(\alpha_dTm^2N)$, such 
that the excitation spectrum is given approximately by

\begin{equation}
\label{specECFT}
\widetilde{E}(p)\approx\sqrt{\frac{n}{\alpha_dm^3T}}~p^{(6-d)/2}.
\end{equation}
The above equation reflects the dynamic scaling behavior, 
$\widetilde{E}(p)\sim p^z$ of the excitation spectrum, where $z$ is 
the dynamic critical exponent. Thus, from 
Eq. (\ref{specECFT}) we see that it implies a dynamic 
exponent

\begin{equation}
\label{zECFT}
z=\frac{6-d}{2}.
\end{equation}
At $d=3$ we obtain $z=3/2$, which is the expected result for 
$^4$He. This result was obtained first by 
Patashinskii and Pokrovskii \cite{PatPok}. 
Note that the same order in $1/N$ when 
$T_c$ is approached from {\it above} fails to give a non-trivial 
dynamic scaling behavior. Only after taking into account non-Gaussian 
Gaussian fluctuations, corresponding to the next-to-leading order 
in $1/N$, is possible to obtain a non-trivial dynamic exponent. The 
origin of this non-symmetric critical behavior comes from the 
intrinsic existing asymmetry in the dilute Bose gas with 
respect to the ordered and disordered phases. Indeed, in the ordered 
phase the spectrum has a relativistic-like form, while in the 
disordered phase the non-relativistic behavior dominates the 
physics. The {\it static} critical exponents are not affected 
by this asymmetric behavior of the theory, but the critical dynamics 
properties are. Note that our value of the 
dynamic exponent is independent of $N$, i.e., $z=(6-d)/2$. This 
exponent agrees with model F critical dynamics only at $d=3$. 
There the dynamic exponent is given {\it exactly} by 
$z=d/2$ \cite{Hohenberg}. At $d=3$ the value of $z$ is indeed 
expected to be $z=3/2$ \cite{Ferrell}. However, a word of caution is 
necessary here. Our calculation of the dynamic exponent was made 
assuming that finite temperature dynamics can be derived out of 
a quantum Hamiltonian using equilibrium statistical mechanics. 
This is not quite right. Critical dynamics 
cannot be derived from a Hamiltonian and equilibrium statistical 
mechanics analysis \cite{Hohenberg}.

The full classical theory, including dynamical effects, should 
account for the frequency dependence of $\hat{\Pi}(i\omega_n,{\bf p})$. 
The full expression for $\hat{\Pi}(i\omega_n,{\bf p})$   
in the classical approximation and $2<d<4$ is \cite{Kondor3} 

\begin{equation}
\label{FullPi}
\hat{\Pi}(i\omega_n,{\bf p})=-A_dm^2Tp^{d-4}
\left[\left(1-\frac{2mi\omega_n}{p^2}\right)^{d-3}+
e^{-i\pi(d-2)}\left(1+\frac{2mi\omega_n}{p^2}\right)^{d-3}\right],
\end{equation}
where

\begin{equation} 
A_d=2^{2-d}\pi^{-d/2}e^{i\pi(d-2)/2}\Gamma(d/2-1)\Gamma(3-d).
\end{equation}
The derivation of Eq. (\ref{FullPi}) is made in Appendix \ref{app:bubble}. 
As $\omega_n\to 0$ Eq. (\ref{FullPi}) reduces correctly to 
$\hat{\Pi}(0,{\bf p})=\hat{\Pi}_0({\bf p})=-\alpha_dTm^2p^{d-4}$ and 
we recover Eq. (\ref{ECFTint}) for the effective interaction. 

For $d=3$ Eq. (\ref{FullPi}) becomes

\begin{equation}
\label{FullPi3d}
\hat{\Pi}(i\omega_n,{\bf p})=-i\frac{Tm^2}{2\pi p}\ln\left(
\frac{2mi\omega_n+p^2}{2mi\omega_n-p^2}\right).
\end{equation}
By making the replacement $i\omega_n\to\widetilde{E}(p)+i\delta$ and 
substituting in Eq. (\ref{spectrum1}), we obtain after 
taking the real part of the effective interaction the 
following expression for the excitation 
spectrum

\begin{equation}
\label{spectrum3d}
\widetilde{E}(p)=\frac{p^2}{2m}\left[1+\frac{4mnNg}{p^2
+\left(\frac{Tm^2Ng}{2\pi}
\ln\left|\frac{2m\widetilde{E}(p)+p^2}{2m\widetilde{E}(p)-p^2}
\right|\right)^2}\right]^{1/2}.
\end{equation}

%
%

\subsection{Depletion of the condensate}

At $T=0$ the depletion of the condensate is more easily obtained from the formula

\begin{equation}
n=\frac{1}{N}(|b_0|^2+\langle|\tilde{b}|^2\rangle)=
n_0+\frac{1}{N}\langle|\tilde{b}|^2\rangle,
\end{equation}
where

\begin{equation}
\label{depletion}
\langle|\tilde{b}|^2\rangle=
\lim_{\varepsilon\to 0^+}
\int_{-\infty}^\infty\frac{d\omega}{2\pi}
\int\frac{d^dp}{(2\pi)^d}\hat{\bf G}_{\tilde{b}^*\tilde{b}}(
i\omega,{\bf p})e^{i\omega\varepsilon}.
\end{equation}
Note that for $T=0$ the polarization bubble vanishes and we can set  
$\hat{\Gamma}(i\omega,p)= g$. 
Explicit evaluation of the $\omega$-integral in Eq. (\ref{depletion}) 
yields

\begin{equation}
\label{depletion1}
\langle|\tilde{b}|^2\rangle=\int\frac{d^dp}{(2\pi)^d}
\left[\frac{
\frac{{\bf p}^2}{2m}+nNg}{2\widetilde{E}({\bf p})}-\frac{1}{2}\right],
\end{equation}
where   
$\widetilde{E}({\bf p})=\sqrt{{\bf p}^4/4m^2+nNg{\bf p}^2/m}$.  
The integral in Eq. (\ref{depletion1}) can be easily evaluated 
using dimensional regularization \cite{Andersen}, so that the factor $1/2$ between 
brackets does not contribute. Thus, we can rewrite Eq. (\ref{depletion1}) as

\begin{equation}
 \langle|\tilde{b}|^2\rangle=\frac{1}{2}\left(A+\frac{M^2}{2}B\right),
\end{equation}
where $M^2=4mnNg$ and

\begin{equation}
 A=\int\frac{d^dp}{(2\pi)^d}\frac{p}{\sqrt{p^2+M^2}},
\end{equation}
and

\begin{equation}
 B=\int\frac{d^dp}{(2\pi)^d}\frac{1}{p\sqrt{p^2+M^2}}.
\end{equation}
The integrals $A$ and $B$ are related to the integral

\begin{equation}
I_\alpha(d)= \int\frac{d^dp}{(2\pi)^d}\frac{1}{(p^2+M^2)^\alpha},
\end{equation}
by

\begin{equation}
 A=\frac{2\sqrt{\pi}\Gamma\left(\frac{d+1}{2}\right)}{\Gamma(d/2)}I_{1/2}(d+1),
\end{equation}
and

\begin{equation}
 B=\frac{\Gamma\left(\frac{d-1}{2}\right)}{2\sqrt{\pi}\Gamma(d/2)}I_{1/2}(d-1).
\end{equation}
The integral $I_\alpha$ can be evaluated with the usually tricks already employed 
in Chapter \ref{ch:fm}. We have evaluated it in Appendix \ref{app:int-d-m}. The 
result is given in Eq. (\ref{I-alpha}).  Thus, , 

\begin{equation}
 \langle|\tilde{b}|^2\rangle=\frac{(mnNg/\pi)^{d/2}}{2\sqrt{\pi}\Gamma(d/2)}\left[\Gamma\left(-\frac{d}{2}\right)\Gamma\left(\frac{d+1}{2}\right)+
\frac{1}{2}\Gamma\left(1-\frac{d}{2}\right)\Gamma\left(\frac{d-1}{2}\right)\right]. 
\end{equation}
From the relation $\Gamma(z+1)=z\Gamma(z)$, we have 

\begin{equation}
 \Gamma\left(1-\frac{d}{2}\right)=-\frac{d}{2}\Gamma\left(-\frac{d}{2}\right),
\end{equation}

\begin{equation}
 \Gamma\left(\frac{d+1}{2}\right)=\Gamma\left(\frac{d-1}{2}+1\right)=\frac{d-1}{2}\Gamma\left(\frac{d-1}{2}\right),
\end{equation}
such that 

\begin{equation}
\label{depletion2}
 \langle|\tilde{b}|^2\rangle=\frac{(mnNg/\pi)^{d/2}(d-2)}{8\sqrt{\pi}\Gamma(d/2)}\Gamma\left(-\frac{d}{2}\right)\Gamma\left(\frac{d-1}{2}\right).
\end{equation}
It can be shown that for $2<d<4$ the coupling $g$ is given in terms of the $s$-wave scattering length by \cite{NogKlei}

\begin{equation}
 g=\frac{4\pi^{d/2}a^{d-2}}{\Gamma(d/2-1)m}.
\end{equation}
For $d=3$ this yields the well-known formula $g=4\pi a/m$.  
Setting $d=3$ finally yields

\begin{equation}
\label{cond1}
n_0=n\left(1-\frac{8}{3}\sqrt{\frac{Nna^3}{\pi}}\right), 
\end{equation}
which for $N=1$ is the usual Bogoliubiv's formula for the depletion of the condensate \cite{Bogoliu}. 

\subsection{The superfluid density}

In order to calculate the superfluid density, we have to perform a Galilei boost in the system \cite{Weichman}. A normal fluid is insensitive 
to a Galilei boost but a superfluid is not. This fact imposes strong constraints on the form of the excitation spectrum of a superfluid \cite{Abrikosov}.
For instance, it implies that an ideal Bose gas in its condensed phase is not a superfluid, although we sometimes speak in this case of 
``ideal superfluidity'', since the calculation of the superfluid density using the standard definition implies in this case that it is identical 
to the condensate density. This is absolutely not the case in general. Indeed, superfluidity may occur even when no Bose condensation is possible, 
as for example in the case of two-dimensional interacting Bose systems at finite temperature, where the Hohenberg-Mermin-Wagner 
\cite{Hohenberg1,MW} theorem forbids the appearance of a Bose condensate. In this situation there is a phase transition to a superfluid state 
without spontaneous symmetry breaking, which is the Kosterlitz-Thouless phase transition \cite{KT} discussed 
in Chap. \ref{ch:KT}. Another example comes from 
three-dimensional superfluids. It is easy to show that for a homogeneous superfluid at zero temperature the superfluid density corresponds to the 
whole density \cite{Gavoret}, while the condensate is depleted due to interaction effects \cite{Abrikosov,Bogoliu}. Experimentally it is estimated 
that for $^4$He in three dimensions and at zero temperature the condensate fraction is about 13\% \cite{Enss}, although the whole system is superfluid. 

A Galilei boost of momentum ${\bf k}_0$ changes the Lagrangian to

\begin{equation}
\label{L-nr}
 {\cal L}_{{\bf k}_0}={\cal L}+\frac{k_0^2}{2m}|b|^2+{\bf k}_0\cdot{\bf j},
\end{equation}
where

\begin{equation}
 {\bf j}=-\frac{i}{2m}(b^*\nabla b-b\nabla b^*)
\end{equation}
is the current density. 

The free-energy density in the presence of the Galilei boost is given by the functional integral

\begin{equation}
 F_{{\bf k}_0}=-\frac{T}{V}\left[\ln\int{\cal D}b^*{\cal D}b e^{-\int_0^\beta d\tau\int d^dr{\cal L}_{{\bf k}_0}}\right].
\end{equation}
The superfluid density is defined by \cite{Weichman}

\begin{equation}
\label{rhos-def}
 \rho_s=m^2\lim_{{\bf k}_0\to 0}\frac{\partial^2F_{{\bf k}_0}}{\partial {\bf k}_0^2}.
\end{equation}
It is now easy to show that

\begin{equation}
\label{rhos}
 \rho_s=m\langle|b|^2\rangle-\frac{m^2}{d}\int_0^\beta d\tau\int d^dr C(\tau,{\bf r}),
\end{equation}
where

\begin{equation}
 C(\tau,{\bf r})=\langle{\bf j}(\tau,{\bf r})\cdot{\bf j}(0,0)\rangle-\langle{\bf j}(\tau,{\bf r})\rangle\cdot\langle{\bf j}(0,0)\rangle
\end{equation}
is the connected current correlation function and 
the factor $1/d$ in Eq. (\ref{rhos}) emerges due to rotational invariance. 
Note that $\langle|b|^2\rangle$ is the total density of the fluid and sometimes we define 
$\rho=m\langle|b|^2\rangle$ as the total fluid density. 

As an example, let us calculate the superfluid density for the Lagrangian (\ref{L-nr}) in a regime where the  non-Gaussian fluctuations are small using a Bogoliubov transformation \cite{Bogoliu} for 
the Lagrangian (\ref{L-nr}). We will not consider the effects of vortices, 
which is certainly important when perfoming calculations using directly the phase of the order parameter \cite{Kleinert-GFCM-1,KT}.  

In the Bogoliubov approximation the superfluid density is given by

\begin{equation}
 \rho_s=\rho-\frac{T}{4d}\sum_{n=-\infty}^\infty\int\frac{d^dk}{(2\pi)^d}{\bf k}^2
\left[G^2(\omega_n,{\bf k})-|F(\omega_n,{\bf k})|^2\right],
\end{equation}
where we the 
Green functions are given by the zero temperature counterpart of the Green functions (\ref{G11}) and (\ref{Fs}), which we write as  

\begin{equation}
 G(\omega_n,{\bf k})=\frac{i\omega_n+\frac{{\bf k}^2}{2m}+\frac{g\rho}{m}}{\omega_n^2+\left(\frac{{\bf k}^2}{2m}
\right)^2+\frac{g\rho}{m^2}{\bf k}^2},
\end{equation}
and

\begin{equation}
 F(\omega_n,{\bf k})=\frac{gb_0^2}{\omega_n^2+\left(\frac{{\bf k}^2}{2m}
\right)^2+\frac{g\rho}{m^2}{\bf k}^2}.
\end{equation}
In the above equation $b_0=\langle b\rangle$ and we are 
approximately setting $\langle|b|^2\rangle\approx |b_0|^2$, which is true up to higher 
order effects. 

The above integral and Matsubara sum can be 
calculated exactly if we assume a low-energy approximation to the Bogoliubov spectrum, i.e., 

\begin{equation}
 E({\bf k})\approx c|{\bf k}|,
\end{equation}
where $c^2=g\rho/m^2$. Thus, by rescaling the momenta to remove the $c$ factor, the expression for 
the normal fluid density (note that $\rho_s=\rho-\rho_n$) becomes

\begin{eqnarray}
 \rho_n&=&\frac{T}{c^{d+2}d}\sum_{n=-\infty}^\infty\int\frac{d^dk}{(2\pi)^d}
\frac{{\bf k}^2}{\omega_n^2+{\bf k}^2}\nonumber\\
&-&\frac{2T}{c^{d+2}d}\sum_{n=-\infty}^\infty\int\frac{d^dk}{(2\pi)^d}
\frac{{\bf k}^2\omega_n^2}{(\omega_n^2+{\bf k}^2)^2},
\end{eqnarray}
which can be rewritten as

\begin{eqnarray}
 \rho_n&=&\frac{T}{c^{d+2}d}\int_{0<|{\bf k}|<\Lambda}\frac{d^dk}{(2\pi)^d}
+\frac{T}{c^{d+2}d}\sum_{n\neq 0}^\infty\int\frac{d^dk}{(2\pi)^d}
\frac{{\bf k}^2}{\omega_n^2+{\bf k}^2}\nonumber\\
&-&\frac{2T}{c^{d+2}d}\sum_{n=-\infty}^\infty\int\frac{d^dk}{(2\pi)^d}
\frac{{\bf k}^2\omega_n^2}{(\omega_n^2+{\bf k}^2)^2}.
\end{eqnarray}

It is convenient in this case to perform the integrations first, and only afterwards evaluate the Matsubara sums. 
The first and second integrals in $\rho_n$ can be related to 
the integrals $I_1$ and $I_2$ evaluated in the Appendix \ref{app:int-d-m}, where 
$\omega_n^2$ plays the role of $M^2$. We only need to notice that

\begin{equation}
 \int\frac{d^dk}{(2\pi)^d}\frac{k^2}{(k^2+M^2)^\alpha}
=2\pi d~ I_\alpha(d+2),
\end{equation}
where $\alpha=1,2$. Note that we have to replace $d\to d+2$ in the integrals $I_1$ and 
$I_2$ of Appendix \ref{app:int-d-m}. Thus, 

\begin{equation}
 \int\frac{d^dk}{(2\pi)^d}
\frac{{\bf k}^2}{\omega_n^2+{\bf k}^2}=-\frac{\pi^{d/2}T^d}{d}\Gamma\left(1-\frac{d}{2}\right)|n|^d.
\end{equation}
Now the sum can be easily performed by using the definition of the zeta function (see Appendix \ref{app:be-int}), so 
that we obtain, 

\begin{equation}
 T\sum_{n\neq 0}^\infty\int\frac{d^dk}{(2\pi)^d}
\frac{{\bf k}^2}{\omega_n^2+{\bf k}^2}=-\frac{2\pi^{d/2}T^{d+1}}{d}\Gamma\left(1-\frac{d}{2}\right)\zeta(-d). 
\end{equation}
The remaining integral along with the Matsubara sum in $\rho_n$ is easily evaluated using a similar procedure.  
Therefore, after some simplifications, the final result for the superfluid density is

\begin{eqnarray}
 \rho_s&=&\rho-\frac{2m^{d+2}}{(g\rho)^{1+d/2}d}\left[\frac{\Lambda^dT}{(4\pi)^{d/2}d\Gamma(d/2)}
\right.\nonumber\\
&-&\left.(d+1)\pi^{d/2}\Gamma\left(1-\frac{d}{2}\right)\zeta(-d)T^{d+1}\right],
\end{eqnarray}
where $\Lambda$ is an ultraviolet cutoff which is of the order of the inverse scattering length, 
$\Lambda\sim a^{-1}$. The term linear in the temperature arises also in a mean-field calculation 
of the stiffness in the {\it classical} $XY$ model \cite{Ebner}. The term proportional 
to $T^{d+1}$ is a quantum correction due to phonon excitations of the Bose liquid \cite{Abrikosov}. 
The Landau formula usually only gives the contribution proportional to $T^{d+1}$.

For $d=2$ we have

\begin{equation}
\label{rhos-2D}
 \rho_s=\rho-\frac{m^4}{2\pi(g\rho)^2}\left[\frac{\Lambda^2T}{4}+3\zeta(3)T^3\right],
\end{equation}
while for $d=3$ we obtain
\begin{equation}
 \rho_s=\rho-\frac{2m^5}{9(g\rho)^{5/2}}\left(\frac{\Lambda^3T}{4\pi^2}+\frac{\pi^2}{5}T^4\right).
\end{equation}
It should be noted that for $d=2$ the coupling $g$ itself is density dependent, being given by 
\cite{Schick}

\begin{equation}
 g=-\frac{4\pi}{m\ln\left(\frac{e^{2\gamma}\rho a^2}{4m}\right)},
\end{equation}
where $\gamma$ is the Euler constant and 
the naive diluteness condition $\rho a^d/m\ll 1$ has to be modified for $d=2$ to 
$\ln\ln[m/(\rho a^2)]\gg 1$ \cite{FH}. This is in contrast with the $d=3$ case where 
$g=4\pi a/m$, with no dependence on the the total density.

\chapter{Mott insulators}
\label{ch:Mott}

\section{The Hubbard model}
\label{sect:Hubbard}

The simplest electronic model of a Mott insulator is provided by the so called Hubbard model:

\begin{equation}
 \label{Hubbard}
H=-t\sum_{\langle i,j\rangle}\sum_\sigma f_{i\sigma}^\dagger f_{j\sigma}-\mu\sum_{i,\sigma}n_{i\sigma}
+U\sum_i n_{i\uparrow}n_{i\downarrow},
\end{equation}
where $f_{i\sigma}$ is a fermion in the lattice, $n_{i\sigma}\equiv f_{i\sigma}^\dagger f_{i\sigma}$, and $\sigma=\uparrow,\downarrow$. In the above 
Hamiltonian $t, U>0$, and $\mu$ is the chemical potential. In the kinetic term the lattice sum runs only nearest neighbors only. 
The Hubbard Hamiltonian is exactly solvable in the limits $U=0$ and $t=0$. The non-interacting limit corresponds to band theory in the tight-binding 
approximation, while the $t=0$ limit corresponds to the atomic limit. Although in the atomic limit the theory is interacting, it is easily 
diagonalizable, since

\begin{equation}
 H_{t=0}=\sum_i h_i,
\end{equation}
where 

\begin{equation}
 h_i=-\mu\sum_{\sigma}n_{i\sigma}+U\sum_i n_{i\uparrow}n_{i\downarrow}.
\end{equation}
Thus, 

\begin{equation}
 [h_i,n_{i\alpha}]=0,
\end{equation}
and therefore the eigenstates of $n_{i\alpha}$ are also eigenstates of $h_i$. In this case we can simply omit the lattice sites of any 
calculation, since the sites are decoupled. The eigenstates of the $h$ are $|0\rangle$, $|\uparrow\rangle$, $|\downarrow\rangle$, and 
$|\uparrow\downarrow\rangle$, corresponding to empty, singly occupied (with either up or down spins), and doubly occupied sites, 
respectively. The corresponding eigenenergies are $\varepsilon_0=0$, $\varepsilon_\uparrow=\varepsilon_\downarrow=-\mu$, and 
$\varepsilon_2=U-2\mu$. 

Let us consider the example of a half-filled band, i.e., the total number of fermions in the system equals the number of lattice sites $L$. In 
this case it can be shown that for a bipartite lattice\footnote{This means that the lattice can be viewed as being made of two interpenetrating 
sublattices. One simple example is the cubic lattice} 
$\mu=U/2$ {\it exactly}. This result is straightforwardly checked in the atomic limit. To see that this result is 
also valid when $t\neq 0$, we just perform a particle-hole (ph) transformation $f_{i\sigma}\to e^{i{\bf Q}\cdot{\bf R}_i}f_{i\sigma}^\dagger$, 
$f_{i\sigma}^\dagger\to e^{i{\bf Q}\cdot{\bf R}_i}f_{i\sigma}$, where ${\bf Q}=(\pi,...,\pi)$. When $\mu=U/2$ the Hamiltonian is invariant under 
this particle-hole transformation. 
Note that the factor $e^{i{\bf Q}\cdot{\bf R}_i}$ is necessary in order to maintain the sign of the hopping term, because fermions anticommute. 
Indeed, we have that $e^{i{\bf Q}\cdot({\bf R}_i-{\bf R}_j)}=-1$, since ${\bf R}_i$ and ${\bf R}_j$ are nearest neighbor sites. Thus, the 
ph-transformation yields in general the Hamiltonian

\begin{equation}
\label{Hubbard-ph}
 H'=U-2\mu-t\sum_{\langle i,j\rangle}\sum_\sigma f_{i\sigma}^\dagger f_{j\sigma}+(\mu-U)\sum_{\sigma}n_{i\sigma}+U\sum_i n_{i\uparrow}n_{i\downarrow}
\end{equation}
Note that the Hamiltonians $H$ from Eq. (\ref{Hubbard}) and the one given above coincide for $\mu=U/2$. 
Now let $f$ and $f'$ be the free energy densities of the Hamiltonians $H$ and $H'$ given in Eqs. (\ref{Hubbard}) and (\ref{Hubbard-ph}), 
respectively. We have the particle density is given as usual by the thermodynamical relation 

\begin{equation}
\label{n-1}
 n=-\frac{\partial f}{\partial\mu},
\end{equation}
while from $f'$ we obtain, 

\begin{equation}
\label{n-2}
 2-n=-\frac{\partial f'}{\partial\mu}.
\end{equation}
Since $\mu=U/2$ implies $H=H'$, we also have that $f=f'$ for this value of the chemical potential. Therefore, for 
$\mu=U/2$ the RHS of both Eqs. (\ref{n-1}) and (\ref{n-2}) are the same, so that we obtain $n=2-n$, which yields $n=1$. 

At half-filling the Hubbard Hamiltonian can be rewritten as

\begin{equation}
 H=-t\sum_{\langle i,j\rangle}\sum_\sigma f_{i\sigma}^\dagger f_{j\sigma}-\frac{2U}{3}\sum_i{\bf S}_i^2,
\end{equation}
where

\begin{equation}
 {\bf S}_i=\frac{1}{2}\sum_{\alpha,\beta}f_{i\alpha}^\dagger\sigmab_{\alpha\beta}f_{i\beta},
\end{equation}
with $\sigmab\equiv(\sigma_1,\sigma_2,\sigma_3)$, $\sigma_i$ being the Pauli matrices. For $U\gg t$ doubly occupied sites are strongly 
suppressed and second-order perturbation theory yields the effective Hamiltonian \cite{Fradkin}

\begin{equation}
\label{Heisenberg-1}
 H=\frac{4t^2}{U}{\bf S}_i\cdot{\bf S}_j,
\end{equation}
subjected to the local constraint

\begin{equation}
\label{constraint}
 \sum_\sigma n_{i\sigma}=1.
\end{equation}
Note the subtlety here. The half-filling condition demands that the particle density 

\begin{equation}
 n=\frac{1}{L}\sum_{i,\sigma}\langle n_{i\sigma}\rangle=1,
\end{equation}
which is easily enforced when $\mu=U/2$.  This is a global constraint, that simply demands the average site occupation be the unity. However, when 
$U\gg t$ the average constraint becomes a local one given by the operator equation (\ref{constraint}), i.e., double occupation is strictly forbidden. 

The effective Hamiltonian (\ref{Heisenberg-1}) is the one of a Heisenberg antiferromagnet. It is rotational invariant in spin space, i.e., it has an 
$SU(2)$ symmetry. This means that the total spin operator 

\begin{equation}
\label{Stot}
 {\bf S}=\sum_i{\bf S}_i,
\end{equation}
commutes with the Hamiltonian. However, the ground state of the Heisenberg Hamiltonian breaks this symmetry. The most favorable state at zero 
temperature corresponds to alternating spins in the lattice. This state is pictorially shown for a square lattice in Fig. 2.1.
The antiferromagnetic state shown in the figure is actually a mean-field state for the Heisenberg model, the so called N\'eel state. This 
state is also a mean-field state of the Hubbard model at half-filling. Incidentally, the total spin operator (\ref{Stot}) obviously commutes 
with the Hubbard Hamiltonian, showing that the Hubbard model is $SU(2)$ symmetric, as expected physically. 

\begin{figure}
\begin{center}
 \includegraphics[width=10cm]{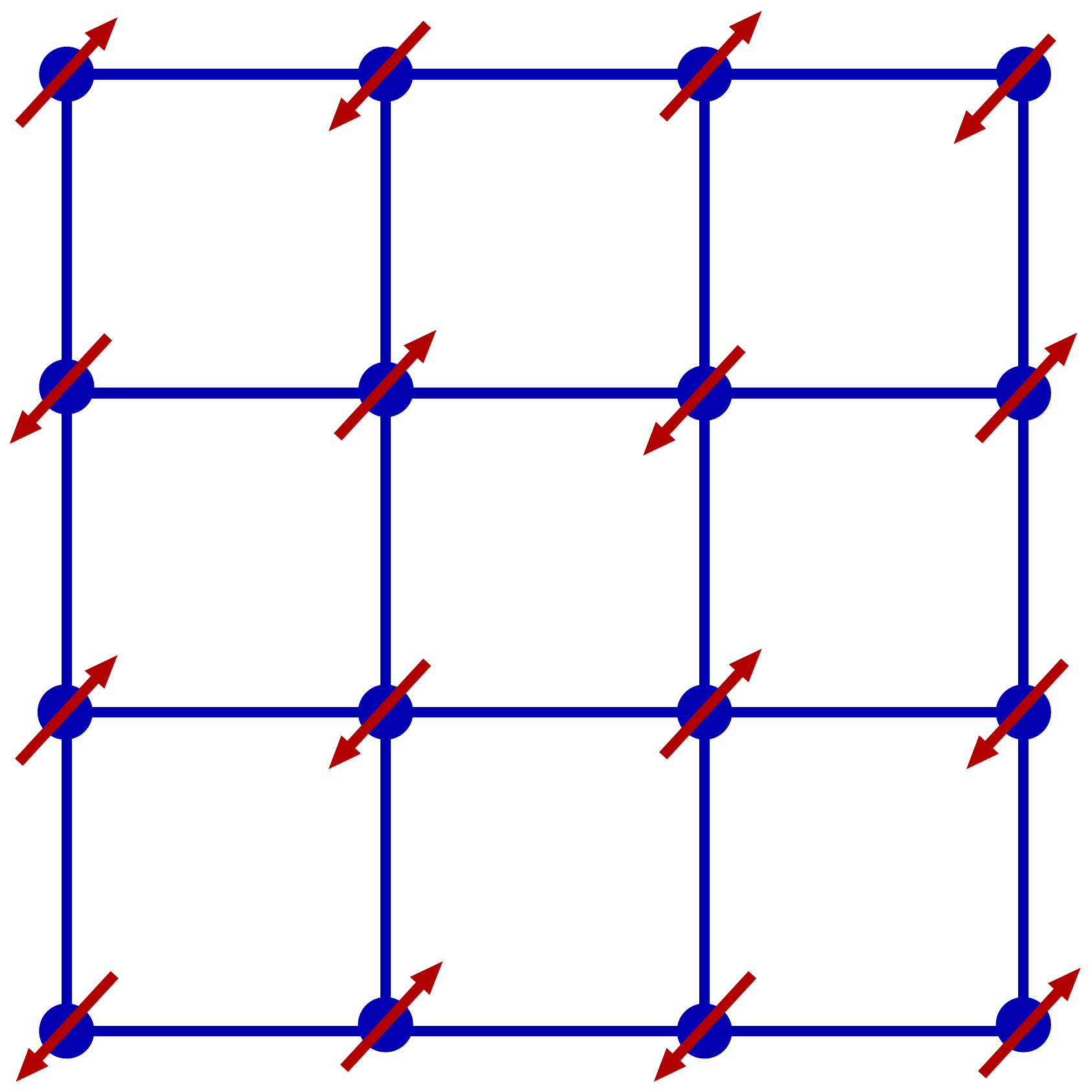}
\caption{Mean-field ground state for a Heisenberg antiferromagnet, the so called N\'eel state.}\label{fig:AF-state} 
\end{center}
\end{figure}  

Let us perform the mean-field theory for the Hubbard model in a $d$-dimensional cubic 
lattice explicitly. In order to do this, we introduce an auxiliary field via a 
Hubbard-Stratonovich transformation:

\begin{equation}
 H=-t\sum_{\langle i,j\rangle}\sum_\sigma f_{i\sigma}^\dagger f_{j\sigma}-U\sum_i{\bf m}_i\cdot{\bf S}_i+\frac{3U}{8}\sum_i{\bf m}_i^2.
\end{equation}
We are looking for a mean-field state with a staggered magnetic moment 

\begin{equation}
 {\bf m}_i=e^{i{\bf Q}\cdot{\bf R}_i}{\bf m},
\end{equation}
where the vector ${\bf m}$ is uniform. Rotational invariance allows us to fix a direction for ${\bf m}$. We will choose the quantization axis to 
be along the $z$-direction. Thus, ${\bf m}=m{\bf e}_z$. Let us define

\begin{equation}
\label{f-sublat}
 f_{i\sigma}=
\left\{
\begin{array}{c}
c_{i\sigma},~~~~~i\in A\\
\noalign{\medskip}
\bar c_{i\sigma},~~~~~i\in B
\end{array}
\right.
\end{equation}
The mean-field Hamiltonian can be written as

\begin{equation}
 H_{\rm MF}=\sum_{{\bf k},\sigma}\psi_{{\bf k}\sigma}^\dagger M_{{\bf k}\sigma}\psi_{{\bf k}\sigma}
+\frac{3UL}{8}m^2,
\end{equation}
where

\begin{equation}
 \psi_{{\bf k}\sigma}=\left[
\begin{array}{c}
 c_{{\bf k}\sigma}\\
\noalign{\medskip}
\bar c_{{\bf k}\sigma}
\end{array}
\right],~~~~~~~~
\psi_{{\bf k}\sigma}^\dagger=\left[
\begin{array}{cc}
 c_{{\bf k}\sigma}^\dagger & \bar c_{{\bf k}\sigma}^\dagger
\end{array}
\right],
\end{equation}
and

\begin{equation}
M_{{\bf k}\sigma}=\left[
\begin{array}{cc}
-\frac{\sigma Um}{2} & \varepsilon_{\bf k}\\
\noalign{\medskip}
\varepsilon_{\bf k} & \frac{\sigma Um}{2}
\end{array}
\right],
\end{equation}
with the tight-binding dispersion

\begin{equation}
 \varepsilon_{\bf k}=-2t\sum_{a=1}^d\cos k_a.
\end{equation}
The mean-field Hamiltonian is easily diagonalized and leads to the energy spectrum

\begin{equation}
 E_{\bf k}^\pm=\pm\sqrt{\varepsilon_{\bf k}^2+\frac{U^2m^2}{4}},
\end{equation}
and we see that the electronic spectrum is gapped.
Thus, the mean-field ground state energy per site is

\begin{equation}
 E_0=-\frac{2}{L}\sum_{\bf k}'E_{\bf k}^++\frac{3U}{8}m^2,
\end{equation}
where the prime on the sum over ${\bf k}$ is to denote that we are summing over the upper 
half of the Brillouin zone. Let us specialize to two dimensions. By minimizing the 
above equation with respect to $m$, we obtain the gap equation,

\begin{equation}
 \frac{3}{2U}=\int_0^\pi\frac{dk_x}{2\pi}\int_0^\pi\frac{dk_y}{2\pi}
\frac{1}{E_{\bf k}^+}.
\end{equation}
In two dimensions it is not a too bad approximation, at least for more qualitative purposes, 
to use a square density of states to evaluate the above momentum integral, 

\begin{equation}
 \rho(\varepsilon)=\frac{1}{W}\theta\left(\frac{W}{2}-|\varepsilon|\right),
\end{equation}
where $W=4t$ is the bandwidth and $\theta(x)$ is the Heaviside function. This yields the 
magnetization,

\begin{equation}
 m=\frac{2W}{U}\frac{e^{-\frac{3W}{2U}}}{1-e^{-\frac{3W}{U}}}.
\end{equation}
Thus, mean-field theory predicts that at half-filling the Hubbard model is an antiferromagnetic insulator for all 
$U>0$. This is not quite accurate, of course. For a small enough $U$ the system should be a metal. Indeed, the Hubbard 
model undergoes a metal insulator transition for a critical value of the interaction typically of the order of the bandwidth $W$ 
\cite{Hubbard-III}. 
The metal-insulator transition in the Hubbard model started to be better understood with the use of the so called dynamical mean-field 
\cite{Kotliar-RMP}, 
which is based on the large dimension limit of the Hubbard model \cite{Metzner}. For a thorough review on the subject, see Ref. \cite{Kotliar-RMP}.

\section{The Bose-Hubbard model}

The Hamiltonian of the so called Bose-Hubbard model \cite{Fisher} is given by
\begin{equation}
\label{BHM}
\hat H=-J\sum_{\langle i,j\rangle}\hat b_i^\dagger \hat b_j
-\mu\sum_i\hat n_i+\frac{U}{2}\sum_i\hat n_i(\hat n_i-1),
\end{equation}
where $U,J>0$, $\hat n_i\equiv \hat b_i^\dagger \hat b_i$ and 
$\mu$ is the chemical potential. The summations are over 
the sites of a cubic lattice and the symbol $\langle i,j\rangle$ means a sum over 
nearest neighbors. The operators $b_i$ obey the usual bosonic commutation relations, 
i.e., $[\hat b_i,\hat b_j^\dagger]=\delta_{ij}$ und $[\hat b_i,\hat b_j]=[\hat b_i^\dagger,\hat b_j^\dagger]=0$.

The aim of this tutorial is to provide an introduction to the theory of the 
Bose-Hubbard model, which 
in the last years gained considerable experimental relevance in the context of 
Bose-Einstein condensation (BEC) \cite{Greiner}. 

What kind of phases we expect for the above model? Firstly, let us note that the 
Hamiltonian (\ref{BHM}) is simply a lattice version of the interacting Bose gas Hamiltonian. 
Indeed, the hopping term is just a lattice derivative. Thus, for small enough 
$U$ we expect to obtain a superfluid featuring the well known Bogoliubov 
spectrum. To see this, just write 
\begin{equation}
\hat b_i=b_0+\delta \hat b_i,~~~~~~~\hat b_i^\dagger=b_0^*+\delta \hat b_i^\dagger,
\end{equation}
where $b_0$ represents the condensate and minimize the Hamiltonian, and 
$\delta \hat b_i$ are small fluctuations around the condensate. By keeping 
just the quadratic fluctuations, it is easy to see that the Hamiltonian 
can be approximately written as $\hat H=LE_0+\delta\hat H$ ($L$ is the number of 
lattice sites), with 
\begin{equation}
E_0=-\left(2dJ+\mu+\frac{U}{2}\right)n_0+\frac{U}{2}n_0^2,
\end{equation} 
\begin{equation}
\delta\hat H=\frac{1}{2}\sum_{{\bf k}\neq 0}\hat \Psi_{\bf k}^\dagger \hat{\bf M}_{\bf k}\hat \Psi_{\bf k},
\end{equation}
where we have performed a Fourier transformation in the lattice and
\begin{equation}
\hat \Psi^\dagger_{\bf k}=\left[
\begin{array}{cc}
\delta\hat{b}^\dagger_{\bf k} & \delta\hat{b}_{-{\bf k}}
\end{array}
\right],~~~~
\hat \Psi_{\bf k}=
\left[
\begin{array}{c}
\delta\hat{b}_{{\bf k}}\\
\noalign{\medskip}
\delta\hat{b}^\dagger_{-{\bf k}}
\end{array}
\right],
\end{equation}
\begin{equation}
\label{matrix}
\hat{\bf M}_{\bf k}=
\left[
\begin{array}{cc}
\varepsilon_{\bf k}-\mu-U/2+2Un_0 & Ub_0^2\\
\noalign{\medskip}
U(b_0^*)^2 & \varepsilon_{\bf k}-\mu-U/2+2Un_0
\end{array}
\right],
\end{equation}
with $\varepsilon_{\bf k}=-2J\sum_{a=1}^d\cos k_a$ and $n_0\equiv|b_0|^2$. Due to the 
minimization condition we have  
\begin{equation}
\label{mu}
\mu=-2dJ-\frac{U}{2}+Un_0.
\end{equation}
The above result follows easily by demanding 
that the linear terms in both $\delta\hat b_i$ and $\delta\hat b_i^\dagger$ vanish. 
Alternatively it may be derived by simply minimizing $E_0$ with respect 
to $n_0$. The energy spectrum can be obtained by simply solving the 
Heisenberg equations of motion. To this end we need the equations
\begin{equation}
i\partial_t\delta\hat b_{\bf k}=[\delta\hat b_{\bf k},\delta\hat H],~~~~~~~~
i\partial_t\delta\hat b_{-{\bf k}}^\dagger=[\delta\hat b_{-{\bf k}}^\dagger,\delta\hat H].
\end{equation}
After straightforward 
evaluation of the commutators, we can rewrite the two equations above as a 
single matrix equation:
\begin{equation}
\label{Heisenberg}
i\hat \sigma_3\partial_t\hat \Psi_{\bf k}=\hat{\bf M}_{\bf k}\hat \Psi_{\bf k},
\end{equation}
where $\hat \sigma_3$ is the third Pauli matrix. The Ansatz
\begin{equation}
\hat \Psi_{\bf k}(t)=e^{-iE_{\bf k}t}\hat \Psi_{\bf k}(0)
\end{equation}
solves Eq. (\ref{Heisenberg}) provided $\det(E_{\bf k}\hat \sigma_3-\hat{\bf M}_{\bf k})=0$, or 
\begin{equation}
\label{Bog}
E_{\bf k}=\pm\sqrt{(\varepsilon_{\bf k}-\mu-U/2+2Un_0)^2-U^2n_0^2}.
\end{equation}
Note that due to the form of the chemical potential (\ref{mu}), the above 
spectrum is gapless, i.e., $E_{{\bf k}=0}=0$, as required by superfluidity. 

The above results are valid for $U$ small, i.e., $U\ll J$. So, what happens now 
in the opposite limit, when $U\gg J$? This is a subtle question. In the 
continuum limit {\it exact} arguments \cite{Gavoret,Hohenberg} 
involving the Ward identities show that at 
zero temperature the superfluid density is identical to the particle density, and 
this for {\it for all} values of $U$. This means that at zero temperature 
the whole system is in a superfluid state, although not every particle is condensed, 
since the condensate is depleted due to interaction effects. Does this exact result also 
holds in the lattice? The answer is: it depends on whether the particle density 
$\langle\hat n\rangle$ 
is integer or not! For noninteger particle density, approaching the strong 
coupling limit from the weak coupling one by varying $J/U$ essentially does not 
change the superfluid characteristics of the system. Thus, in this situation the 
system is still a superfluid for $U\gg J$. However, for 
$\langle\hat n\rangle=n\in\mathbb{N}$ the situation is different. In this case 
the bosons will localize for large enough $U$ and the system will become an 
insulator, whose ground state has $n$ particles per site. In order to better 
understand how it actually works, let us consider the eigenstates of the 
number operator (we omit the site index for simplicity)
\begin{equation}
|n\rangle=\frac{1}{\sqrt{n!}}(\hat b^\dagger)^n|0\rangle, 
\end{equation}
and the coherent state 
\begin{eqnarray}
|z\rangle&=&e^{-|z|^2/2}\exp(z\hat b^\dagger)|0\rangle\nonumber\\
&=&e^{-|z|^2/2}\sum_{n=0}^{\infty}\frac{z^n}{\sqrt{n!}}|n\rangle.
\end{eqnarray}
Recall that for a coherent state $\hat b|z\rangle=z|z\rangle$. Let us consider 
for simplicity a two-site problem. For this case let us assume that the 
system is in the coherent state
\begin{equation}
|\Phi\rangle=|z_1,z_2\rangle.
\end{equation}
The expectation value of the Hamiltonian in this state is the energy  
\begin{equation}
\label{E-coherent}
E(z_1,z_2)\equiv\langle\Phi|\hat H|\Phi\rangle=-J(z_1^*z_2+z_2^*z_1)+\sum_{i=1,2}
\left[-\mu|z_i|^2+\frac{U}{2}|z_i|^2(|z_i|^2-1)\right].
\end{equation}
In terms of $z_i=\sqrt{\bar n_i}e^{i\varphi_i}$, where $\bar n_i$ is the mean 
particle number at the site $i$, Eq. (\ref{E-coherent}) becomes
\begin{equation}
\label{E-coherent-1}
E_{\bar n_1,\bar n_2}(\varphi_1,\varphi_2)=-2J\sqrt{\bar n_1\bar n_2}\cos(\varphi_1-\varphi_2)
+\sum_{i=1,2}
\left[-\mu\bar n_i+\frac{U}{2}\bar n_i(\bar n_i-1)\right].
\end{equation}
The above energy is similar to the classical Hamiltonian of a Josephson junction \cite{Anderson}. In 
order to explore this similarity further, we rewrite Eq. (\ref{E-coherent-1}) as
%
%
%
\begin{equation}
\label{Jos}
E(\Delta\bar n,\Delta\varphi)=-J\sqrt{N^2-(\Delta\bar n)^2}\cos\Delta\varphi
+\frac{U}{4}(\Delta\bar n)^2+E_N,
\end{equation}
where $N=\bar n_1+\bar n_2$ is the total number of particles of the system, 
$\Delta\bar n\equiv\bar n_1-\bar n_2$, $\Delta\varphi\equiv\varphi_1-\varphi_2$, and
\begin{equation}
E_N=-\left(\mu+\frac{U}{2}\right)N+\frac{U}{4}N^2.
\end{equation}
The energy (\ref{Jos}) corresponds  
precisely to the Hamiltonian describing a two-level 
Bose-Einstein condensate via a Josephson junction as 
discussed by Leggett \cite{Leggett}. In this case 
the variable $\Delta\bar n$ plays the role of the momentum conjugated to 
$\Delta\varphi$. Therefore, the Josephson current is given 
by 
\begin{equation}
\partial_t\Delta\bar n=-J\sqrt{N^2-(\Delta\bar n)^2}
\sin\Delta\varphi.
\end{equation}
The Josephson effect implies superfluidity. Semiclassically, since $\Delta\bar n$ and 
$\Delta\varphi$ are canonically conjugated, we have the uncertainty relation \cite{Anderson} 
\begin{equation}
\Delta\bar n\Delta\varphi\sim 1.
\end{equation}
The so called Josephson ``phase-voltage'' relation is here generalized to

\begin{equation}
\label{Jos-2}
 \partial_t\Delta\varphi=\frac{U}{2}\Delta\bar n+\frac{J\Delta\bar n}{\sqrt{N^2-(\Delta\bar n)^2}}\cos\Delta\varphi.
\end{equation}
Recall that in the case of superconductors we have $\partial_t\Delta\varphi=2eV/\hbar$, where the voltage $V$ the same as the 
difference of chemical potential across the junction. In the above equation the role of $2eV/\hbar$ is played by 
$U\Delta\bar n/2$. The second term is absent in the Josephson relation. This term appears in the context of Josephson junctions in 
Bose-Einstein condensates \cite{Smerzi}.

If $U\gg J$, the second term in Eq. (\ref{Jos}) will constraint $\bar n_1\approx\bar n_2$ and 
a commensurate situation $\bar n_1=\bar n_2=n=1,2,3,\dots$ will be favored. It is 
then clear that the current will be zero and we have an insulator. 

Now we will solve the full model approximately using a Green function method. The aim is 
to compute the Green function
\begin{equation}
G_{ij}(t)=-i\langle T[\hat b_i(t)\hat b_j^\dagger(0)]\rangle,
\end{equation}
where $T[\hat b_i(t_1)\hat b_j^\dagger(t_2)]=\theta(t_1-t_2)\hat b_i(t_1)\hat b_j^\dagger(t_2)
+\theta(t_2-t_1)\hat b_j^\dagger(t_2)\hat b_i(t_1)$ and $\theta(t)$ is the 
Heaviside function. The Green function can be calculated exactly in both limit cases 
$J=0$ and $U=0$. We will consider a solution that corresponds to a perturbation 
expansion in $J/U$, i.e., around the limit case $J=0$. Thus, we have to compute the 
exact Green function in this limit in order to proceed. 

The $J=0$ limit is actually a single-site system, since the Hamiltonian can be written as 
a sum of single-site Hamiltonians:
\begin{equation}
\hat H_{J=0}=\sum_i\hat h_i,
\end{equation}
where
\begin{equation}
\hat h_i=-\mu\hat n_i+\frac{U}{2}\hat n_i(\hat n_i-1).
\end{equation}
In this case it is enough to compute the Green function for the single-site 
Hamiltonian $\hat h_i$ and the site index can even be omitted. 
Since $[\hat h,\hat n]=0$, the eigenstates $|n\rangle$ of $\hat n$ are 
the exact eigenstates of $\hat h$ with eigenvalues
\begin{equation}
E_n=-\mu n+\frac{U}{2}n(n-1).
\end{equation}

We want 
to compute
\begin{eqnarray}
{\cal G}(t)&=&-i\langle T[\hat b(t)\hat b^\dagger(0)]\rangle
\nonumber\\
&=&-i[\theta(t)\langle\hat b(t)\hat b^\dagger(0)\rangle+
\theta(-t)\langle\hat b^\dagger(0)\hat b(t)\rangle].
\end{eqnarray}
Let us assume a ground state with $n\in\mathbb{N}$ particles per site. Then  
we have
\begin{eqnarray}
\langle\hat b(t)\hat b^\dagger(0)\rangle&=&\langle n|
\hat b(t)\hat b^\dagger(0)|n\rangle\nonumber\\
&=&\sqrt{n+1}\langle n|e^{i\hat h t}\hat b(0)e^{-i\hat h t}|n+1\rangle
\nonumber\\
&=&(n+1)e^{i(E_{n}-E_{n+1})t}
\end{eqnarray}
Similarly we find
\begin{equation}
\langle\hat b(0)^\dagger\hat b(t)\rangle=ne^{i(E_{n-1}-E_{n})t}.
\end{equation}
Therefore, 
\begin{equation}
\label{gt}
{\cal G}(t)=-i[\theta(t)(n+1)e^{i(E_{n}-E_{n+1})t}
+\theta(-t)ne^{i(E_{n-1}-E_{n})t}],
\end{equation}
such that the Fourier transformation
\begin{equation}
{\cal G}(\omega)=\int_{-\infty}^\infty dt e^{i\omega t}{\cal G}(t)
\end{equation}
reads
\begin{eqnarray}
\label{atomic-g}
{\cal G}(\omega)&=&\frac{n+1}{\omega+E_{n}-E_{n+1}+i\delta}
-\frac{n}{\omega+E_{n-1}-E_{n}-i\delta}\nonumber\\
&=&\frac{n+1}{\omega+\mu-Un+i\delta}
-\frac{n}{\omega+\mu-U(n-1)-i\delta},
\end{eqnarray}
where $\delta\to 0^+$. Note that in order to perform the Fourier 
transformation of (\ref{gt}) a convergence factor $e^{-\delta t}$ 
was used for the first term, while for the second term a 
convergence factor $e^{\delta t}$ was needed. 

The approximation we are going to employ is equivalent to a well known mean-field theory 
approach to solve the Bose-Hubbard model \cite{Sachdev}. Instead performing a mean-field 
theory in the Hamiltonian, we will compute the Green function approximately using an expansion 
on the hopping. Thus, the unperturbed Green function will be given by Eq. (\ref{atomic-g}). In order 
to better motivate the method of solution, let us show how it can be used to solve the 
exactly solvable limit $U=0$. For this particular case the Hamiltonian is easily diagonalized 
by a Fourier transformation:
\begin{equation}
\hat H_{U=0}=\sum_{\bf k}(\varepsilon_{\bf k}-\mu)\hat n_{\bf k},
\end{equation}
where $\hat n_{\bf k}=\hat b_{\bf k}^\dagger\hat b_{\bf k}$. The exact Green function is  
obviously given by
\begin{equation}
G({\bf k},\omega)=\frac{1}{\omega+\mu-\varepsilon_{\bf k}+i\delta}.
\end{equation}
Let us derive the above Green function in a less direct way, namely, via the hopping expansion. 
It will be a more or less complicate way to derive a straightforward result, but it will serve 
the purpose of illustrating the strategy to solve the Bose-Hubbard model approximately. 

The $U=0$ Hamiltonian can be decomposed in the following way:
\begin{equation}
\hat H_{U=0}=\hat H_0+\hat H_1,
\end{equation}
where
\begin{equation}
\hat H_0=-\mu\sum_i\hat n_i,
\end{equation}
is the unperturbed Hamiltonian, and 
\begin{equation}
\hat H_1=-J\sum_{\langle i,j\rangle}\hat b_i^\dagger\hat b_j,
\end{equation}
is the perturbation. The unperturbed Green function is given by
\begin{equation}
\label{atomic-g-0}
{\cal G}_0(\omega)=\frac{1}{\omega+\mu+i\delta}.
\end{equation}
The exact Green function can be obtained by performing the infinite sum of diagrams shown in 
Fig. 2.2. The continuum line represents the local Green function (\ref{atomic-g-0}) at a lattice site 
$i$, while the dashed line represents a hopping process between two neighboring sites. The exact 
Green function gives the response of the system to a boson propagating from a site $i$ at 
some time $t$ to another site $j$ at an earlier time $t'$. The diagrams of the figure illustrate 
all the possible processes for a quadratic Hamiltonian. In this case only tree diagrams 
contribute are nonzero, since $U=0$. The perturbation expansion reads simply
\begin{equation}
G_{ij}(\omega)={\cal G}_0(\omega)\delta_{ij}+{\cal G}_0(\omega)J_{ij}{\cal G}_0(\omega)
+{\cal G}_0(\omega)\sum_l J_{il}{\cal G}_0(\omega)J_{lj}{\cal G}_0(\omega)+\dots,
\end{equation}
where $J_{ij}=-J$ if $(i,j)$ are nearest neighbors and zero otherwise. This has the 
structure of a geometric series and can be rewritten as
\begin{eqnarray}
\label{G-series}
G_{ij}(\omega)&=&{\cal G}_0(\omega)\delta_{ij}+{\cal G}_0(\omega)\sum_l J_{il}
\left[\delta_{lj}{\cal G}_0(\omega)+{\cal G}_0(\omega)J_{lj}{\cal G}_0(\omega)+\dots\right]
\nonumber\\
&=&{\cal G}_0(\omega)\delta_{ij}+{\cal G}_0(\omega)\sum_lJ_{il}G_{lj}(\omega).
\end{eqnarray}
The Fourier representations 
\begin{equation}
J_{ij}=\frac{1}{L}\sum_{\bf k}e^{i{\bf k}\cdot({\bf R}_i-{\bf R}_j)}\varepsilon_{\bf k},
\end{equation}
\begin{equation}
G_{ij}(\omega)=\frac{1}{L}\sum_{\bf k}e^{i{\bf k}\cdot({\bf R}_i-{\bf R}_j)}G({\bf k},\omega),
\end{equation}
\begin{equation}
\delta_{ij}=\frac{1}{L}\sum_{\bf k}e^{i{\bf k}\cdot({\bf R}_i-{\bf R}_j)},
\end{equation}
allow us to rewrite Eq. (\ref{G-series}) in the simpler form:
\begin{equation}
G({\bf k},\omega)={\cal G}_0(\omega)+ {\cal G}_0(\omega)\varepsilon_{\bf k}G({\bf k},\omega),
\end{equation}
which can easily be solved to obtain once more the exact Green function for $U=0$: 
\begin{eqnarray}
G({\bf k},\omega)&=&\frac{1}{{\cal G}_0^{-1}(\omega)-\varepsilon_{\bf k}}
\nonumber\\
&=&\frac{1}{\omega+\mu-\varepsilon_{\bf k}+i\delta}.
\end{eqnarray}
\begin{figure}
\includegraphics[width=12cm]{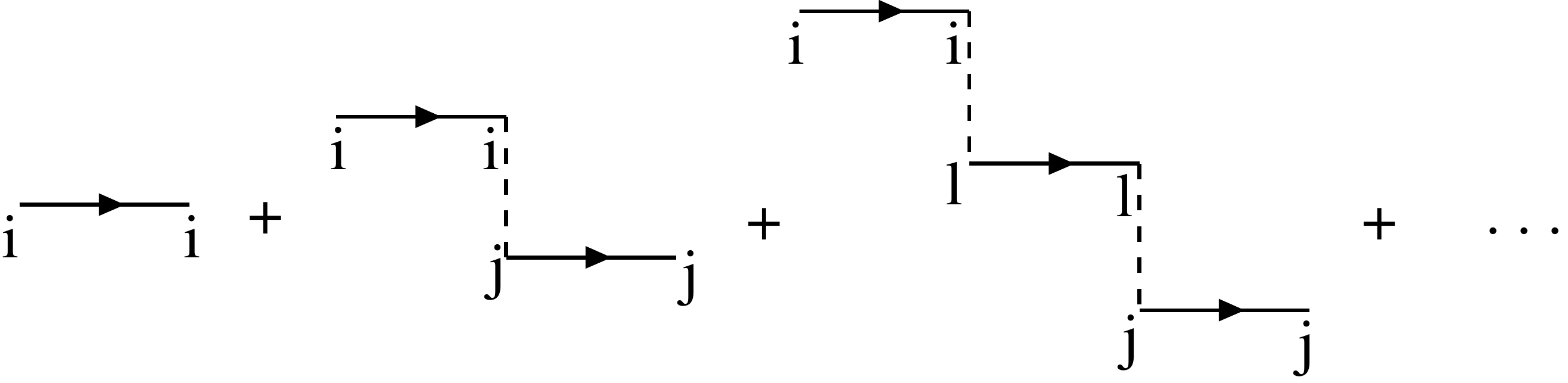}
\caption{Hopping expansion for the Green function.}\label{fig:hop-exp}
\end{figure}

\begin{figure}
\includegraphics[width=12cm]{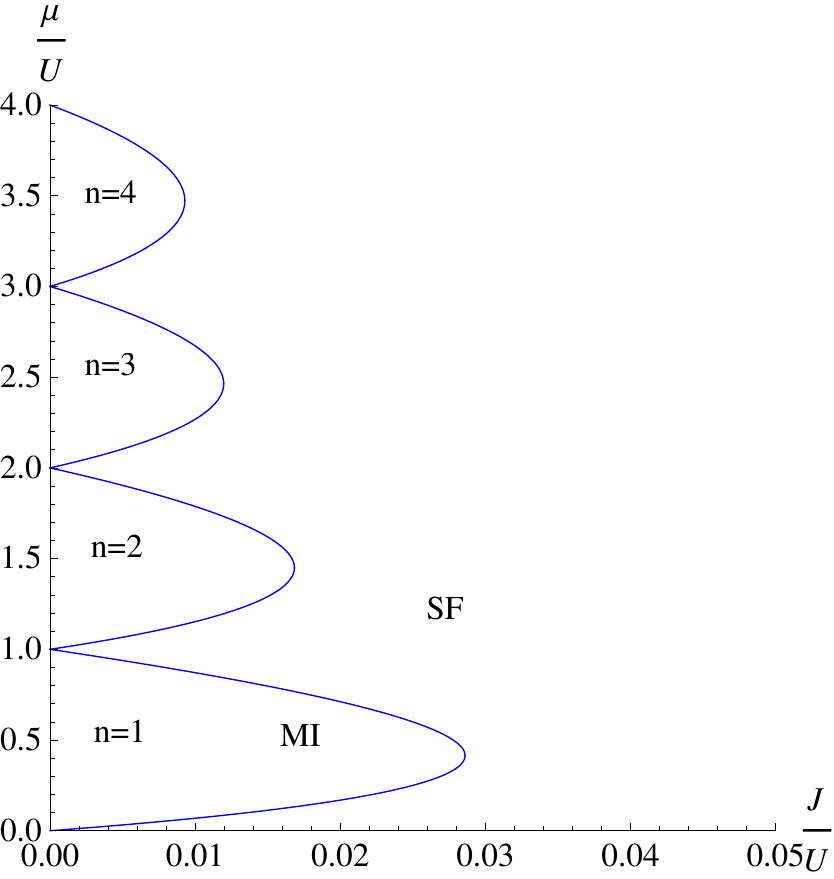}
\caption{Phase diagram of the Bose-Hubbard model.}\label{fig:pt-bhm}
\end{figure}

Unfortunately, when $U\neq 0$ the series of diagrams given in Fig. \ref{fig:hop-exp} do not lead 
to the exact Green function, since {\it loop} diagrams containing higher order 
local Green functions (higher order cumulants) are missing. The latter vanish when $U=0$. 
Nevertheless, the diagrams of Fig. \ref{fig:hop-exp} still give a good approximation, especially in three 
dimensions. Thus, we can approximate the Green function for the Bose-Hubbard model using 
Eq. (\ref{G-series}) with ${\cal G}_0(\omega)$ replaced by ${\cal G}(\omega)$ given 
in Eq. (\ref{atomic-g}), i.e., 
\begin{equation}
\label{Hubbard-I}
G({\bf k},\omega)=\frac{1}{{\cal G}^{-1}(\omega)-\varepsilon_{\bf k}}.
\end{equation}
The energy spectrum is given by the poles of the above Green function:
\begin{equation}
E_\pm({\bf k})=-\mu+\frac{1}{2}[\varepsilon_{\bf k}+(2n-1)U]
\pm\frac{1}{2}\sqrt{\varepsilon_{\bf k}^2+2(2n+1)U\varepsilon_{\bf k}+U^2}.
\end{equation}
For large enough $U$ 
there is an energy gap $\Delta$ between the $+$ and $-$ branches of the spectrum, which is given by
\begin{eqnarray}
\Delta&=&E_+(0)-E_-(0)\nonumber\\&=&\sqrt{(2dJ)^2-4d(2n+1)UJ+U^2}.
\end{eqnarray}
The presence of the energy gap indicates that the system is an insulator. As $U$ gets 
smaller, it will eventually attains a critical value $U_c$ below which the system becomes 
a superfluid. This critical value of $U$ is found by demanding that the gap vanishes for 
$U=U_c$. The condition $\Delta=0$ yields
\begin{equation}
U_c^\pm=2dJ\left[2n+1\pm2\sqrt{n(n+1)}\right].
\end{equation}
Note that both solutions are positive. In order to know what is the right one we have to plot 
the phase diagram and study it more carefully. The transition from the insulating phase 
to the superfluid phase occurs when the bosons condense. This happens when the Green function 
is singular for $\omega=0$ and ${\bf k}=0$, since the bosons condense at ${\bf k}=0$. Thus, the 
phase diagram is given by the equation
\begin{equation}
{\cal G}(0)=\frac{1}{\varepsilon_{{\bf k}=0}},
\end{equation}
or
\begin{equation}
\label{phase-diag}
\frac{n}{\mu/U+1-n}-\frac{n+1}{\mu/U-n}=\frac{U}{2dJ}.
\end{equation}
In Fig. \ref{fig:pt-bhm} we plot the phase diagram for $d=3$ and $n=1,2,3,4$. It features the so called 
Mott lobes \cite{Fisher}. The phase inside the Mott lobes is an insulating one. Outside them 
we have a superfluid phase. Each Mott lobe corresponds to a Mott-insulating phase with 
$n$ particles per site. Thus, the largest lobe, corresponding to $0\leq\mu/U\leq 1$ 
has $n=1$. The next one, in the interval $1\leq\mu/U\leq 2$ has $n=2$, and so on. The 
tips of the lobes correspond to the points where the upper and lower bands of the spectrum 
meet for ${\bf k}=0$, thus closing the gap. The coordinates of the tips can be 
easily obtained by extremizing Eq. (\ref{phase-diag}) with respect to $\mu/U$. The tip of a 
Mott lobe corresponds to the maximum value of $J/U$ for a given value of $n$. 
Extremization gives the results:
\begin{equation}
\left(\frac{\mu}{U}\right)_c=\sqrt{n(n+1)}-1,
\end{equation}
\begin{equation}
\label{Uc}
U_c=2dJ\left[2n+1+2\sqrt{n(n+1)}\right],
\end{equation}
and we see that $U_c$ corresponds to the solution $U_c^+$ of the equation $\Delta=0$.     

\begin{figure}
\begin{center}
\includegraphics[width=12cm]{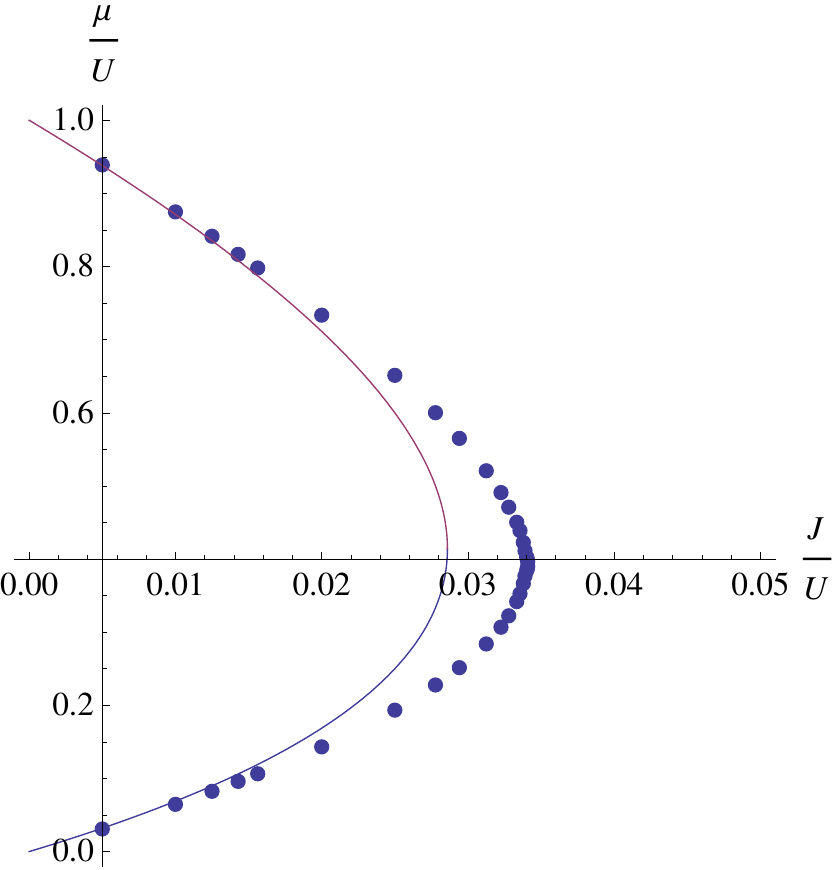}
\caption{Phase diagram for $n=1$. The Black circles are Monte Carlo points from 
Ref. \cite{MC}. The continuous line is the mean-field result.}\label{fig:mc-bhm}
\end{center}
\end{figure}

Let us compare the above results with recent highly precise Monte Carlo (MC) simulations for the $n=1$ case 
\cite{MC}. There it is obtained that $J/U_c\approx 0.03408$. On the 
other hand, our formula (\ref{Uc}) for $d=3$ and $n=1$ gives
\begin{equation}
\frac{J}{U_c}=\frac{1}{6(3+2\sqrt{2})}\approx 0.0286.
\end{equation}
Near the tip of the Mott lobe, which corresponds to the quantum critical regime, our approximation 
is very bad. This is to be expected, since our approach is equivalent to mean-field theory. 
In Fig. \ref{fig:mc-bhm} we compare our result for $n=1$ with the MC phase 
diagram of Ref. \cite{MC}. Indeed, the MC points 
agree with the mean-field curve only far away of the critical point, i.e., for small $J/U$. 

\begin{figure}
\begin{center}
\includegraphics[width=14cm]{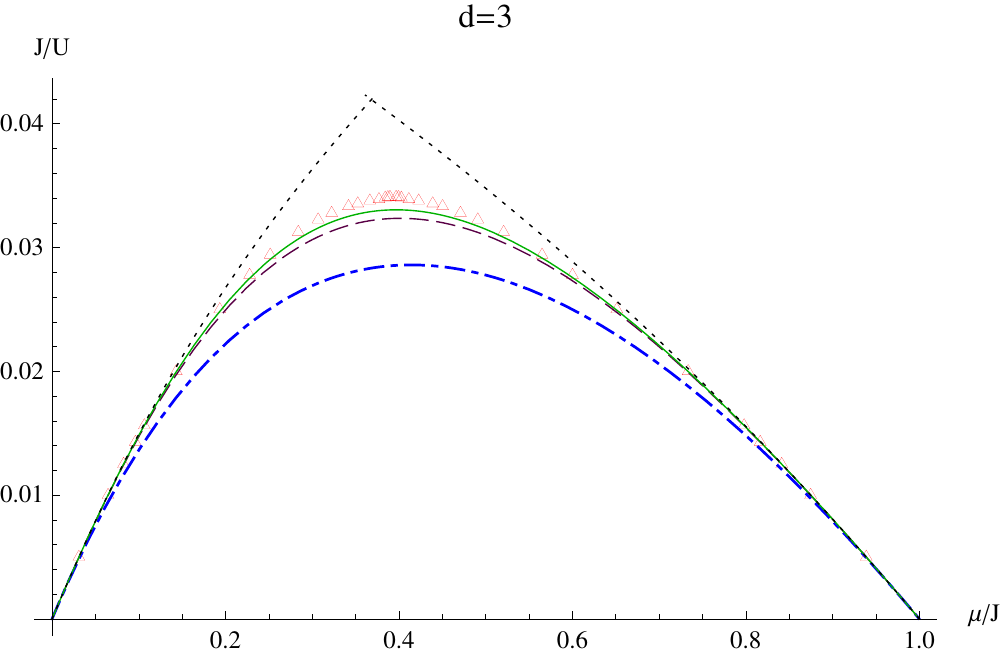}
\caption{Phase diagram of the Bose-Hubbard model 
in three dimensions beyond 
the mean-field approximation discussed in the text. The result 
is from Ref. \cite{dosSantos-1}. The solid green line corresponds to 
the effective action approach employed by dos Santos and 
Pelster \cite{dosSantos-1}. The dashed purple line corresponds 
to a variational approach used by the same authors. Both 
are compared with the MC results from Ref. \cite{MC} (red triangles), 
mean-field theory (dotted dashed blue line), and third-order 
 strong-coupling expansion (dotted black line) \cite{Monien-1996}. }\label{fig:pd-edie} 
\end{center}
\end{figure}

A better approximation, where higher order cumulants are included, so that also loop 
diagrams appear in the hopping expansion, allows to go beyond mean-field theory and 
approach better the MC result. This was done recently in a remarkable paper 
by dos Santos and Pelster \cite{dosSantos-1}, where a quantum Landau-Ginzburg formalism 
including higher order cumulants was developed. Their result closely agrees 
with the MC simulations. Furthermore, in contrast with the mean-field approach discussed 
here, their analysis shows how the behavior of the phase 
diagram changes with the dimensionality. Their phase diagram is shown for 
$d=3$ in Fig. \ref{fig:pd-edie}. 

\section{The Heisenberg antiferromagnet}

\subsection{Antiferromagnetic spin waves}

We have seen in Chap. \ref{ch:fm} that the spin waves in a ferromagnet have a dispersion $\omega\sim k^2$. For an 
antiferromagnetic system the situation is different, due to sublattice magnetization (see Fig. \ref{fig:AF-state}). Thus, 
in order to derive the spin-wave excitation of a Heisenberg antiferromagnet we have to consider the lattice sites 
explicitly. We will perform the calculations in the semi-classical limit. The Heisenberg model for an antiferromagnet will be written 
as

\begin{equation}
 H=JS^2\sum_{\langle i,j\rangle} {\bf n}_i\cdot{\bf n}_j,
\end{equation}
where ${\bf n}_i^2=1$. We can easily write down the corresponding lattice Landau-Lifshitz equation if we rewrite the above 
Hamiltonian as

\begin{equation}
 H=JS^2\sum_i{\bf n}_i\cdot{\bf B}_i,
\end{equation}
where the effective local magnetic field is given by

\begin{equation}
 {\bf B}_i=\sum_j'{\bf n}_j,
\end{equation}
where the prime in the sum reminds us that the sum is over the sites which are nearest neighbors of $i$ (that is why the effective field is 
$i$-dependent). For example, in one dimension we have

\begin{equation}
 {\bf B}_i={\bf n}_{i-1}+{\bf n}_{i+1}.
\end{equation}
Therefore, the LL equation is given for the lattice model as

\begin{equation}
 \partial_t{\bf n}_i=JS^2({\bf n}_i\times{\bf B}_i).
\end{equation}

For simplicity, we will derive the spin-wave spectrum in one dimension. The generalization to higher dimensions is 
straightforward. 

In one dimension the staggered magnetization occurs between even and odd lattice sites, such that we have

\begin{equation}
\label{dn1}
 {\bf n}_{2m}=n_0{\bf e}_3+\delta{\bf n}^{\perp}_{2m},
\end{equation}
and 

\begin{equation}
\label{dn2}
 {\bf n}_{2m+1}=-n_0{\bf e}_3+\delta{\bf n}^{\perp}_{2m+1},
\end{equation}
where $\delta{\bf n}^{\perp}_{i}$ is a small transverse fluctuation that will be used to linearize the LL equation. Thus, the calculation is 
similar to the one for ferromagnets in Chap. \ref{ch:fm}, except that here we have to consider a staggered magnetization and 
work on the lattice. Thus, we have the coupled equations, 

\begin{equation}
 \partial_t{\bf n}_{2m}=JS^2[{\bf n}_{2m}\times({\bf n}_{2m-1}+{\bf n}_{2m+1})],
\end{equation}

\begin{equation}
 \partial_t{\bf n}_{2m+1}=JS^2[{\bf n}_{2m+1}\times({\bf n}_{2m}+{\bf n}_{2(m+1)})].
\end{equation}
Substituting Eqs. (\ref{dn1}) and (\ref{dn2}) in the above equations and neglecting terms of order 
higher than two in the fluctuations, we obtain, 

\begin{equation}
\label{lin-LL-1}
 \partial_t\delta n_{2m}^+={\rm i}n_0JS^2(\delta n_{2m-1}^++\delta n_{2m+1}^++2\delta n_{2m}^+),
\end{equation}
and 

\begin{equation}
\label{lin-LL-2}
 \partial_t\delta n_{2m+1}^+=-{\rm i}n_0JS^2(\delta n_{2m}^++\delta n_{2(m+1)}^++2\delta n_{2m+1}^+),
\end{equation}
where we have defined $\delta n_{j}^+=\delta n_{j}^1+{\rm i}\delta n_{j}^2$. 

We will solve the Eqs. (\ref{lin-LL-1}) and (\ref{lin-LL-2}) via the Ans\"atze:

\begin{equation}
 \delta n_{2m}^+=ue^{{\rm i}(2km-\omega t)},~~~~~~~~~~~\delta n_{2m+1}^+=ve^{{\rm i}[(2m+1)k-\omega t]}.
\end{equation}
Thus, Eqs. (\ref{lin-LL-1}) and (\ref{lin-LL-2}) can be put in matrix form, 

\begin{equation}
 \omega\left[
\begin{array}{c}
 u \\
\noalign{\medskip} 
v
\end{array}
\right]=\omega_0\left[
\begin{array}{cc}
 -1 & -\cos k\\
\noalign{\medskip}
\cos k & 1
\end{array}
\right]\left[
\begin{array}{c}
 u \\
\noalign{\medskip} 
v
\end{array}
\right],
\end{equation}
where $\omega_0=2n_0S^2J$. The frequency $\omega$ is therefore determined by the eigenvalue equation, 

\begin{equation}
 \det\left[
\begin{array}{cc}
 -(\omega+\omega_0 )& -\omega_0\cos k\\
\noalign{\medskip}
\omega_0\cos k & \omega_0-\omega
\end{array}
\right]=0,
\end{equation}
which yields the spin-wave spectrum:

\begin{equation}
\omega(k)=\omega_0\sqrt{1-\cos^2k}=\omega_0|\sin k|. 
\end{equation}

The generalization to higher dimensions is straightforward, leading to the result, 

\begin{equation}
\omega({\bf k})=\omega_0\sqrt{d^2-\left(\sum_{a=1}^d\cos k_a\right)^2}. 
\end{equation}
In the long wavelength limit, the above equation becomes

\begin{equation}
 \omega({\bf k})\approx\sqrt{d}~\omega_0 |{\bf k}|.
\end{equation}
Therefore, in contrast with a ferromagnet, which has a quadratic dispersion, the antiferromagnet has a linear dispersion.

\subsection{The quantum $O(n)$ non-linear $\sigma$ model}

In Chap. \ref{ch:fm} we have studied the classical $O(n)$ non-linear $\sigma$ model in detail. In that Chapter the model was 
classical because it was being used to describe the thermal fluctuations in a ferromagnet. However, 
in the context of an antiferromagnet, the classical $O(n)$ non-linear $\sigma$ model is used to describe the quantum dynamics of 
an antiferromagnet at $T=0$. In this case, the temperature of the classical model studied in Chap. \ref{ch:fm} is replaced by 
a coupling constant $g$ and the dimension $d$ should be interpreted as $D+1$, such that $d$ corresponds to the number of 
dimensions of spacetime with time being imaginary. In this context $D$ is the dimension of space. 

The action of the  quantum $O(n)$ non-linear $\sigma$ model is given by

\begin{equation}
 S=\frac{1}{2g}\int_0^\beta d\tau \int d^Dx [(\partial_\tau{\bf n})^2+(\partial_i{\bf n})\cdot(\partial_i{\bf n})+
i\lambda({\bf n}^2-1)].
\end{equation}

The zero temperature case follows directly from the results for the classical model, with $T$ replaced by $g$, and $d$ being interpreted 
as the dimension of spacetime. So, let us consider the so called quantum critical finite temperature case. In this regime, we have 
that $g=g_c$, but the temperature is not zero, so that the mass gap $m$ is non-vanishing. In fact, in this case the temperature is the only energy scale available (up to the energy cutoff) and 
we actually expect that $m^2=aT^2$, where $a$ is a number to be determined. For $g=g_c$ and $T>0$ we still have $s=0$, so that the relevant gap equation is 
(\ref{gap-m2}) (with $T$ replaced by $g$), which at finite temperature becomes

\begin{equation}
 T\sum_{n=-\infty}^\infty\int\frac{d^Dk}{(2\pi)^D}\frac{1}{\omega_n^2+k^2+m^2}=\frac{1}{ng_c}=\int\frac{d^{D+1}p}{(2\pi)^{D+1}}\frac{1}{p^2},
\end{equation}
where we have used Eq. (\ref{gc}) with $T$ replaced by $g$. {\it We stress that the temperature $T$ appearing in 
the above equation has nothing to do with the temperature $T$ in Chap. \ref{ch:fm}, as 
in the present case $T$ arises from the integration over $\tau\in(0,\beta=1/T)$ in the 
action}. 

We can perform the Matsubara sum straightforwardly (see Appendix \ref{app:sums}) to obtain

\begin{equation}
\int\frac{d^{D+1}p}{(2\pi)^{D+1}}\frac{1}{p^2}= \int\frac{d^Dk}{(2\pi)^D}\frac{1}{2\sqrt{k^2+m^2}}+\frac{1}{(2\pi)^D}\int\frac{d^Dk}{\sqrt{k^2+m^2}}\frac{1}{e^{\sqrt{k^2+m^2}/T}-1}.
\end{equation}
Note that by considering the $T=0$ limit, we have from the above equation that

\begin{equation}
 \int\frac{d^{D+1}p}{(2\pi)^{D+1}}\frac{1}{p^2}=\int\frac{d^Dk}{(2\pi)^D}\frac{1}{2|{\bf k}|}.
\end{equation}  

The integrals can be easily evaluated for $D=2$ and we obtain

\begin{equation}
 m=-2T\ln(1-e^{-m/T}),
\end{equation}
whose solution is

\begin{equation}
 m(T)=T\ln\left(\frac{3+\sqrt{5}}{2}\right).
\end{equation}
The above result has been obtained before by Chubukov {\it et al.} \cite{Chubukov}. 

\subsection{The CP$^{N-1}$ model}

\begin{figure}
\begin{center}
 \includegraphics[width=10cm]{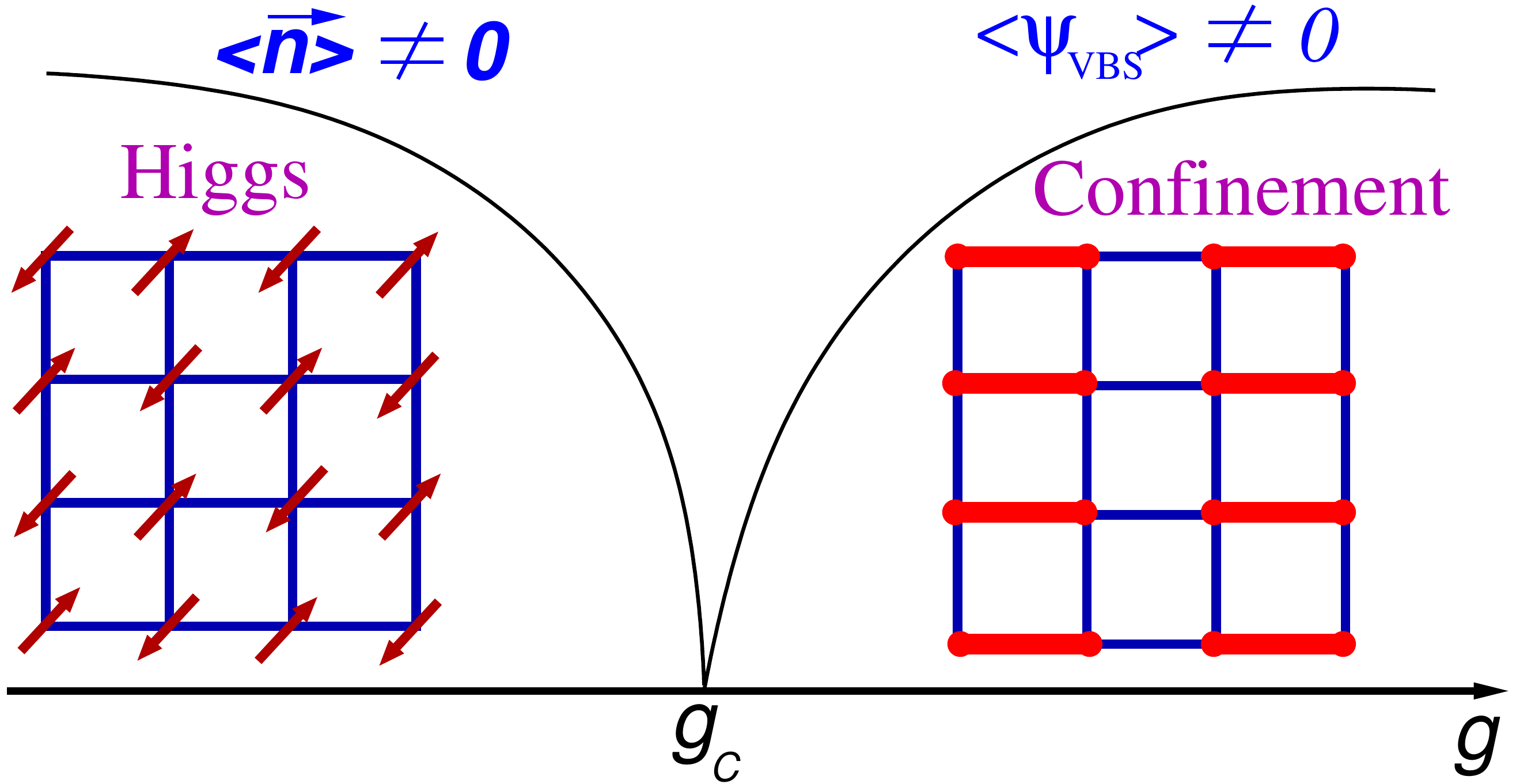}
\end{center} 
\caption{Schematic phase diagram showing a (second-order) quantum phase transition between 
a N\'eel state and a VBS as a function of a 
dimensionless coupling $g$. The different order parameters are shown. In the 
Higgs phase the spinons condense due to a spontaneous $U(1)$ symmetry breaking. 
This Higgs mechanism produces an antiferromagnetic state. In the confinement 
phase the excitations are gapped and the spinons are confined. While in the 
N\'eel phase the emergent photon is gapped, in the confined phase it is the 
dual of the photon which is gapped.}
\label{Fig:AF-VBS}
\end{figure}

Many two-dimensional Mott insulators feature competing orders. One example is the competition 
between a N\'eel state and a valence-bond solid (VBS) state \cite{RS}, as illustrated in Fig. \ref{Fig:AF-VBS}. In a 
Landau-Ginzburg-Wilson (LGW) framework, competing orders usually feature a first-order phase transition. 
This is because the order parameters involved are the most fundamental fields in a LGW 
type of theory. Furthermore, within this so called LGW paradigm \cite{Sachdev-Review,Senthil-2004,Senthil-2004a} 
second-order phase transitions are characterized by a very small anomalous dimension of the order 
parameter. In quantum phase transitions, on the other hand, there are examples of order 
parameters which are themselves made of more elementary building blocks. This is precisely the 
case of the N\'eel-VBS transition illustrated in Fig. \ref{Fig:AF-VBS}. In this case both 
order fields, the staggered magnetization orientation field ${\bf n}$ and 
the valence-bond order field $\psi_{\rm VBS}$, are made of spinons, which in this case 
are more conveniently represented as a bosonic excitation.  

Mott insulators featuring competing ordered states have been matter of intensive  
discussions in recent years \cite{Sachdev-Review}. In particular, there has been some 
controversy on the nature of the N\'eel-VBS phase transition  
\cite{Motrunich_2008,Kuklov_2008-comment,Kuklov_2008,Sandvik_2007,Melko-Kaul_2008,Jiang_2008}. 
The LGW point of 
view ignores the topological order of the Mott insulator. This topological order is essential in 
the VBS phase, since degenerate VBS ground states are connected by tunneling events characterized by 
a topological charge \cite{RS}. This important aspect of Mott insulators is more easily seen by  
performing a CP$^1$ map of the unit vector ${\bf n}$ giving the direction of the magnetization into 
a larger space defined by some set of complex fields, i.e., 

\begin{equation}
 {\bf n}=z^*_a\sigmab_{ab} z_b,
\end{equation}
where $|z_1|^2+|z_2|^2=1$ and $\sigmab=(\sigma_1,\sigma_2,\sigma_3)$ is a Pauli matrix vector.
Since ${\bf n}^2=1$, we have a map of the two-dimensional sphere $S_2$ into a 
three-dimensional sphere $S_3$ (since 
$|z_1|^2+|z_2|^2=\alpha_1^2+\beta_1^2+\alpha_2^2+\beta_2^2=1$, with  
$z_a=\alpha_a+i\beta_a$). It represents ${\bf n}$ as a composite field having the elementary 
constituents $z_a$, the so called spinons. One important aspect of the CP$^1$ map is that it 
introduces a gauge structure in the system. Indeed, the field ${\bf n}$ written in terms of 
spinons fields is a gauge invariant object, since the local phase transformation 
$z_a(x)\to e^{i\theta(x)}z_a(x)$ leaves ${\bf n}$ invariant. This gauge 
symmetry can be better understood by considering the topological charge

\begin{equation}
\label{hedgehog-af}
Q=\frac{1}{8\pi}\oint_{S_2}dS_\mu\epsilon_{\mu\nu\lambda}{\bf n}\cdot(\partial_\nu{\bf n}
\times\partial_\lambda{\bf n}),
\end{equation}
where $Q\in\mathbb{N}$. We have encountered the above topological charge before in a different context in Chap. \ref{ch:fm}, 
Eq. (\ref{hedgehog-2}). There it was a spatial topological object that we called a ``hedgehog'', a magnetic monopole-like 
excitation.  The main difference between the topological charge in (\ref{hedgehog-af}) and the one in Eq. (\ref{hedgehog-2}) is that 
the former lives in a $(2+1)$-dimensional spacetime, while the latter arises in three-dimensional space. The spacetime hedgehog is 
often called instanton. 

In order to obtain the gauge field from the CP$^1$ representation we need the formula

\begin{equation}
 \epsilon_{abc}\sigma_ {\mu\nu}^a\sigma_{\alpha\beta}^b\sigma_{\gamma\delta}^c=2i(\delta_{\mu\delta}\delta_{\alpha\nu}\delta_{\beta\gamma}
-\delta_{\mu\beta}\delta_{\nu\gamma}\delta_{\alpha\delta}),
\end{equation}
which can be easily derived by using the commutation relation $[\sigma^a,\sigma^b]=2i\epsilon_{abc}\sigma^c$ and the completeness relation

\begin{equation}
 \sigma^a_{\alpha\beta}\sigma^a_{\gamma\delta}=2\delta_{\alpha\delta}\delta_{\beta\gamma}-\delta_{\alpha\beta}\delta_{\gamma\delta}.
\end{equation}
Therefore, we obtain from the CP$^1$ map that the topological charge is given by the magnetic flux

\begin{equation}
Q=\frac{1}{2\pi}\oint_{S_2}dS_\mu B_\mu,
\end{equation}
where $B_\mu=\epsilon_{\mu\nu\lambda}\partial_\nu A_\lambda$ and

\begin{equation}
\label{A_mu}
A_\mu=\frac{i}{2}(z_a^*\partial_\mu z_a-z_a\partial_\mu z_a^*).
\end{equation}

By coupling the above gauge field minimally to the spinons we obtain the 
Lagrangian for the CP$^1$ model

\begin{equation}
\label{L-CP1}
{\cal L}=\frac{1}{g}|(\partial_\mu-iA_\mu)z_a|^2.
\end{equation}
In view of Eq. (\ref{A_mu}) and the form of the above Lagrangian together 
with ${\bf n}=z^*_a\sigmab_{ab} z_b$ along with the corresponding constraint, it is not difficult to see that the Lagrangian 
(\ref{L-CP1}) is equivalent to the Lagrangian of the $O(3)$ nonlinear $\sigma$ model.

\subsubsection{The $J-Q$ model: A lattice model for the AF-VBS transition}

\begin{figure}
\begin{center}
 \includegraphics[width=10cm]{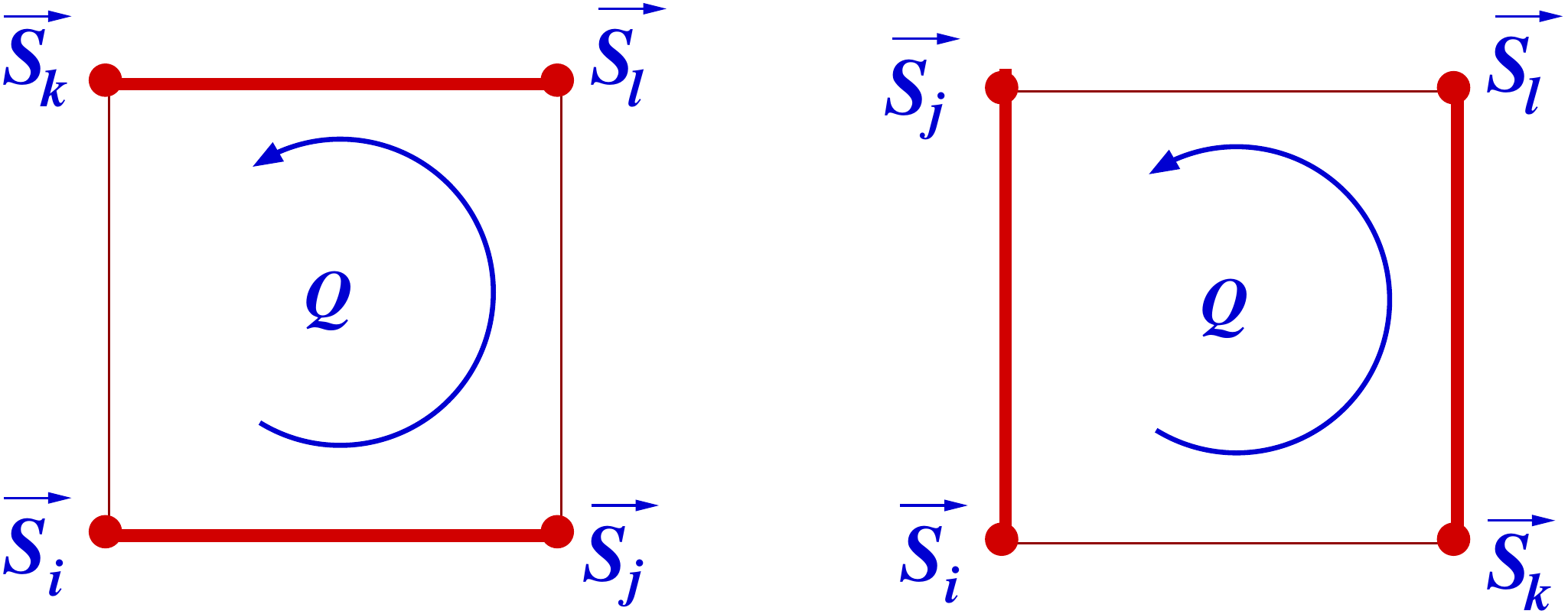}
\end{center} 
\caption{Schematic representation of the plaquette term in the $J-Q$ model, Eq. (\ref{J-Q}). 
The sum goes over the pairs of spins indicated by the red bonds at left and 
right panels.}\label{fig:Q-exch}
\end{figure}

Competing orders need competing interactions. For instance, the AF-VBS transition can be obtained from 
the following lattice model due to Sandvik \cite{Sandvik_2007}:

\begin{equation}
\label{J-Q}
 H=J\sum_{\langle i,j\rangle}{\bf S}_i\cdot{\bf S}_j -Q\sum_{\langle ijkl\rangle}\left({\bf S}_i\cdot{\bf S}_j-\frac{1}{4}\right)
\left({\bf S}_k\cdot{\bf S}_l-\frac{1}{4}\right),
\end{equation}
where the sum over $ijkl$ is around a plaquette, as shown in Fig. \ref{fig:Q-exch}. The phases of the model are as shown 
in Fig. \ref{Fig:AF-VBS}. When $Q/J\ll 1$ the Heisenberg term dominates over the $Q$-term and a N\'eel state is 
favored. For  $Q/J\gg 1$, on the other hand, the $Q$-term determines the ground state, favoring a crystalline pattern of singlet 
valence bonds. Recall that

\begin{equation}
 P_{ij}=\frac{1}{4}-{\bf S}_i\cdot{\bf S}_j,
\end{equation}
is a projection operator for singlet states. Indeed, we can rewrite the above operator as

\begin{equation}
 P_{ij}=\frac{1}{4}+\frac{1}{2}[{\bf S}_i^2+{\bf S}_j^2-({\bf S}_i+{\bf S}_j)^2].
\end{equation}
Note that when the total spin of the $(i,j)$-bond is zero, $P_{ij}$ yields the unit as eigenvalue. For 
a triplet bond it vanishes.  

There is some discussion in the literature on whether the $J-Q$ model really features a quantum critical point or, in other words, 
on whether it undergoes a second-order phase transition \cite{Melko-Kaul_2008,Jiang_2008,Sandvik-2010}. A related issue refers to 
the gauge theory description of the $J-Q$ model in terms of a CP$^1$ model with instanton suppression 
via a Maxwell term \cite{Sachdev-Review,Senthil-2004a,Motrunich_2008,Kuklov_2008-comment,Kuklov_2008}. While Kuklov {\it et al.} 
\cite{Kuklov_2008} numerically find a weak first-order phase transition for the CP$^1$ model in $2+1$ dimensions with a Maxwell term, a recent paper 
by Sandvik \cite{Sandvik-2010} shows strong evidence that a second-order phase transition occurs in the $J-Q$ model. 
However, logarithmic violations of scaling are also found,  which may cast some doubts on whether CP$^1$-like models 
can really fully provide a description of quantum criticality in the $J-Q$ model. 

\subsubsection{The CP$^{N-1}$ model in the large $N$ limit}

The CP$^1$ model can be generalized to include $N$ spinons instead of two. This yields the so called 
CP$^{N-1}$ model, which can be studied in the large $N$ limit. 

In order to facilitate the analysis, we write the CP$^{N-1}$ model as

\begin{eqnarray}
{\cal L}&=&\frac{1}{4e_0^2}F_{\mu\nu}^2+\frac{\Lambda^{d-2}}{g}\left[|(\partial_\mu-iA_\mu)z_a|^2
\right.\nonumber\\
&+&\left.i\sigma(|z_a|^2-1)\right],
\end{eqnarray}
such that the constraint is accounted for by the introduction of a Lagrange multiplier field 
$\sigma$. Note that $g$ here is a dimensionless coupling. We have also introduced a Maxwell term in 
the Lagrangian. With this Maxwell term the model at $N=2$ is no longer equivalent to the $O(3)$ nonlinear $\sigma$ model. 
This occurs only in the $e_0\to\infty$ limit. By integrating out $N-2$ spinon fields 
we obtain the effective action

\begin{eqnarray}
S_{\rm eff}&=&(N-2){\rm Tr}\ln[-(\partial_\mu-iA_\mu)^2+i\sigma]\nonumber\\
&+&\sum_{a=1,2}\int d^3x\left\{\frac{1}{4e_0^2}F_{\mu\nu}^2+\frac{\Lambda^{d-2}}{g}\left[|(\partial_\mu-iA_\mu)z_a|^2\right.\right.\nonumber\\
&+&\left.\left.i\sigma(|z_a|^2-1)\right]\right\}.
\end{eqnarray}   
The saddle-point equations are obtained in a standard way. We will consider 
a saddle-point with $A_\mu=0$, $i\sigma=\sigma_0$, and $z_a=\zeta_a$, where 
$\sigma_0$ and $\zeta_a$ are constants. This means that no classical instanton solution is being 
considered here. 

The calculation below is very similar to the one made for the classical $O(n)$ nonlinear $\sigma$ model in 
Chap. \ref{ch:fm}, so the reader may be interested in going back to that Chapter again to recall the main 
features of the calculation.  

The VBS phase occurs for $g>g_c$, where $g_c$ is a critical coupling. In this case 
we have $\sigma_0\neq 0$ and $\zeta_a=0$, leading to a gap equation (for large $N$)

\begin{equation}
\label{gap-eq+}
Ng\Lambda^{2-d}\int\frac{d^dp}{(2\pi)^d}\frac{1}{p^2+\sigma_0}=1.
\end{equation}

For $g<g_c$, on the other hand, we are in the N\'eel phase and $\sigma_0= 0$ and $\zeta_a\neq 0$. 
Thus, we have 

\begin{equation}
\label{gap-eq-}
Ng\Lambda^{2-d}\int\frac{d^dp}{(2\pi)^d}\frac{1}{p^2}=1-\sum_{a=1,2}|\zeta_a|^2.
\end{equation}
For $g=g_c$ we have $\sigma_0=0$ and $\zeta_a=0$, which determines the critical 
coupling: $g_c=(d-2)(2\pi)^d/(NS_d)$, where $S_d=2\pi^{d/2}/\Gamma(d/2)$ is the surface of 
a unit sphere in $d$ dimensions. 

It is easy to see that in general we should have

\begin{equation}
\sum_{a=1,2}|\zeta_a|^2+\frac{g}{g_c}-1=Ng\Lambda^{2-d}
\sigma_0\int\frac{d^dp}{(2\pi)^d}\frac{1}{p^2(p^2+\sigma_0)}.
\end{equation}
Therefore, we obtain for $g\geq g_c$,

\begin{equation}
\sigma_0=\Lambda^2\left[\frac{1}{\Gamma(d/2)\Gamma(2-d/2)}\left(1-\frac{g_c}{g}\right)
\right]^{2/(d-2)}.
\end{equation}  
It is clear that $\sigma_0$ should be identified with $\xi^{-2}$, where $\xi$ is the 
correlation length. This leads to the well known large $N$ critical exponent for 
this quantity, $\nu=1/(d-2)$. The limit $d\to 2$, on the other hand, yields

\begin{equation}
 \sigma_0=\Lambda^2\exp\left(-\frac{4\pi}{Ng}\right),
\end{equation}
and we see that in this case there is a gap for all $g>0$.

For $g<g_c$ we obtain $\sum_{a=1,2}|\zeta_a|^2=1-g/g_c$, which implies that the 
critical exponent of the N\'eel field order parameter
 
\begin{equation}
\langle{\bf n}\rangle=\zeta_a^*\sigmab_{ab}\zeta_b 
\end{equation}
is $\beta_N=1$. 
Thus, from the 
hyperscaling relation $\beta_N=\nu(d-2+\eta_N)/2$ it follows that the anomalous dimension is  

\begin{equation}
\eta_N=d-2 
\end{equation}
for large $N$. 

Now we consider the gauge field fluctuations to lowest order for $g\geq g_c$. This correction is 
obtained from the expansion of 

\begin{equation}
 {\rm Tr}\ln[-(\partial_\mu-iA_\mu)^2+\sigma_0]
\end{equation}
up to quadratic order in the gauge field. In order to do this, we first note that the covariant differential operator 
appearing in the inside the logarithm can expanded as

\begin{equation}
 -(\partial_\mu-iA_\mu)^2=-\partial^2+i\partial_\mu A_\mu+2iA_\mu\partial_\mu+A^2,
\end{equation}
so that 

\begin{eqnarray}
\label{trlog}
 {\rm Tr}\ln[-(\partial_\mu-iA_\mu)^2+\sigma_0]&\approx&{\rm Tr}\ln(-\partial^2+\sigma_0)+G(0)\int d^d xA^2(x)
\nonumber\\
&-&\frac{1}{2}\int d^d x\int d^d x'[i\partial_\mu A_\mu(x)+2iA_\mu(x)\partial_\mu]G(x-x')
\nonumber\\
&\times& [i\partial_\mu' A_\mu(x')+2iA_\mu(x')\partial_\mu']G(x'-x),
\end{eqnarray}
where

\begin{equation}
 G(x)=\int\frac{d^dp}{(2\pi)^d}e^{ip\cdot x}G(p),
\end{equation}
with
\begin{equation}
 G(p)=\frac{1}{p^2+\sigma_0}.
\end{equation}
The first term on the RHS of (\ref{trlog}) contributes to the saddle-point approximation. The second term can be written 
in momentum space as

\begin{equation}
 \frac{1}{2}\int\frac{d^dp}{(2\pi)^d}\Sigma_{\mu\nu}(p)A_\mu(p)A_\nu(-p),
\end{equation}
where

\begin{eqnarray}
\label{Sigma-A}
 \Sigma_{\mu\nu}(p)&=&2\delta_{\mu\nu}\int\frac{d^dk}{(2\pi)^d}G(k)
-\int\frac{d^dk}{(2\pi)^d}(2k-p)_\mu(2k-p)_\nu G(k-p)G(k)\nonumber\\
&=&2\delta_{\mu\nu}\int\frac{d^dk}{(2\pi)^d}\frac{1}{k^2+\sigma_0}
-\int\frac{d^dk}{(2\pi)^d}\frac{(2k-p)_\mu(2k-p)_\nu}{[(k-p)^2+\sigma_0](k^2+\sigma_0)}.
\nonumber\\
\end{eqnarray}
Next we set $M^2=\sigma_0$ and employ some tricks of dimensional regularization. 
First, by applying the operator $Md/dM$ to both sides of Eq. (\ref{I1-1}) from Appendix \ref{app:int-d-m}, we obtain

\begin{equation}
 \int\frac{d^dk}{(2\pi)^d}\frac{1}{k^2+M^2}=-\frac{2M^2}{d-2}\int\frac{d^dk}{(2\pi)^d}\frac{1}{(k^2+M^2)^2}.
\end{equation}
Second, we can decompose the second integral appearing in the second line of Eq. (\ref{Sigma-A}) into its transversal and 
longitudinal parts, i.e., 

\begin{equation}
 \int\frac{d^dk}{(2\pi)^d}\frac{(2k-p)_\mu(2k-p)_\nu}{[(k-p)^2+M^2](k^2+M^2)}=
D_t(p)\left(\delta_{\mu\nu}-\frac{p_\mu p_\nu}{p^2}\right)+D_l(p)\frac{p_\mu p_\nu}{p^2}.
\end{equation}
By taking the trace of both sides of the above equation on one hand and contracting with $p_\mu p_\nu$ on the other hand, we 
can determine $D_t(p)$ and $D_l(p)$:

\begin{eqnarray}
 D_t(p)&=&-\frac{1}{d-1}\left\{\frac{4M^2}{d-2}\int_k\frac{1}{(k^2+M^2)^2}\right.
\nonumber\\
&+&\left.(p^2+4M^2)\int_k\frac{1}{[(k-p)^2+M^2](k^2+M^2)}\right\},\nonumber\\
\end{eqnarray}
and

\begin{equation}
 D_l(p)=-\frac{4M^2}{d-2}\int_k\frac{1}{(k^2+M^2)^2},
\end{equation}
where we have  used the shorthand notation for the $d$-dimensional integrals of Chap. \ref{ch:fm}, Eq. (\ref{shorthand}). 
Thus, we obtain, 

\begin{eqnarray}
 \Sigma_{\mu\nu}(p)&=&-\left[-D_t(p)+\frac{4M^2}{d-2}\int_k\frac{1}{(k^2+M^2)^2}\right]\delta_{\mu\nu}\nonumber\\
&+&[D_t(p)-D_l(p)]\frac{p_\mu p_\nu}{p^2}\nonumber\\
&=&\Sigma(p)\left(\delta_{\mu\nu}-\frac{p_\mu p_\nu}{p^2}\right),
\end{eqnarray}
where 

\begin{eqnarray}
\label{Sigma-A-1}
 \Sigma(p)&=&\frac{1}{d-1}\left\{-4M^2\int_k\frac{1}{(k^2+M^2)^2}\right.
\nonumber\\
&+&\left.(p^2+4M^2)\int_k\frac{1}{[(k-p)^2+M^2](k^2+M^2)}\right\}.\nonumber\\
\end{eqnarray}
For large distances or, equivalently, in the infrared (i.e., $p$ small), we can write $\Sigma(p)\approx {\rm const} ~p^2$, such 
that in real spacetime this fluctuation correction generates a Maxwell term. In order to obtain this result, we have to expand 
$[(k-p)^2+M^2]^{-1}$ up to second order in $p$. Thus, we have

\begin{equation}
 \frac{1}{(k-p)^2+M^2}=\frac{1}{k^2+M^2}+\frac{2p\cdot k}{(k^2+M^2)^2}+\frac{p_\mu p_\nu(4k_\mu k_\nu-\delta_{\mu\nu})}{(k^2+M^2)^3}+\dots
\end{equation}
The above expression has to inserted in Eq. (\ref{Sigma-A-1}). In order to perform the integrals with help of the 
results from Appendix \ref{app:int-d-m}, some simplifications are needed. First of all, rotational invariance implies

\begin{equation}
 \int_k\frac{k_\mu k_\nu}{(k^2+M^2)^4}=\frac{\delta_{\mu\nu}}{d}\int_k\frac{k^2}{(k^2+M^2)^4}.
\end{equation}
Furthermore, the integral on the RHS of the above equation can be rewritten as

\begin{equation}
 \int_k\frac{k^2}{(k^2+M^2)^4}=\int_k\frac{1}{(k^2+M^2)^3}-M^2\int_k\frac{1}{(k^2+M^2)^4}.
\end{equation}
It should also be noted that

\begin{equation}
 \int_k\frac{k_\mu}{(k^2+M^2)^2}=0.
\end{equation}

Now we are ready to express $\Sigma(p)$ in terms of the integrals evaluated in Appendix \ref{app:int-d-m} to obtain 

\begin{eqnarray}
 \Sigma(p)&=&\frac{p^2}{d-1}\left[I_2(d)+\frac{4(4-d)M^2}{d}I_3(d)-\frac{16M^4}{d}I_4(d)\right]
\nonumber\\
&=&\frac{p^2M^{d-4}}{(d-1)(4\pi)^{d/2}}\left[\Gamma\left(2-\frac{d}{2}\right)+\frac{2(4-d)}{d}\Gamma\left(3-\frac{d}{2}\right)
\right.\nonumber\\
&-&\left.\frac{8}{3d}\Gamma\left(4-\frac{d}{2}\right)\right]
\nonumber\\
&=&\frac{p^2M^{d-4}}{3(4\pi)^{d/2}}\Gamma\left(2-\frac{d}{2}\right),
\end{eqnarray}
where in the simplifications repeated use of the identity $\Gamma(z+1)=z\Gamma(z)$ was made. 
Therefore, after replacing back $M=\sqrt{\sigma_0}$, we obtain at large 
distances the following low-energy contribution to the effective Lagrangian:

\begin{equation}
{\cal L}_{\rm Maxwell}=\frac{1}{4}\left(\frac{1}{e_0^2}+Nc_d\sigma_0^{(d-4)/2}\right)F_{\mu\nu}^2,
\end{equation}
where $F_{\mu\nu}=\partial_\mu A_\nu-\partial_\nu A_\mu$ and

\begin{equation}
c_d=\frac{1}{3(4\pi)^{d/2}}\Gamma\left(2-\frac{d}{2}\right).
\end{equation}
We have 
$c_2=1/(12\pi)$ and $c_3=1/(24\pi)$ for $d=2$ and $d=3$, respectively.
Note that even if a Maxwell term were not be present in the Lagrangian (i.e., $e_0\to\infty$), it would be 
generated by fluctuations in the paramagnetic phase.   

By using $M=\sqrt{\sigma_0}$ as a mass scale, we can define a dimensionless gauge coupling constant 
$f=M^{d-4}e^2$, where

\begin{equation}
\label{e2}
 e^2(M)=\frac{e_0^2}{1+c_dNe_0^2M^{d-4}}.
\end{equation}
Therefore, the RG $\beta$ function for $f$ is

\begin{equation}
 M\frac{df}{dM}=-(4-d)f+(4-d)c_dNf^2.
\end{equation}
Note that by expanding around $d=4$ we obtain exactly the one-loop $\beta$ function for the Abelian 
Higgs model in the minimal subtraction scheme \cite{ZJ}, although we have considered $N$ large. 

We see that within the large $N$ approach the presence of a bare Maxwell term does not spoil the 
quantum critical point. Indeed, the behavior of $e^2$ for $e_0^2$ large is the same as for $M$ small 
(or $g\to g_c$), provided $2<d<4$.  However, as we have already mentioned, for the CP$^1$ (i.e., for $N=2$) model with a 
Maxwell term  there is strong numerical evidence against the existence of a quantum critical point \cite{Kuklov_2008-comment,Kuklov_2008}. 
However, there are indications that a doped version of the model exhibits quantum criticality \cite{Kaul,Nogueira-2008}. 
 
\subsubsection{Finite temperature at criticality}

The finite temperature analysis at large $N$ is similar to the one made for 
the non-linear $\sigma$ model, since for the leading order in $1/N$ we can set 
the gauge field to zero. This leads again to a thermal gap

\begin{equation}
\label{m(T)-CP(N-1)}
 m(T)\equiv\sqrt{\sigma_0}=T\ln\left(\frac{3+\sqrt{5}}{2}\right).
\end{equation}

Let us next calculate the uniform susceptibility $\chi_u$ at the QCP. In the case of the $O(3)$ non-linear $\sigma$ model it is obtained by considering 
the response to an 
external field twisting in the time direction of the direction field ${\bf n}$. Thus, the following replacement holds

\begin{equation}
 (\partial_\tau{\bf n})^2\to (\partial_\tau{\bf n}-i{\bf H}\times{\bf n})^2; 
\end{equation}
see for example Ref. \cite{Sachdev}. The CP$^{N-1}$ model for $N=2$ is equivalent to the $O(3)$ non-linear $\sigma$ model. In this case we replace 
\cite{Melko-Kaul_2008}

\begin{equation}
 \partial_\tau z_a\to\partial_\tau z_a-iH\sigma^z_{ab}z_b/2,
\end{equation}
for a field ${\bf H}=H{\bf e}_z$.  We can generalize the above to the CP$^{N-1}$ model by considering the generators of $SU(N)$ instead of 
the Pauli matrices.  
Thus, at leading order the large $N$ free energy density at the presence of $H$ becomes for $D=2$

\begin{eqnarray}
 f\equiv F/V&=&\frac{N}{2}\sum_{n=-\infty}^\infty\int\frac{d^2p}{(2\pi)^2}[\ln(\omega_n^2-H\omega_n+p^2+H^2/4+m^2)
\nonumber\\
&+&\ln(\omega_n^2+H\omega_n+p^2+H^2/4+m^2)]-\frac{\Lambda}{g_cT}m^2.
\end{eqnarray}
The uniform susceptibility is then defined by \cite{Sachdev,Melko-Kaul_2008}

\begin{equation}
 \chi_u=\frac{1}{N}\left.\frac{\partial^2 f}{\partial H^2}\right|_{H=0}.
\end{equation}
Therefore, 

\begin{equation}
 \chi_u=\frac{T}{2}\sum_{n=-\infty}^\infty\int\frac{d^2p}{(2\pi)^2}\frac{1}{\omega_n^2+p^2+m^2}-T\sum_{n=-\infty}^\infty\int\frac{d^2p}{(2\pi)^2}\frac{\omega_n^2}{(\omega_n^2+p^2+m^2)^2}
\end{equation}
The Matsubara sums and the integrals are straightforwardly done to obtain

\begin{equation}
\label{chi_u-1}
\chi_u=\frac{1}{4\pi}\left[\frac{m}{e^{m/T}-1}-T\ln(1-e^{-m/T})\right].
\end{equation}
Upon substituting $m(T)$ from Eq. (\ref{m(T)-CP(N-1)}), we obtain

\begin{equation}
 \chi_u=\frac{T}{4\pi}\sqrt{5}\ln\left(\frac{1+\sqrt{5}}{2}\right).
\end{equation}

\subsection{Quantum electrodynamics in $2+1$ dimensions}

We have seen in Section \ref{sect:Hubbard} that for $U\gg t$ the Hamiltonian of the Hubbard model 
becomes the Heisenberg Hamiltonian for an antiferromagnet [Eq. (\ref{Heisenberg-1})]. In this 
Section we will consider a generalization of it in which the $SU(2)$ symmetry is enlarged 
to $SU(N)$. The aim of this generalization is to provide effective field-theoretic models that can be 
studied systematically within a $1/N$ expansion. This approach was pioneered by Affleck and Marston 
\cite{Affleck,Affleck-1} long time ago with the physical motivation of better understanding the Mott-insulating 
physics underlying the cuprates superconductors. 

The $SU(N)$ generalization of the Heisenberg model is given, up to irrelevant constant terms by

\begin{equation}
H=-\frac{J}{N}\sum_{\langle i,j\rangle}f_{i\alpha}^\dagger f_{j\alpha}
f_{j\beta}^\dagger f_{i\beta},
\end{equation}
where from now on summation over repeated Greek indices is implied,  
$J=4t^2/U$ [see Eq. (\ref{Heisenberg-1})], and we have rescaled $J\to J/N$ in order to 
facilitate the large $N$ approach. In addition, the fermions must satisfy the local constraint:
 
\begin{equation}
 f_{i\alpha}^\dagger f_{i\alpha}=N/2.
\end{equation}
 
By performing a Hubbard-Stratonovich transformation, we can rewrite the Hamiltonian as

\begin{equation}
 H=\frac{N}{J}|\chi_{ij}|^2-\sum_{\langle i,j\rangle}\chi_{ij}
f_{i\alpha}^\dagger f_{j\alpha}+{\rm h.c.}.
\end{equation}
The saddle-point solution at large $N$ is given by the so called $\pi$-flux phase, 

\begin{eqnarray}
 \chi_{ij}&=&\chi_0 e^{i\theta_{ij}},\nonumber\\
\sum_{{\cal P}=\{ijkl\}}\theta_{\cal P}&=&\pi,
\end{eqnarray}
corresponding to a flux amount of $\pi$ around a plaquette $\cal P$. This gives the spectrum at 
large $N$:

\begin{equation}
\label{pi-spec}
 E_k=4\chi_0\sqrt{\cos^2k_x+\cos^2k_y}.
\end{equation}
The above spectrum is gapless at the points $(\pm \pi/2,\pm\pi/2)$. 

The large $N$ saddle-point approximation we just described corresponds to a Mott insulator without any broken 
symmetries. Thus, the solution is paramagnetic and preserves the symmetries of the square lattice. The obtained solution 
constitutes one of the simplest examples of spin liquid. 

Let us now consider the fluctuations around the saddle-point. First, we would like to show that the elementary excitations near 
the nodes of the large $N$ spectrum (\ref{pi-spec}) are Dirac fermions. Here we will closely 
follow Refs. \cite{Morinari} and \cite{Richert}. Let us define

\begin{eqnarray}
 \chi_1 &=& \chi_{j,j+{\hat e}_x},~~~~~~~\chi_2 = \chi_{j+{\hat e}_x,j+{\hat e}_y},\nonumber\\
\chi_3 &=& \chi_{j+{\hat e}_x+{\hat e}_y,j+{\hat e}_y},~~~~~~~\chi_4 = \chi_{j+{\hat e}_y,j}.
\end{eqnarray}
Since the square lattice is bipartite, we can describe it as being composed by two interpenetrating sublattices, $A$ and $B$, similarly 
to the mean-field theory for the Hubbard model at half-filling discussed in Sect. \ref{sect:Hubbard}. Therefore, using 
Eq. (\ref{f-sublat}), we can write 

\begin{eqnarray}
H&=& -\sum_{j\in A} \sum_{\sigma}
\left( 
\chi_1 \bar c_{j+{\hat e}_x,\sigma}^{\dagger} c_{j\sigma}
+\chi_4^* \bar c_{j+{\hat e}_y,\sigma}^{\dagger} c_{j\sigma} + {\rm h.c.}
\right)
\nonumber\\
&-&\sum_{j\in B}
\left( 
\chi_3^* c_{j+{\hat e}_x,\sigma}^{\dagger} \bar c_{j\sigma}
+\chi_2 c_{j+{\hat e}_y,\sigma}^{\dagger} \bar c_{j\sigma} + {\rm h.c.}
\right) \\
& & +\frac{NL}{J} \left(
|\chi_1|^2+|\chi_2|^2+|\chi_3|^2+|\chi_4|^2
\right).
\nonumber
\end{eqnarray}
Therefore, the Hamiltonian can be rewritten as

\begin{equation}
H= {\sum_{\bf k}}^{\prime}
\left[ 
\begin{array}{cc}
c_{{\bf k}\sigma}^{\dagger} & \bar c_{{\bf k}\sigma}^{\dagger} 
\end{array} 
\right]
\left[
\begin{array}{cc}
0  & |\chi| \left( \cos k_1 + i\cos k_2 \right) \\
|\chi| \left( \cos k_1 - i\cos k_2 \right)& 0
\end{array}
\right]
\left[ 
\begin{array}{c}
c_{{\bf k}\sigma} \\
\bar c_{{\bf k}\sigma}
\end{array}
\right],
\end{equation}
where, as before in Sect. \ref{sect:Hubbard}), the prime on the sum implies that 
the sum is carried over the half of the Brilouin zone.  
Let us denote $c_{1{\bf k}\sigma}$ ($\bar c_{1{\bf k}\sigma}$) as the fermionic operator $c_{{\bf k}\sigma}$ ($\bar c_{{\bf k}\sigma}$) near the node $(\pi/2,\pi/2)$. 
Similarly, $c_{2{\bf k}\sigma}$ ($\bar c_{2{\bf k}\sigma}$) will denote the fermionic operator $c_{{\bf k}\sigma}$ ($\bar c_{{\bf k}\sigma}$) near the 
node $(-\pi/2,\pi/2)$. Thus, the Hamiltonian becomes approximately 

\begin{eqnarray}
H &\simeq &
{\sum_{\bf k}}'
\left[ 
\begin{array}{cccc}
c_{1{\bf k}\sigma}^{\dagger} &
\bar c_{1{\bf k}\sigma}^{\dagger} &
c_{2{\bf k}\sigma}^{\dagger} &
\bar c_{2{\bf k}\sigma}^{\dagger}
\end{array} \right]
\left\{
-|\chi| k_1 
\left[ \begin{array}{cc} \sigma_1 & 0 \\
0 & -\sigma_1 \end{array} \right]
\right. \nonumber \\
& & \left.
+|\chi| k_2
\left[ \begin{array}{cc} \sigma_2 & 0 \\
0 & \sigma_2 \end{array} \right]
\right\}
\left[
\begin{array}{c}
c_{1{\bf k}\sigma}\\
\bar c_{1{\bf k}\sigma}\\
c_{2{\bf k}\sigma}\\
\bar c_{2{\bf k}\sigma}
\end{array}
\right],
\end{eqnarray}
where $\sigma_1$ and $\sigma_2$ are the usual Pauli matrices.
The above Hamiltonian is the one of four-component Dirac fermions in two spatial dimensions. If in addition we include the phase 
fluctuations of the link field $\chi_{ij}$, i.e., 

\begin{equation}
 \chi_{lm}=\chi e^{iA_{lm}},
\end{equation}
the Dirac fermions will get gauged. The resulting effective theory that emerges in the continuum is quantum electrodynamics in 
$2+1$ dimensions (QED$_{2+1}$), with a Lagrangian given by

\begin{equation}
\label{L-fermion}
{\cal L}_f=\sum_{\alpha=1}^{N}\bar \psi_\alpha(\slashchar{\partial}+ie_0\slashchar{A})\psi_\alpha, 
\end{equation}
where we have 
introduced a bare ``electric charge'' $e_0$ to set the energy scale, and 
the standard notation $\slashchar{a}\equiv\gamma_\mu a_\mu$ is used. 
The Dirac matrices are in this case given by

\begin{equation}
\gamma_0=\left(
\begin{array}{cc}
\sigma_3 & 0\\
\noalign{\medskip}
0 & -\sigma_3
\end{array}
\right),~~~~~~~~~
\gamma_1=\left(
\begin{array}{cc}
\sigma_2 & 0\\
\noalign{\medskip}
0 & -\sigma_2
\end{array}
\right),
\nonumber
\end{equation}
\begin{equation}
\gamma_2=\left(
\begin{array}{cc}
\sigma_1 & 0\\
\noalign{\medskip}
0 & -\sigma_1
\end{array}
\right).
\end{equation}
The gamma matrices above satisfy the anticommutation relation

\begin{equation}
 \gamma_\mu\gamma_\nu+\gamma_\nu\gamma_\mu=2\delta_{\mu\nu}I,
\end{equation}
where $I$ is a $4\times 4$ identity matrix. The above relation implies

\begin{equation}
\label{trace-gamma}
{\rm tr}(\gamma_\mu\gamma_\nu)=4\delta_{\mu\nu}. 
\end{equation}

It is well known that massless QED$_{2+1}$ featuring four-component Dirac spinors has a chiral symmetry \cite{Pisarski}, in contrast with 
QED$_{2+1}$ involving two-component spinors, which even admits a mass term breaking parity symmetry. Indeed, for the QED$_{2+1}$ above there 
are two $\gamma_5$-like matrices which anticommute with all the other Dirac matrices, namely, 

\begin{equation}
\label{gamma35}
\gamma_3=\left(
\begin{array}{cc}
0 & I\\
\noalign{\medskip}
I & 0
\end{array}
\right),~~~~~~~~~~~~~~~
\gamma_5=\left(
\begin{array}{cc}
0 & I\\
\noalign{\medskip}
-I & 0
\end{array}
\right),
\end{equation}
with $I$ being a $2\times 2$ identity matrix. The chiral symmetry of the Lagrangian (\ref{L-fermion}) is then 

\begin{equation}
\psi\to e^{i\gamma_{3,5}\theta}\psi.
\end{equation}

Spontaneous breaking of the chiral symmetry leads to a dynamical mass generation in  QED$_{2+1}$ \cite{Appelq-2}. This mass is proportional to 
the chiral condensate, i.e.,

\begin{equation}
 m\sim\sum_{\alpha=1}^N\langle\bar\psi_\alpha\psi_\alpha\rangle.
\end{equation}
The dynamical mass generation implies the spontaneous breaking of the $SU(2)$ symmetry in the Heisenberg model, leading in 
this way to a staggered magnetization (for more details on this point, see Ref. \cite{Kim}). A similar result \cite{Herbut-QED3} holds also 
in the case of effective QED theories for $d$-wave superconductivity \cite{FT-1,FT-2}. 

The dynamical mass generation in QED$_{2+1}$ is more easily obtained by studying the Schwinger-Dyson equations \cite{Appelq-2}. 
First of all, we need the photon propagator in the large $N$ limit. This is obtained in a functional integral formalism by performing 
the Gaussian integral over the Dirac fermions exactly. In this case Dirac fields are Grassmann variables (i.e., they anticommute), so 
that the effective action reads 

\begin{equation}
 S_{\rm eff}=-N{\rm Tr}\ln(\slashchar{\partial}+ie_0\slashchar{A}).
\end{equation}
In order to obtain the photon propagator we simply expand the above effective action up to quadratic order in $A_\mu$. 
Although we are interested in the $d=3$ case, we will evaluate $\Sigma_{\mu\nu}(p)$ in $d$ spacetime dimensions and set $d=3$ at the end. 
In momentum space we obtain in this way the quadratic form, 

\begin{equation}
  S_{\rm eff}\approx\frac{1}{2}\int\frac{d^dp}{(2\pi)^d}\Sigma_{\mu\nu}(p)A_\mu(p)A_\nu(-p),
\end{equation}
where

\begin{equation}
 \Sigma_{\mu\nu}(p)=-\alpha\int\frac{d^dk}{(2\pi)^d}\gamma_\mu G_0(k)\gamma_\nu G_0(p-k),
\end{equation}
with the fermionic propagator

\begin{equation}
 G_0(p)=-\frac{i\slashchar{p}}{p^2},
\end{equation}
and $\alpha=Ne_0^2$. Gauge invariance implies $p_\mu\Sigma_{\mu\nu}(p)=0$, so that $\Sigma_{\mu\nu}(p)$ is transverse, so we 
can write

\begin{equation}
 \Sigma_{\mu\nu}(p)=p^2\Pi(p)\left(\delta_{\mu\nu}
-\frac{p_\mu p_\nu}{p^2}\right).
\end{equation}
Taking the trace over the vector indices yields

\begin{equation}
\label{pi-int}
(d-1)p^2\Pi(p)=\alpha\int\frac{d^dk}{(2\pi)^d}\frac{{\rm tr}[\gamma_\mu\slashchar{k}
\gamma_\mu(\slashchar{k}-\slashchar{p})]}{k^2(p-k)^2}.
\end{equation}
Next we use the identity

\begin{equation}
\label{ident-Dirac-1}
 \gamma_\mu\slashchar{p}\gamma_\mu=(2-d)\slashchar{p},
\end{equation}
and the trace formula (\ref{trace-gamma}) to rewrite Eq. (\ref{pi-int}) as

\begin{equation}
(d-1)p^2\Pi(p)=4(d-2)\alpha\int\frac{d^3k}{(2\pi)^3}\frac{k\cdot(p-k)}{k^2(p-k)^2}.
\end{equation}
Since in the numerator we have

\begin{equation}
 k\cdot p-k^2=\frac{p^2-k^2-(p-k)^2}{2},
\end{equation}
and assuming the rules of dimensional regularization,\footnote{In writing the following 
equation we have used $\int d^dkk^{-2}=\int d^dk(p-k)^{-2}=0$, which 
is true in dimensional regularization; see Appendix \ref{app:int-d-m}.} we obtain

\begin{equation}
(d-1)\Pi(p)=2(d-2)\alpha\int\frac{d^dk}{(2\pi)^d}\frac{1}{k^2(p-k)^2}.
\end{equation}
The integral in the equation above was evaluated in Chap. \ref{ch:fm}, Sect. \ref{sec:nlsm}. It is the same as 
the one in Eq. (\ref{Ib-0}), with the result of the integration given in Eq. (\ref{Ib}). Therefore, 

\begin{equation}
 \Pi(p)=\frac{2(d-2)c(d)}{d-1}|p|^{d-4},
\end{equation}
with $c(d)$ given in Eq. (\ref{c(d)}). 
Thus, the gauge field propagator has 
the form

\begin{equation}
\label{A-prop}
D_{\mu\nu}(p)=\frac{1}{p^2\Pi(p)}\left(\delta_{\mu\nu}
-\frac{p_\mu p_\nu}{p^2}\right),
\end{equation}
where $\Pi(p)$ is given for $d=3$ by

\begin{equation}
\label{Pi}
\Pi(p)=\frac{\alpha}{8|p|}.
\end{equation} 

Let us insert the propagator (\ref{A-prop}) in the one loop correction to 
the fermionic full propagator, 

\begin{equation}
 G^{-1}(p)=G_0^{-1}(p)+\int\frac{d^3k}{(2\pi)^3}\gamma_\mu G(p-k)\gamma_\nu D_{\mu\nu}(k),
\end{equation}
with a dressed fermion propagator

\begin{equation}
 G(p)=\frac{1}{i\slashchar{p}Z(p)+\Sigma(p)}=\frac{\Sigma(p)-i\slashchar{p}Z(p)}{Z^2(p)p^2+\Sigma^2(p)}.
\end{equation}
Explicitly, we have, 

\begin{eqnarray}
\label{sc}
i\slashchar{p}Z(p)+\Sigma(p)&=&i\slashchar{p}+\frac{8}{N}
\int\frac{d^3k}{(2\pi)^3}\frac{\gamma_\mu
[\Sigma(k-p)+i(\slashchar{k}-\slashchar{p})Z(k-p)]\gamma_\mu}
{[Z^2(k-p)(k-p)^2+\Sigma^2(k-p)]|k|}
\nonumber\\
&-&\frac{8}{N}
\int\frac{d^3k}{(2\pi)^3}\frac{\slashchar{k}[\Sigma(k-p)+i(\slashchar{k}-
\slashchar{p})Z(k-p)]\slashchar{k}}{[Z^2(k-p)(k-p)^2+\Sigma^2(k-p)]|k|^3}.
\end{eqnarray}
Thus, we obtain the 
selfconsistent equations:

\begin{equation} 
\Sigma(p)=\frac{16}{N}\int\frac{d^3k}{(2\pi)^3}\frac{\Sigma(k)}
{[Z^2(k)k^2+\Sigma^2(k)]|k+p|},
\end{equation}

\begin{equation}
Z(p)=1-\frac{8}{Np^2}\int\frac{d^3k}{(2\pi)^3}\frac{[k^2-p^2+(k+p)^2](k+p)\cdot p 
Z(k)}{[Z^2(k)k^2+\Sigma^2(k)]|k+p|^3}.
\end{equation}
In writing the above equations we have used the identity (\ref{ident-Dirac-1}) and 

\begin{equation}
\slashchar{k}\slashchar{p}\slashchar{k}=2p\cdot k\slashchar{k}-k^2\slashchar{p}.
\end{equation}
After integrating over the angles, we obtain 

\begin{eqnarray}
\label{sc-int-sig}
\Sigma(p)&=&\frac{4}{N\pi^2 p}\int_0^\alpha dk
\frac{k\Sigma(k)(k+p-|k-p|)}{Z^2(k)k^2+\Sigma^2(k)}
\nonumber\\
&=&\frac{8}{N\pi^2 p}\left[\int_0^pdk\frac{k^2\Sigma(k)}{Z^2(k)k^2+\Sigma^2(k)}
+p\int_p^\alpha dk\frac{k\Sigma(k)}{Z^2(k)k^2+\Sigma^2(k)}\right],
\nonumber\\
\end{eqnarray}

\begin{equation}
\label{sc-int-Z}
Z(p)=1-\frac{8}{3\pi^2N}\left[\int_p^\alpha dk\frac{kZ(k)}{Z^2(k)k^2+
\Sigma^2(k)}+\frac{1}{p^3}\int_0^p dk\frac{k^4Z(k)}{Z^2(k)k^2+
\Sigma^2(k)}\right].
\end{equation}
From Eq. (\ref{sc-int-sig}) we obtain

\begin{equation}
p^2\frac{d\Sigma(p)}{dp}=-\frac{8}{N\pi^2}\int_0^p dk\frac{k^2\Sigma(k)}{Z^2(k)k^2+\Sigma^2(k)},
\end{equation}
and therefore, 

\begin{equation}
\label{d-Sig}
\frac{d}{dp}\left[p^2\frac{d\Sigma(p)}{dp}\right]=
-\frac{8}{\pi^2N}\frac{p^2\Sigma(p)}{Z^2(p)p^2+\Sigma^2(p)},
\end{equation}
A similar calculation using Eq. (\ref{sc-int-Z})  yields

\begin{equation}
\label{d-Z}
\frac{d}{dp}\left[p^4\frac{dZ(p)}{dp}\right]=
\frac{8}{\pi^2N}\frac{p^4Z(p)}{Z^2(p)p^2+\Sigma^2(p)}.
\end{equation}
In Refs. \cite{Appelq-1} and \cite{Appelq-2} $Z(p)$ is considered as being approximately one, 
so that Eq. (\ref{d-Z}) is absent in their treatment.

From Eq. (\ref{sc-int-sig}) it follows the boundary condition

\begin{equation}
\lim_{p\to 0}p\Sigma(p)=0.
\end{equation}
Another important boundary 
condition is obtained from Eq. (\ref{sc-int-sig}), since

\begin{equation}
p\frac{d\Sigma(p)}{dp}=-\Sigma(p)+\frac{8}{N\pi^2}\int_p^\alpha 
dk\frac{k\Sigma(k)}{Z^2(k)k^2+\Sigma^2(k)},
\end{equation}
implying 

\begin{equation}
\label{bc-sig}
\left.p\frac{d\Sigma(p)}{dp}\right|_{p=\alpha}=-\Sigma(\alpha),
\end{equation}
which should be considered in addition to the boundary condition 
$0\leq\Sigma(0)<\infty$. Similarly, from Eq. (\ref{sc-int-Z}) follows 
the boundary condition

\begin{equation}
\label{bc-Z}
\left.p\frac{dZ(p)}{dp}\right|_{p=\alpha}=3[1-Z(\alpha)],
\end{equation}
while positivity of the spectral represention implies 
$0<Z(0)\leq 1$.

In Ref. \cite{Pisarski} it is found that a fermion mass is dynamically generated 
for all values of $N$. Such a result 
can be obtained by considering the limit 
$p\to 0$ of Eq. (\ref{sc-int-sig}) and replacing $\Sigma(k)$ and 
$Z(k)$ in the integrand by their lowest order expansion in $1/N$,  
$\Sigma(0)$ and unity, respectively. Doing this  
we obtain the gap equation
\begin{equation}
\label{gap-eq}
\Sigma(0)=\frac{8~\Sigma(0)}{N\pi^2}\int_0^\alpha dk\frac{k}{k^2+\Sigma^2(0)}.
\end{equation}
If we assume $\Sigma(0)\neq 0$ we can easily solve the above gap equation to obtain  
\begin{equation}
\label{Sigma}
\Sigma(0)=\alpha e^{-N\pi^2/8},
\end{equation} 
reproducing the result of Ref. \cite{Pisarski}. 
Under the same approximation used 
to derive Eq. (\ref{gap-eq}), we obtain from Eq. (\ref{sc-int-Z}) that 
\begin{eqnarray}
\label{Z}
Z(p)&=&1+\frac{8}{3\pi^2N}
\left\{\ln\left[\frac{\sqrt{p^2+\Sigma^2(0)}}{\alpha}\right]-\frac{1}{3}
\right.
\nonumber\\
&+&\left.\frac{\Sigma^2(0)}{p^2}-\frac{\Sigma^3(0)}{p^3}\arctan
\left[\frac{p}{\Sigma(0)}\right]
\right\}.
\end{eqnarray}
If we insert in Eq. (\ref{Z}) the trivial solution $\Sigma(0)=0$ of the gap equation 
(\ref{gap-eq}), we obtain 
\begin{equation}
\label{Z-1/N}
Z(p)=1+\frac{8}{3\pi^2N}\left[\ln\left(\frac{p}{\alpha}\right)-\frac{1}{3}\right].
\end{equation}
The above equation implies the existence of an anomalous 
dimension $\eta$ defined through
\begin{equation}
\label{eta-usual}
\eta=-\lim_{p\to 0}p\frac{d\ln Z(p)}{dp}, 
\end{equation}
and we obtain the known $1/N$ result \cite{Atkinson} 
\begin{equation}
\label{eta1}
\eta=-\frac{8}{3\pi^2N}.
\end{equation}
On the other hand, 
the $p\to 0$ limit gives
\begin{equation}
\label{Z0-1}
Z(0)=1+\frac{8}{3\pi^2N}\ln\left[\frac{\Sigma(0)}{\alpha}\right].
\end{equation}
Substitution of the result (\ref{Sigma}) in (\ref{Z0-1}) yields 
$Z(0)=2/3$, which contradicts  the value $Z(0)=3/4$ obtained 
in Ref. \cite{Pisarski} through an 
argument involving the Ward identities \footnote{This result is discussed in Ref. [19] of Ref. \cite{Pisarski}.}. The contradiction is removed if instead 
of Eq. (\ref{gap-eq}) the following gap equation is used:
\begin{equation}
\label{gap-eq-1}
\Sigma(0)=\frac{8~\Sigma(0)}{N\pi^2}\int_0^\alpha dk\frac{k}{Z^2(0)k^2+\Sigma^2(0)}.
\end{equation}
In this case we obtain
\begin{equation}
\Sigma(0)=\alpha Z(0)e^{-N\pi^2Z^2(0)/8},
\end{equation} 
\begin{equation}
Z(0)=1+\frac{8}{3\pi^2NZ(0)}\ln\left[\frac{\Sigma(0)}{\alpha Z(0)}\right],
\end{equation}
implying that
\begin{equation}
Z(0)=\frac{3}{4}.
\end{equation}
Note that such a modification affects the value of the anomalous 
dimension, since in the limit of a vanishing gap we have
\begin{eqnarray}
Z(p)&=&1+\frac{8}{3\pi^2NZ(0)}\left[\ln\left(\frac{p}{\alpha}\right)-\frac{1}{3}\right]
\nonumber\\
&=&1+\frac{32}{9\pi^2N}\left[\ln\left(\frac{p}{\alpha}\right)-\frac{1}{3}\right],
\end{eqnarray}
instead of the standard $1/N$ result (\ref{Z-1/N}). Thus, we obtain instead of 
Eq. (\ref{eta1}) the result   
\begin{equation}
\label{eta2}
\eta=-\frac{32}{9\pi^2N}.
\end{equation}

Thus, revisiting the earlier approach of Ref. \cite{Pisarski} taught us an 
important lesson: in order to not violate the Ward identities when 
linearizing the problem, we have to 
account not only for $\Sigma(0)$ but also for $Z(0)$. 

We consider next an improved linearized problem 
given by the approximated differential equations
\begin{equation}
\label{d-Sig-1}
\frac{d}{dp}\left[p^2\frac{d\Sigma(p)}{dp}\right]=
-\frac{8}{\pi^2N}\frac{p^2\Sigma(p)}{Z^2(0)p^2+\Sigma^2(0)},
\end{equation}
\begin{equation}
\label{d-Z-1}
\frac{d}{dp}\left[p^4\frac{dZ(p)}{dp}\right]=
\frac{8}{\pi^2N}\frac{p^4Z(p)}{Z^2(0)p^2+\Sigma^2(0)}.
\end{equation}
The solutions obeying the initial conditions $\Sigma'(0)=0$ and 
$Z'(0)=0$, while $\Sigma(0)\neq 0$ and $Z(0)\neq 0$, as required by 
the boundary conditions, are
\begin{equation}
\label{sol-Z}
Z(p)=Z(0) {}_2F_1\left[\frac{3}{4}-\frac{3}{4}\zeta,\frac{3}{4}+\frac{3}{4}\zeta;
\frac{5}{2};-\frac{Z^2(0)p^2}{\Sigma^2(0)}\right],
\end{equation}
\begin{equation}
\label{sol-Sig}
\Sigma(p)=\Sigma(0) {}_2F_1\left[\frac{1}{4}-\frac{\rm i}{4}\gamma,\frac{1}{4}+\frac{\rm i}{4}\gamma;
\frac{3}{2};-\frac{Z^2(0)p^2}{\Sigma^2(0)}\right],
\end{equation}
where $_2F_1$ is a hypergeometric function and 
%
\begin{equation}
\label{ag}
\zeta=\sqrt{1+\frac{32}{9\pi^2Z^2(0)N}},
~~~~~~~~~~~~~~~~
\gamma=\sqrt{\frac{32}{\pi^2Z^2(0)N}-1}.
\end{equation}
The solution (\ref{sol-Sig}) with $Z(0)=1$ was obtained before in Ref. \cite{Kondo}. 

The CSB is more easily analysed in the regime where $Z^2(0)p^2\gg\Sigma^2(0)$, in which case the 
above solutions simplify to 
\begin{equation}
\label{sol-Z-a}
Z(p)\approx\frac{9\sqrt{\pi}}{8}Z(0)B\left[\frac{Z(0)p}{\Sigma(0)}\right]
^{-3/2}\cosh\left\{\frac{3\zeta}{2}\ln\left[\frac{Z(0)p}{\Sigma(0)}\right]
+\varphi\right\},
\end{equation}
\begin{equation}
\label{sol-Sig-a}
\Sigma(p)\approx\frac{|C|}{4}\sqrt{\frac{\pi\Sigma^3(0)}{Z(0)p}}\cos
\left\{\frac{\gamma}{2}\ln\left[\frac{Z(0)p}{\Sigma(0)}\right]+\theta\right\},
\end{equation}
where
\begin{equation}
A_\pm=\frac{\Gamma(\pm\zeta)(1\pm\zeta)}{\Gamma^2\left(\frac{7}{4}\pm\frac{3\zeta}{4}
\right)},~~~~~~~~~B=\sqrt{|A_+A_-|},~~~~~~~~~
C=\frac{\Gamma({\rm i}\gamma/2)(1+{\rm i}\gamma)}{\Gamma^2\left(\frac{5}{4}
+{\rm i}\frac{\gamma}{4}\right)},
\end{equation}
and 
\begin{equation}
\theta=\arccos\left(\frac{C+C^*}{2|C|}\right),~~~~~~~~~~
\varphi=\frac{1}{2}\ln\left|\frac{A_+}{A_-}\right|.
\end{equation}
The boundary conditions can be used 
to determine $\Sigma(0)$ and simplify the 
above equations. For $\Sigma(0)$ we obtain 
\begin{equation}
\label{spectrum-new}
\Sigma(0)=\alpha Z(0)\exp\left\{-\frac{2}{\gamma}\arccos\left[
\frac{\pi Z(0)}{4}\sqrt{\frac{N}{2}}\left(\sin\theta+\gamma\cos\theta
\right)\right]\right\}.
\end{equation}
Note that the critical number of flavors is now modified due to the wave function 
renormalization at $p=0$ and given by demanding that $\gamma$ vanishes, i.e., 
\begin{equation}
\label{Nch}
N_{\rm ch}=\frac{32}{\pi^2Z^2(0)}.
\end{equation}
Therefore, Eq. (\ref{spectrum-new}) can be rewritten as
\begin{equation}
\label{Sigma(0)}
\Sigma(0)=\alpha Z(0)\exp\left\{-\frac{2}{\gamma}\arccos\left[
\sqrt{\frac{N}{N_{\rm ch}}}\left(\sin\theta+\gamma\cos\theta
\right)\right]\right\}.
\end{equation}
Note that $\theta$ is non-singular as $N\to N_{\rm ch}$, approaching the 
value $(2m+1)\pi/2$, with $m=1,2,\dots$. Thus, near $N=N_{\rm ch}$ 
Eq. (\ref{Sigma(0)}) becomes 
\begin{equation}
\label{Sigma(0)-1}
\Sigma(0)=\alpha Z(0)\exp\left(-\frac{2\pi n}{\gamma}\right),
\end{equation}
with $n=1,2,\dots$, which  
agrees with Ref. \cite{Appelq-2} for $Z(0)=1$. Interestingly, the functional form of the generated mass gap $\Sigma(0)$ resembles 
the one in Eq. (\ref{gap-KT}) for the KT transition. In the present case the number of fermionic degrees of freedom is playing 
the role of the temperature. 

Inserting (\ref{Sigma(0)-1}) with $n=1$ in (\ref{sol-Z-a}) yields
\begin{equation}
Z(p)=\frac{9\sqrt{\pi}}{8}Z(0)Be^{-3\pi/\gamma}\left
(\frac{p}{\alpha}\right)^{-3/2}\cosh\left[\frac{3\zeta}{2}\ln\left(\frac
{p}{\alpha}\right)+\frac{3\pi\zeta}{\gamma}+\varphi\right].
\end{equation}
Using Eq. (\ref{bc-Z}) this can be further rewritten as
\begin{equation}
Z(p)=\frac{2(p/\alpha)^{-3/2}}{\cosh(3\pi\zeta/\gamma+\varphi)
+\zeta\sinh(3\pi\zeta/\gamma+\varphi)}\cosh\left[\frac{3\zeta}{2}\ln\left(\frac
{p}{\alpha}\right)+\frac{3\pi\zeta}{\gamma}+\varphi\right].
\end{equation}
For $N\geq N_{\rm ch}$ the mass gap $\Sigma(0)$ vanishes and the above equation 
becomes simply
\begin{equation}
Z(p)=\frac{2}{1+\zeta}\left(\frac{p}{\alpha}\right)^{-\widetilde \eta},
\end{equation}
which defines the anomalous dimension
\begin{equation}
\label{eta-tilde}
\widetilde \eta=\frac{3}{2}(1-\zeta).
\end{equation}
Precisely at $N=N_{\rm ch}$ we have $\widetilde \eta=-0.08$. 
On the other hand, for $N\gg N_{\rm ch}$ we obtain the following 
large $N$ result:
\begin{equation}
\label{eta-tilde1}
\widetilde \eta\approx-\frac{N_{\rm ch}}{12N}=-\frac{8}{3\pi^2Z^2(0)N}.
\end{equation}

The result (\ref{eta-tilde}) can be more easily obtained by considering 
Eq. (\ref{d-Z}) at the critical point, in which case it becomes
\begin{equation}
\label{d-Z-crit}
\frac{d}{dp}\left[p^4\frac{dZ(p)}{dp}\right]=
\frac{8}{\pi^2N}\frac{p^2}{Z(p)}.
\end{equation}
Eq. (\ref{d-Z-crit}) can be rewritten as a set of two RG-like equations:
\begin{equation}
p\frac{d\ln Z(p)}{dp}=-\widetilde \gamma(p),
\end{equation}
\begin{equation}
p\frac{d\widetilde \gamma}{dp}=\widetilde \gamma(\widetilde \gamma-3)-\frac{8}{\pi^2NZ^2}.
\end{equation}
Thus, $\widetilde \eta=\lim_{p\to 0}\widetilde \gamma(p)$. This limit is attained at the 
fixed point, which in this case is given by
\begin{equation}
\widetilde \gamma_\pm=\frac{3}{2}(1\pm\zeta).
\end{equation}
Only the solution $\widetilde \gamma_-$ makes sense since $\tilde \gamma_+>3$, which would 
lead to a vanishing of the fermion propagator as $p\to 0$. Therefore, 
we obtain that $\widetilde \eta=\widetilde \gamma_-$, in agreement with Eq. (\ref{eta-tilde}).

Within the present calculation the 
determination of $Z(0)$ is not so straightforward, as the approximation we have used implies an 
anomalous dimension.  On the 
other hand, since the anomalous dimension arises only for vanishing gap, it is 
reasonable to assume that $\widetilde \eta$ coincides with the anomalous dimension we 
have computed before, Eq. (\ref{eta2}). Thus, we demand that 
$\widetilde \eta=\eta$, such that 
$Z(0)=\sqrt{3}/2$ is obtained, contrasting with the $Z(0)$ obtained before, which 
is given by the square of the present result. Therefore, we obtain from Eq. (\ref{Nch}) 
the critical value $N_{\rm ch}=128/(3\pi^2)\approx 4.32$, in full agreement with the result of 
Ref. \cite{Kondo1} and the one of 
Ref. \cite{Nash} in the case where the anomalous dimension is computed 
up to order $1/N$.  

The physical interpretation of the above results is as follows. The number $N_{\rm ch}$ separates two different 
insulating regimes, namely, a magnetically ordered one at $N<N_{\rm ch}$ and another one having no broken symmetries at 
$N>N_{\rm ch}$. The latter can be identified with a $U(1)$ spin liquid state. The physically interesting case is $N=2$, which corresponds to an 
$SU(2)$ Heisenberg antiferromagnet. Thus, the above Schwinger-Dyson calculation indicates that there is no spin liquid 
state in this case. However, there is numerical evidence \cite{Kogut-2008} that the Schwinger-Dyson approach overestimates the 
value of $N_{\rm ch}$. The numerical analysis made in Ref. \cite{Kogut-2008} indicates that $N_{\rm ch}\approx 1.5$. This 
result is in agreement with an earlier prediction \cite{Appel} based on mathematical inequalities and symmetry arguments. 
If this result is correct, it would imply that the two-dimensional Heisenberg antiferromagnet has a spin liquid ground state. 

\chapter*{Acknowledgements}

I would like to the thank the organizers of the VIII School of Physics of the CBPF (Brazilian Center for Research in Physics in the Portuguese 
abreviation), especially the director of the school, Luiz Sampaio, and the colleagues who helped to make the invitation possible, like 
Adolfo P. C. Malbouisson and Itzhak Roditi, with whom I had many interesting discussions during 
my stay.

I have discussed with several colleagues 
on the topics of these lecture notes in the last years. I would 
like to thank in particular my collaborators Hagen Kleinert, Asle Sudb{\o},    
Karl Bennemann, Ilya Eremin, Ren\'e Tarento, Zohar Nussinov, Dirk Manske, Steinar Kragset, 
Joakim Hove, Eivind Sm{\o}rgrav, and Jo Smiseth. 

I had many illuminating discussions with the following 
colleagues. For the topic on quantum phase transitions in Mott insulators, 
I benefited in the last years from discussions I had with 
Anders Sandvik, Subir Sachdev, Anatoly Kuklov, 
Nikolay Prokof'ev, Boris Svistunov, Zi Cai, and Ying Jiang. 
For topics around QED in $2+1$ dimensions I had many interesting 
discussions with Igor Herbut, Zlatko Tesanovic, J\"org Schmalian,  
Guo-Zhu Liu, Petr Jizba, and  J\"urgen Dietel.  
I also benefited from discussions I had on Bose-Einstein condensation with   
Axel Pelster, Anna Posazhennikova, 
Adriaan Schakel, Victor Bezerra, Ednilson Santos, 
and Aristeu Lima.

\begin{appendices}
\chapter{The surface of a sphere in $d$ dimensions}
\label{app:s_d}

In this appendix we will derive the formula for the surface of a sphere whose radius is 
unity in $d$ dimensions. Such a factor occurs in many calculations in the text and we will 
derive it here as a simple exercise involving $d$-dimensional integrals. The trick is 
to consider the $d$-dimensional Gauss integral 

\begin{equation}
 I=\int d^dx e^{-{\bf x}^2}
\end{equation}
in two ways. In the first way we simply evaluate the above integral directly, since it 
is just an ordinary one-dimensional Gauss integral to the power $d$. Thus, 

\begin{equation}
\label{gauss-d}
 I=\pi^{d/2}.
\end{equation}
Next, we use spherical coordinates in $d$ dimensions and formally perform the integral over 
the $d-1$ angular variables. We call the result of such an integration $S_d$, which is 
precisely the surface of a unit sphere in $d$ dimensions. In this way, we can also write, 

\begin{equation}
 I=S_d\int_0^\infty dr r^{d-1}e^{-r^2}.
\end{equation}
Now we make the change of variables $u=r^2$, such that,

\begin{eqnarray}
 I&=&\frac{S_d}{2}\int_0^\infty du u^{d/2-1}e^{-u}
\nonumber\\
&=&\frac{S_d\Gamma(d/2)}{2},
\end{eqnarray}
where we have used the definition of the gamma function. From Eq. (\ref{gauss-d}) we obtain 
immediately,

\begin{equation}
 S_d=\frac{2\pi^{d/2}}{\Gamma(d/2)},
\end{equation}
which is the desired result.

\chapter{Dimensional regularization and evaluation of some simple integrals in $d$ dimensions}
\label{app:int-d-m}

In this Appendix we evaluate the following simple $d$-dimensional integrals and, at the same time, 
introduce the method of dimensional regularization of integrals:

\begin{equation}
 I_1=\int\frac{d^dk}{(2\pi)^d}\frac{1}{k^2+M^2},
\end{equation}
and 

\begin{equation}
 I_2=\int\frac{d^dk}{(2\pi)^d}\frac{1}{(k^2+M^2)^2}.
\end{equation}
Both integrals can be obviously be done in $2<d\leq 4$ if an ultraviolet cutoff $\Lambda$ is introduced. 
In the limit $\Lambda\to\infty$ divergencies appear. Dimensional regularization converts logarithmic divergencies due 
to the infinite cutoff into poles for some values of $d$. Moreover, divergencies involving positive powers of 
$\Lambda$ ``disappear''. In order to understand the consistence of the technique, let us consider the integral 
$I_1$ as a function of $M$ and $d$, while demanding that it should be cutoff independent. This can be done by considering 
$d$ somewhere in the complex plane, such that the integral is convergent. The result of the integration, as a function of 
$d$, can then be analytically continued in a larger domain (for more details, see Refs. \cite{ZJ} and \cite{KSF}). 
In this way, simple dimensional analysis yields

\begin{equation}
\label{I1-1}
 I_1(M,d)=c_1(d)M^{d-2},
\end{equation}
where $c(d)$ is a dimensionless function of $d$. We obtain similarly for $I_2$, 

\begin{equation}
 I_2(M,d)=c_2(d)M^{d-4}.
\end{equation}
Note that if it is assumed that no cutoff is needed, the only scale available is $M$. This way of doing things is not always useful and 
sometimes we may need the explicit cutoff dependence. Note that Eq. (\ref{I1-1}) implies that $I_1$ vanishes for all $d>2$ if $M=0$, which 
is certainly not true if we regularize the integral with a cutoff. Indeed, if a cutoff is introduced, we have

\begin{equation}
 I_1(0,d)=\frac{S_d\Lambda^{d-2}}{d-2},
\end{equation}
thus leading to a positive power of $\Lambda$ when $d>2$. Thus, by demanding that $I_1(M,d)$ is independent of $\Lambda$ for all $d$, we obtain 
that this condition actually imposes

\begin{equation}
\int\frac{d^dk}{(2\pi)^d}\frac{1}{k^2}=0. 
\end{equation}
The above equation is one of the properties required by dimensional regularization. In fact, an even more general formula should be used, namely, 

\begin{equation}
\int\frac{d^dk}{(2\pi)^d}\frac{1}{k^\alpha}=0, 
\end{equation}
where $\alpha\in\mathbb{C}$.

We now proceed with the explicit evaluation of the constants $c_1(d)$ and $c_2(d)$. To this end, note that the integral 
$I_1$ can be written as

\begin{equation}
 I_1=\int_0^\infty d\lambda\int\frac{d^dk}{(2\pi)^d} e^{-\lambda(k^2+M^2)}.
\end{equation}
The $d$-dimensional Gauss integral is easily performed to obtain, 

\begin{equation}
 I_1=\frac{1}{(4\pi)^{d/2}}\int_0^\infty d\lambda \lambda^{-d/2}e^{-\lambda M^2}.
\end{equation}
By changing the variable as $u=M^2\lambda$, we obtain, 

\begin{equation}
 I_1=\frac{M^{d-2}}{(4\pi)^{d/2}}\int_0^\infty du u^{-d/2}e^{-u}
=\frac{M^{d-2}}{(4\pi)^{d/2}}\Gamma\left(1-\frac{d}{2}\right),
\end{equation}
where we have used the integral definition of the gamma function. 

The integral $I_2$ is obtained from $I_1$ via simple differentiation of 
$I_1$ with respect to $M^2$. We obtain, 

\begin{equation}
 I_2=-\frac{\partial I_1}{\partial M^2}
=\frac{M^{d-4}}{(4\pi)^{d/2}}\Gamma\left(2-\frac{d}{2}\right).
\end{equation}

Note that the integral $I_1$ has poles for $d=2$ and for $d=4$, while $I_2$ has only a pole at $d=4$.  

Now we will consider the most general case, i.e., we will evaluate the integral 

\begin{equation}
 I_\alpha(d)=\int\frac{d^dk}{(2\pi)^d}\frac{1}{(k^2+M^2)^\alpha}.
\end{equation}
Using 
the identity (\ref{ident}), we have, 

\begin{eqnarray}
\label{I-alpha}
 I_\alpha(d)&=&\frac{1}{\Gamma(\alpha)}\int_0^\infty d\lambda \int\frac{d^dk}{(2\pi)^d}e^{-\lambda(k^2+M^2)}\lambda^{\alpha-1}
\nonumber\\
&=&\frac{1}{(4\pi)^{d/2}\Gamma(\alpha)}\int_0^\infty d\lambda\lambda^{\alpha-1-d/2}e^{-\lambda M^2}
\nonumber\\
&=&\frac{M^{d-2\alpha}}{(4\pi)^{d/2}\Gamma(\alpha)}\int_0^\infty du u^{\alpha-1-d/2}e^{-u}
\nonumber\\
&=&\frac{M^{d-2\alpha}}{(4\pi)^{d/2}\Gamma(\alpha)}\Gamma\left(\alpha-\frac{d}{2}\right),
\end{eqnarray}
where from the first to the second line we have performed the $d$-dimensional Gauss integral in ${\bf k}$ and from the second to the 
third line we made the substitution $u=M^2\lambda$. From the third to the last line we used the definition of the gamma function. 

Of course, for $\alpha=1$ and $\alpha=2$ the integral (\ref{I-alpha}) reproduces the results 
$I_1$ and $I_2$, respectively. 

\chapter{Evaluation of the integral over a Bose-Einstein distribution in
$d$ dimensions}
\label{app:be-int}

Let us consider the following integral:

\begin{eqnarray}
 I_{\rm BE}(z,D)&=&\int\frac{d^dk}{(2\pi)^d}\frac{1}{e^{c|{\bf k}|^z}-1}
\nonumber\\
&=&\frac{S_d}{(2\pi)^d}\int_0^\infty dk\frac{k^{d-1}}{e^{ck^z}-1},
\end{eqnarray}
where $c>0$ and $S_d$ is given in Appendix \ref{app:s_d}. The exponent $z>0$ defines 
the power law for the spectrum of the corresponding Bose particle. For example, 
a free particle will have $z=2$, while phonons have $z=1$. 

After performing the change of variables $u=ck^z$, the integral becomes, 

\begin{equation}
\label{int-bec}
 I_{\rm BE}(z,D)=\frac{S_d}{(2\pi)^dzc^{d/z}}\int_0^\infty\frac{u^{d/z-1}}{e^u-1}.
\end{equation}
In order to calculate the integral appearing in the expression above, we consider the 
integral representation of the gamma function, 

\begin{equation}
 \Gamma(s)=\int_0^\infty dt t^{s-1}e^{-t},
\end{equation}
and make the replacement $t\to nt$, where $n$ is a positive integer. This leads to 

\begin{equation}
 \frac{\Gamma(s)}{n^s}=\int_0^\infty dt t^{s-1}e^{-nt}.
\end{equation}
Now we sum over $n$ from $1$ to $\infty$, 

\begin{equation}
 \Gamma(s)\sum_{n=1}^\infty\frac{1}{n^s}=\int_0^\infty dt t^{s-1}\sum_{n=1}^\infty e^{-nt}. 
\end{equation}
On the LHS of the above equation appears the definition of the zeta function,

\begin{equation}
 \zeta(s)=\sum_{n=1}^\infty\frac{1}{n^s},
\end{equation}
while on the RHS features a trivial geometric sum, 

\begin{equation}
 \sum_{n=1}^\infty e^{-nt}=\frac{1}{1-e^{-t}}-1=\frac{1}{e^t-1}.
\end{equation}
Therefore, 

\begin{equation}
 \zeta(s)\Gamma(s)=\int_0^\infty dt\frac{t^{s-1}}{e^t-1}.
\end{equation}
Using the above result in Eq. (\ref{int-bec}), we obtain finally, 

\begin{equation}
\label{int-bec-1}
 I_{\rm BE}(z,D)=\frac{S_d}{(2\pi)^dzc^{d/z}}\zeta\left(\frac{d}{z}\right)
\Gamma\left(\frac{d}{z}\right).
\end{equation}

\chapter{Matsubara sums}
\label{app:sums}

The computation of Matsubara sums is a standard method in both relativistic and 
non-relativistic quantum field theory; see for example Ref. \cite{Mahan}. 

Let us evaluate here two bosonic Matsubara sums frequently used in the text, namaly, 

\begin{equation}
\label{s1}
 S_1=-\frac{1}{\beta}\sum_{n=-\infty}^\infty\frac{1}{i\omega_n-\epsilon},
\end{equation}
and

\begin{equation}
\label{s2}
 S_2=\frac{1}{\beta}\sum_{n=-\infty}^\infty\frac{1}{\omega_n^2+\epsilon^2},
\end{equation}
where $\omega_n=2\pi n/\beta$.

Although, strictly speaking, the first sum does not converge, it can nevertheless be done 
using regularization methods. Moreover, as we will see, it is related to the second sum, which 
is convergent. 

Since $S_2$ is convergent, let us calculate this one first. 

\begin{figure}
\begin{center}
\includegraphics[width=10cm]{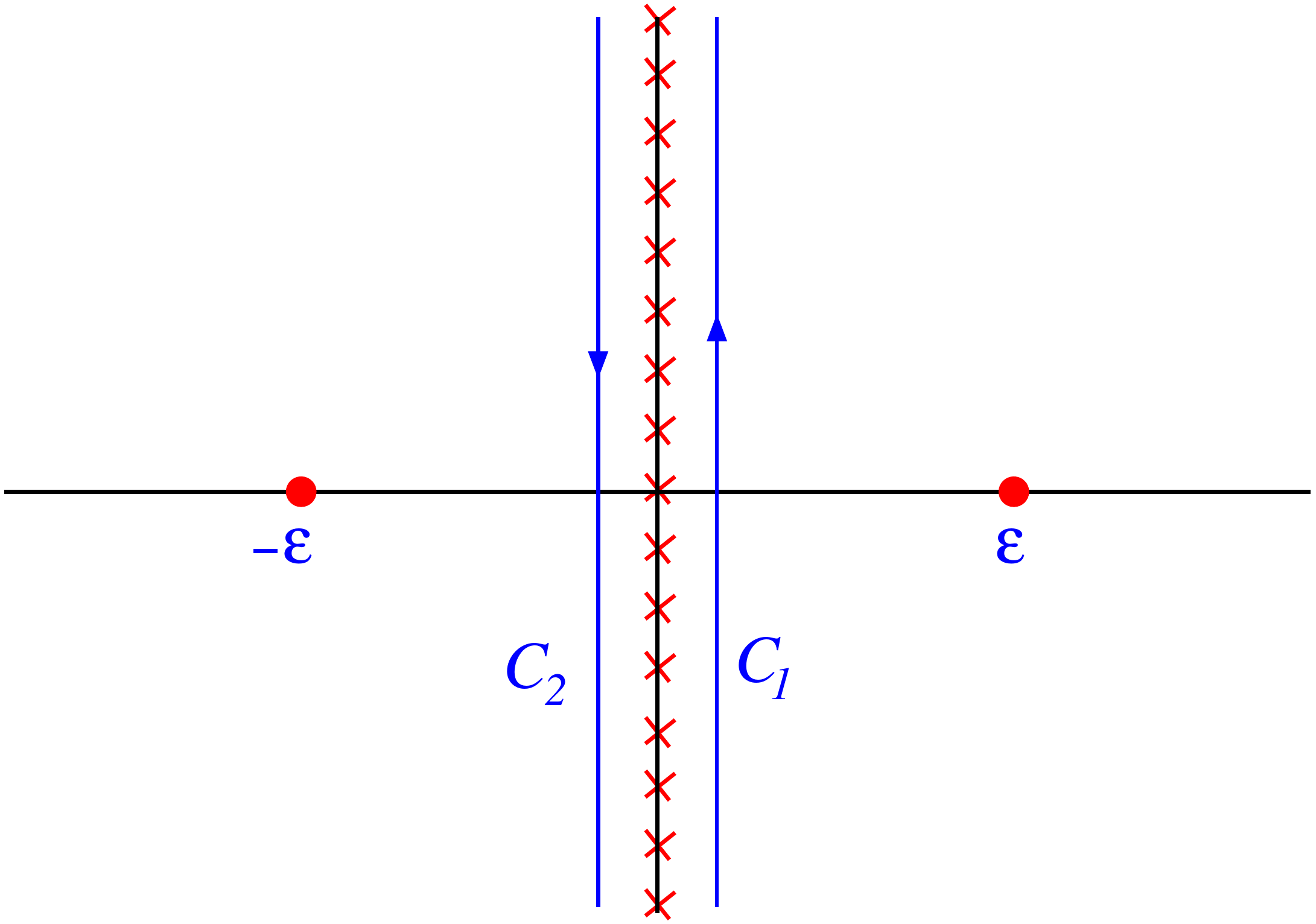}
\caption{Initial contour integration in complex plane 
used for performing the Matsubara sum $S_2$. The 
red crosses indicate the position of the Matsubara modes.}\label{fig:contour-s2-1}
\end{center}
\end{figure} 

\begin{figure}
\begin{center}
\includegraphics[width=10cm]{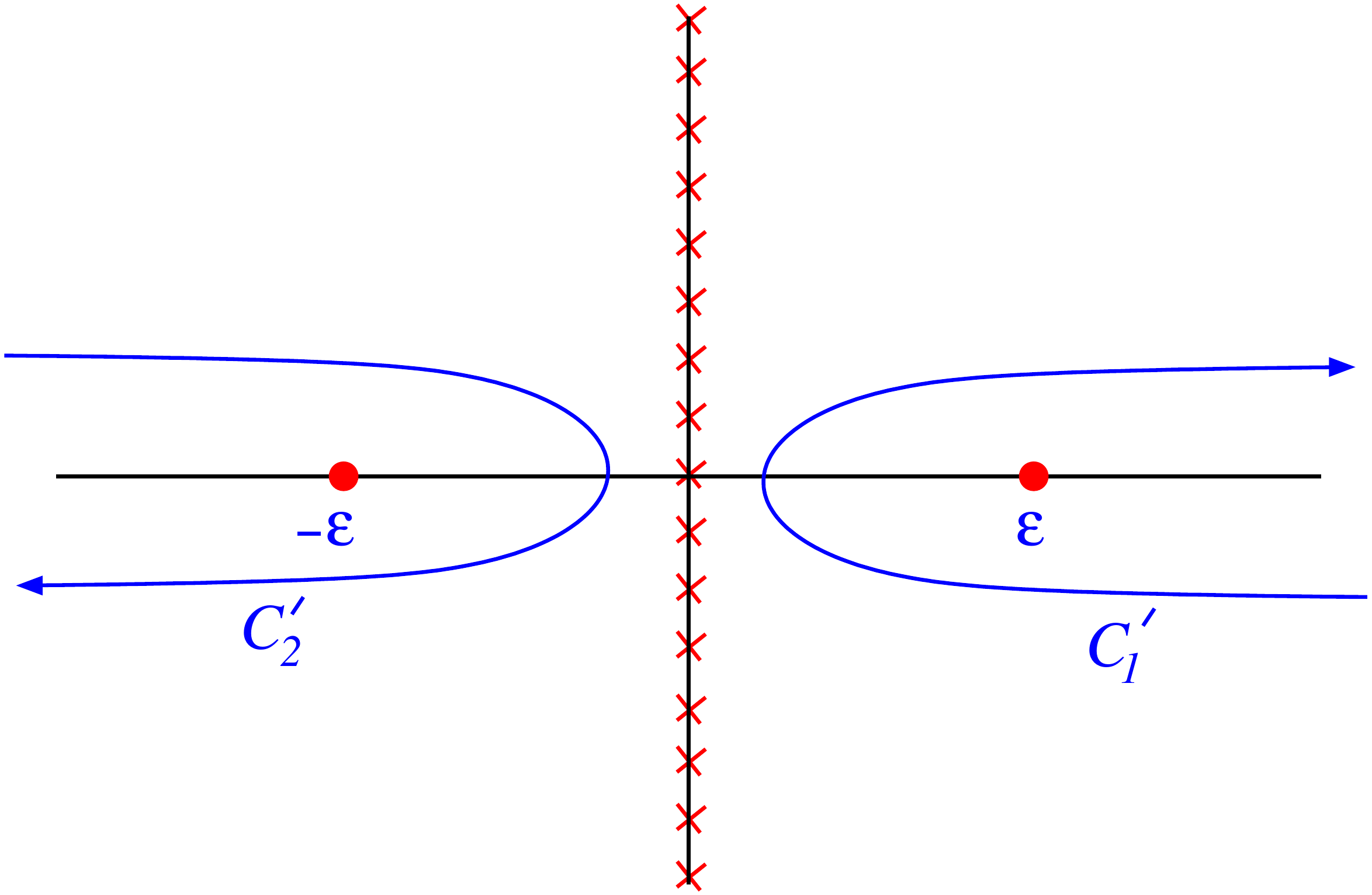}
\caption{New contour integration in complex plane 
used for performing the Matsubara sum $S_2$. The contours 
$C_1'$ and $C_2'$ enclose the poles $z=\pm \epsilon$.}\label{fig:contour-s2-2}
\end{center}
\end{figure}

In order to calculate $S_2$, we rewrite the sum in terms of a contour integral over an appropriate 
contour in such a way as to set the sum of the residues at poles given by the Matsubara 
frequency modes. The two straight lines shown in Fig. \ref{fig:contour-s2-1} enclose 
all Matsubara modes. We can imagine a finite number of modes with a closed curve around all 
of them. In the limit of an infinite number of modes, the closed curve becomes two 
parallel lines having the Matsubara modes between them. This picture explains why the 
arrows are pointing the way they do: if to the closed curve a counterclockwise orientation 
is given, in the limit of an infinite number of Matsubara modes the resulting straight line on 
the right will point upwards, while the one on the left will point downwards. Thus, the 
sum $S_2$ is equivalent to

\begin{equation}
 S_2=-\int_{C_1\cup C_2}\frac{dz}{2\pi i}\frac{1}{z^2-\epsilon^2}\frac{1}{e^{\beta z}-1}.
\end{equation}
The contour $C_1\cup C_2$ encloses the poles of the function $1/(e^{\beta z}-1)$, but not the 
two poles of $1/(z^2-\epsilon^2)$. The straight lines $C_1$ and $C_2$ can be deformed into 
the curves $C_1'$ and $C_2'$ shown in Fig. \ref{fig:contour-s2-2} without encountering any 
singularity. Note that the deformed contours go clockwise around the poles 
$z=\pm \epsilon$, which introduces an extra minus sign. Thus, the residue theorem yields, 

\begin{equation}
 S_2=\frac{1}{2\epsilon}\coth\left(\frac{\beta\epsilon}{2}\right).
\end{equation}

\begin{figure}
\begin{center}
\includegraphics[width=10cm]{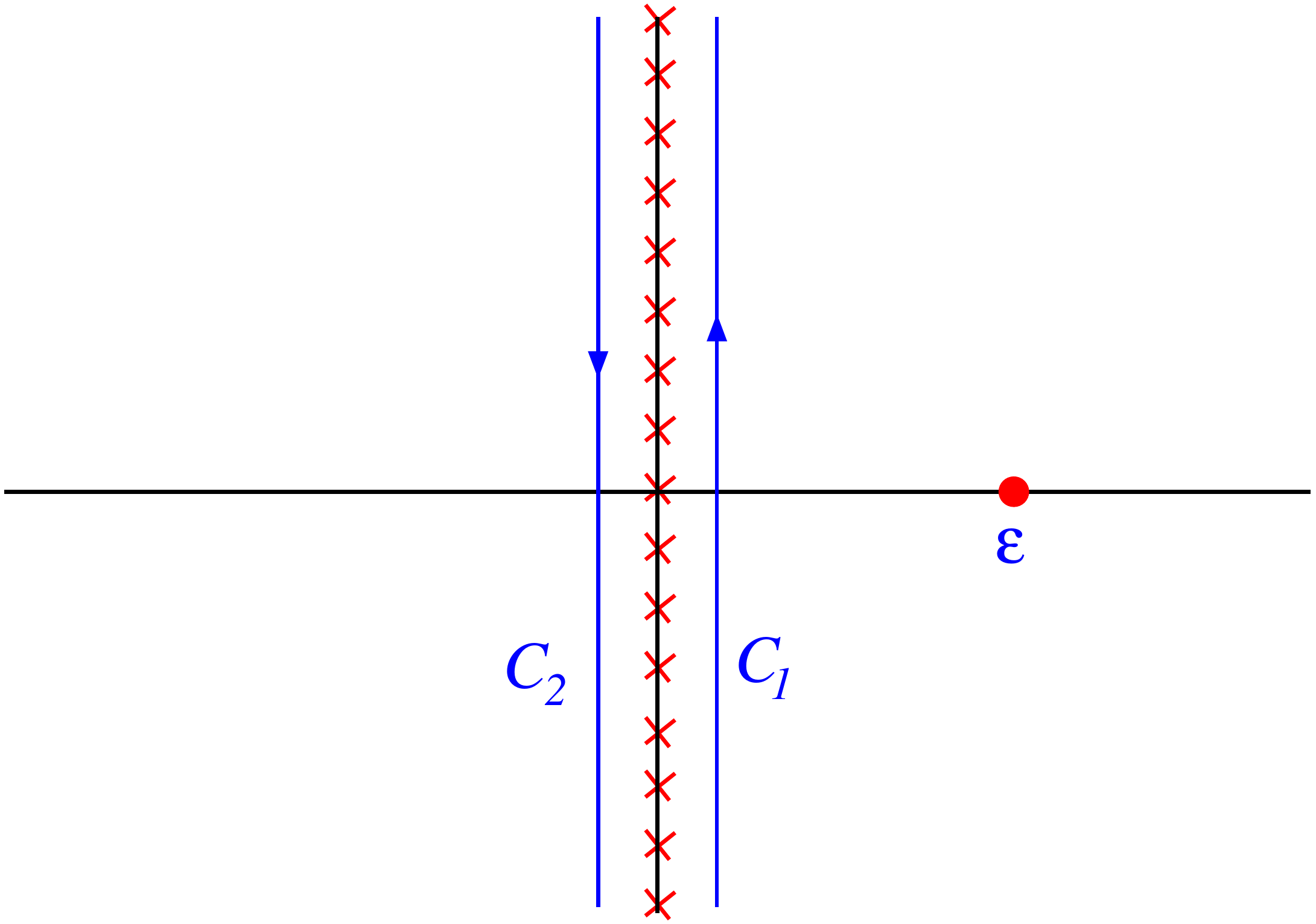}
\caption{Initial contour integration in the complex plane 
used for performing the Matsubara sum $S_1$.}\label{fig:contour-s1-1}
\end{center}
\end{figure}

\begin{figure}
\begin{center}
\includegraphics[width=10cm]{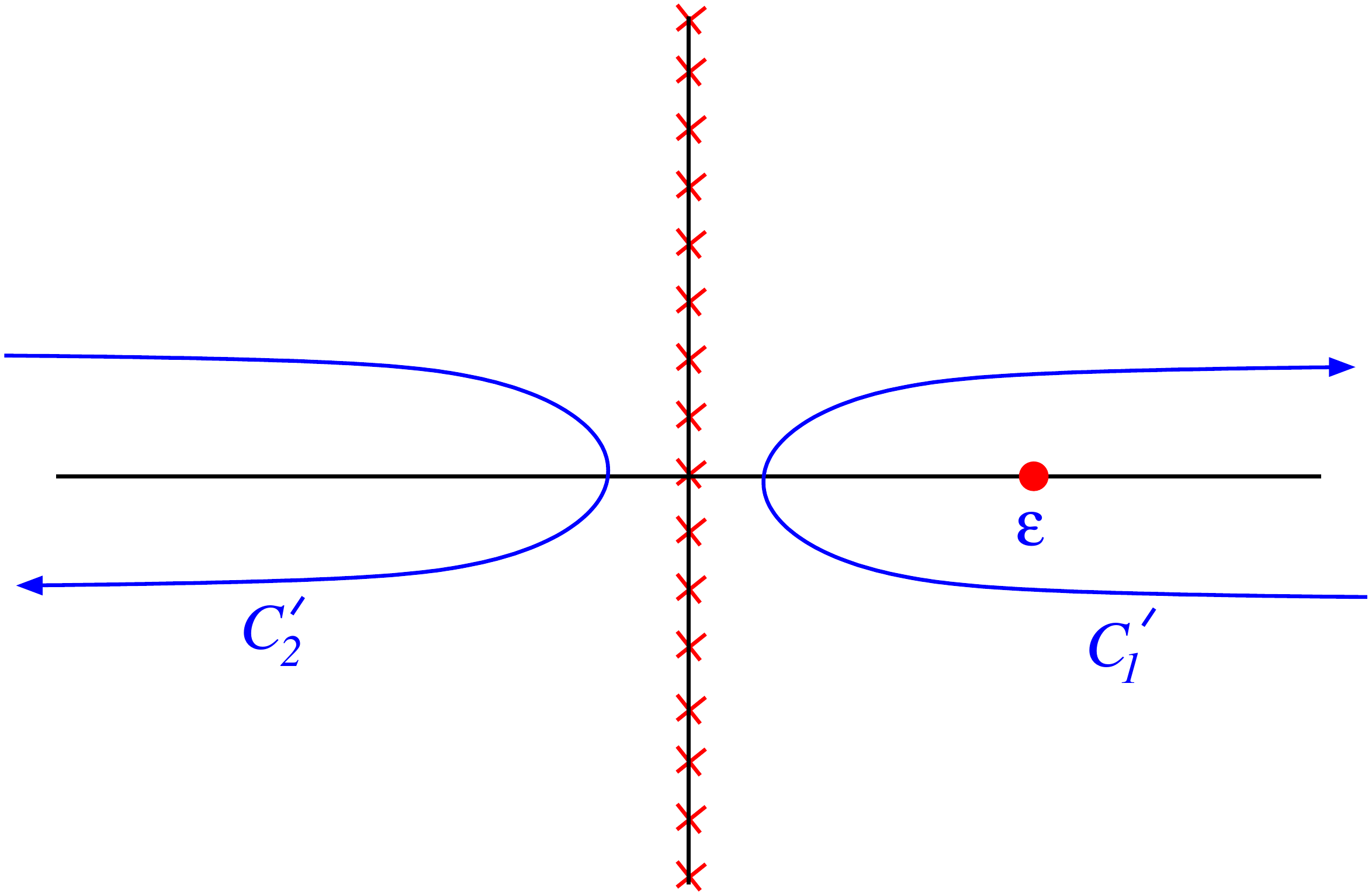}
\caption{Contour integration in complex plane 
used for performing the Matsubara sum $S_1$. The contour 
$C_1'$ encloses the pole $z=\epsilon$, while no pole is enclosed by 
the contour $C_2'$.}\label{fig:contour-s1-2}
\end{center}
\end{figure}

Now we will consider the sum $S_1$. In contrast to the case of $S_2$, this sum is related to only one 
pole at $z=\epsilon$ in the contour integration. The initial contour is shown in Fig. \ref{fig:contour-s1-1}. 
Now by deforming the contours $C_1$ and $C_2$ into the new contours $C_1'$ and $C_2'$ illustrated in 
Fig. \ref{fig:contour-s1-2}, we see that $C_2'$ does not enclose any pole and thus contributes nothing in 
the evaluation of the sum. Therefore, we obtain,  

\begin{equation}
\label{s1-2}
 S_1=\frac{1}{e^{\beta\epsilon}-1}.
\end{equation}

The sum $S_1$ can be related to $S_2$ in the following way, 

\begin{eqnarray}
\label{s1-1}
 S_1&=&\frac{1}{\beta}\sum_{n=-\infty}^\infty\frac{\epsilon+i\omega_n}{\omega_n^2+\epsilon^2}
\nonumber\\
&=&\epsilon S_2+\frac{1}{\beta}\sum_{n=-\infty}^\infty\frac{i\omega_n}{\omega_n^2+\epsilon^2}.
\end{eqnarray}
The second term in the second line of the above equation can be thought to vanish, as to every positive term 
there is a corresponding negative term of identical magnitude. Thus, it seems that we 
just have to plug the result of the sum $S_2$ into the above result and we are done. 
However, this is not quite correct. The point is that $S_2$ involves two poles, $z=\pm \epsilon$, 
corresponding to a ``particle'' and an ``anti-particle'' mode, while $S_1$ involves just a single 
pole at $z=\epsilon$. 
Actually the second sum in Eq. (\ref{s1-1}) is  not convergent. This is consistent with the the already 
mentioned fact, that the sum (\ref{s1}) is not convergent.  
Since, 

\begin{equation}
 S_1=\frac{1}{2}\coth\left(\frac{\beta\epsilon}{2}\right)+S_3=
\frac{1}{2}+\frac{1}{e^{\beta\epsilon}-1}+S_3,
\end{equation}
where 

\begin{equation}
 S_3=\frac{1}{\beta}\sum_{n=-\infty}^\infty\frac{i\omega_n}{\omega_n^2+\epsilon^2},
\end{equation}
and using Eq. (\ref{s1-2}), we obtain that 

\begin{equation}
 \label{s3}
S_3=-\frac{1}{2}.
\end{equation}

The factor $1/2$ subtracted via the sum $S_3$ reflects the ``vacuum'' contribution associated to the particle-antiparticle 
pair. In $S_2$ this leads to a finite zero-temperature contribution, which is absent in the 
non-relativistic Matsubara sum. 

\chapter{Classical limit of the polarization bubble of the dilute Bose gas}
\label{app:bubble}

After setting $\lambda_0=\mu$ Eq. (\ref{bubble}) can be rewritten 
as

\begin{equation}
\label{bubble-rw}
\Pi(i\omega,{\bf p})=4m^2\int\frac{d^dq}{(2\pi)^d}n_B\left(
\frac{q^2}{2m}\right)\left(\frac{1}{2mi\omega-p^2-2{\bf p}\cdot
{\bf q}}-\frac{1}{2mi\omega+p^2-2{\bf p}\cdot{\bf q}}\right).
\end{equation}
In the classical approximation we write $n_b(x)\approx T/x$ and 
the polarization bubble can be rewritten as 

\begin{equation}
\label{picl}
\Pi(i\omega,{\bf p})=4m^2T(I_+-I_-),
\end{equation}
where

\begin{equation}
I_\pm=-i\int\frac{d^dq}{(2\pi)^d}\frac{1}{2m\omega+i(2{\bf p}\cdot{\bf q}
\pm p^2)}\frac{1}{q^2}.
\end{equation}
The integrals $I_\pm$ can be evaluated using the Feynman parameters 
\cite{KSF}, 

\begin{equation}
I_\pm=-i\int_0^\infty d\lambda_1\int_0^\infty d\lambda_2 
\frac{d^dq}{(2\pi)^d}e^{-\lambda_1(2m\omega\pm ip^2+2i{\bf p}\cdot{\bf q})}
e^{-\lambda_2 q^2}.
\end{equation}
After evaluating the Gaussian integral over ${\bf q}$ we 
obtain

\begin{eqnarray}
I_\pm&=&-\frac{i}{(2\pi)^{d}}\int_0^\infty d\lambda_1\int_0^\infty d\lambda_2 
\left(\frac{\pi}{\lambda_2}\right)^{d/2}e^{-\lambda_1(2m\omega
\pm ip^2)} e^{-p^2\lambda_1^2/\lambda_2}\nonumber\\
&=&\mp\frac{(\pm i)^{d-2}}{2^d\pi^{d/2}}\Gamma(d/2-1)\Gamma(3-d)
p^{d-4}\left(1\mp\frac{2mi\omega}{p^2}\right)^{d-3}.
\end{eqnarray}
Substituting the above expression back into (\ref{picl}) we obtain 
Eq. (\ref{FullPi}).

\end{appendices}

\end{document}